%% file: archive20240712.tex
\documentclass[12pt]{article}
\newcommand{\blind}{1}

\usepackage{calc}
\usepackage[dvipsnames]{xcolor}
\usepackage{amsmath, amsthm, amssymb}
\usepackage{algorithm}
\usepackage{algpseudocode}
\usepackage{spacing}
\usepackage{rotating}
\usepackage{subcaption}
\usepackage{graphics}
\usepackage{graphicx}
\usepackage{natbib,upgreek}
\usepackage{hyperref}
\iffalse
\usepackage{xr-hyper, hyperref}
\else
  \usepackage{hyperref, nameref, zref-xr}
  \zxrsetup{toltxlabel}
  \zexternaldocument*[supp-]{supp2024}
\fi
\newtheorem{theorem}{Theorem}%  meant for continuous numbers
\newtheorem{proposition}[theorem]{Proposition}%

\renewcommand{\baselinestretch}{1.5}

\begin{document}

\def\spacingset#1{\renewcommand{\baselinestretch}%
{#1}\small\normalsize} \spacingset{1}

\newcommand{\rrtitle}[1]{Exploring Differences between Two Decades of Mental Health
Related Emergency Department Visits by Youth via Recurrent Events Analyses}

\if1\blind {
  \title{\bf \rrtitle}
  \author{Yi Xiong\thanks{Correspondence to: Yi Xiong, \href{yxiong3@buffalo.edu}{yxiong3@buffalo.edu}} \hspace{.2cm}\\
    Department of Biostatistics, University at Buffalo\\
    and \\
    X.\ Joan Hu\\
    Department of Statistics and Actuarial Science, Simon Fraser
    University\\
    and\\
    Rhonda J.\ Rosychuk\\
    Department of Pediatrics, University of Alberta
    }
  \maketitle
} \fi

\bigskip
\begin{abstract}
We aim to develop a tool for understanding how the mental health of
youth aged less than 18 years evolve over time through
administrative records of mental health related emergency department
(MHED) visits in two decades. Administrative health data usually
contain rich information for investigating public health issues;
however, many restrictions and regulations apply to their use.
Moreover, the data are usually not in a conventional format since
administrative databases are created and maintained to serve
non-research purposes and only information for people who seek
health services is accessible. Analysis of administrative health
data is thus challenging in general. In the MHED data analyses, we
are particularly concerned with (i) evaluating dynamic patterns and
impacts with doubly-censored recurrent event data, and (ii) re-calibrating estimators developed based on truncated data by leveraging summary statistics from the population. The findings are
verified empirically via simulation. We have established the asymptotic properties
of the inference procedures.
The contributions of this paper are twofold. We present innovative
strategies for processing doubly-censored recurrent event data, and
overcoming the truncation induced by the data collection. In addition,
through exploring the pediatric MHED
visit records, we provide new insights into children/youths mental
health changes over time.
\end{abstract}

\noindent%
 \vfill

\maketitle
\newpage

\spacingset{1.9}
\section{Introduction}
%Background and Motivation, Literature Review, and
%Outline

\indent Children and youths are important to our society. There has been great interest
in understanding children and youths, for example their culture, development, and psychology
\citep[e.g.,][]{janssen1999psychological,damon2004positive}.
Mental health refers to a broad array of functions and poor mental health in children and youths can have negative immediate and long-term consequences \citep{Bitsko}.
If services for mental health care are limited, children and youths may seek care at emergency departments (EDs).
To understand how children/youths differ from one decade to the
other from the perspective of their age-varying mental health care trajectories, we herein collect and analyze records from
an administrative health database on mental health related
presentations to EDs made by children and youths
(aged younger than 18 years old at the time of the ED visit).

Two collections of pediatric MHED (Mental Health Emergency Department) visit
records, one from 2002--2010 and the other from 2010--2017,
are available to us by two different data extractions
from the population-based
administrative data sources: the National Ambulatory Care Reporting System
(\href{https://www.cihi.ca/en/national-ambulatory-care-reporting-system-metadata-nacrs}{NACRS})
and the Alberta Health Care Insurance Plan (\href{https://www.alberta.ca/ahcip.aspx}{AHCIP}).
The MHED data include both basic demographic characteristics
(e.g., \texttt{Sex} and \texttt{Region})
and the visit characteristics
(e.g., \texttt{Date} and \texttt{Diagnosis} of the ED visit, and {\texttt{Age} in integer years at the ED visit})
on all the MHED visits made by residents under 18 years old of the province
of Alberta, Canada. Viewing an ED visit as the event of interest, we
formulate the MHED records into recurrent event data. Analyses of the recurrent event data are conducted to compare mental
health related emergency department visits by children
and youths of the two decades with respect to the frequency in age and
how it is associated with risk factors/exposures such as \texttt{Sex} and geographic \texttt{Region}.
In the rest of this paper, the
two datasets are referred to as MHED data of ``early" and ``late"
groups, short for the decades starting in 2000 and 2010, respectively.

Administrative health data in general contain rich information for
investigating public health issues such as infectious diseases,
disease surveillance, and health service use, cross-sectionally and
longitudinally. These data can be particularly useful if the data
are population-based (i.e., including all residents in a defined geographic region), collected over time, at
individual-level, and accessible at low cost \citep{Burgun}. The
data may include records of physician visits, laboratory
investigations, health service use, diagnoses, and treatments
\citep{Gavrielov}. Disciplines like pharmacoepidemiology
\citep{Tamariz}, epidemiologic surveillance \citep{Ferrante} and
health services research \citep{Penfold} frequently use
administrative health data. Administrative health data can be more
cost-effective than resource-demanding studies to investigate, say,
trends in infectious diseases \citep{Gavrielov}. Recently,
administrative health data have been used to study trends in the
coronavirus disease 2019 (COVID-19) pandemic \citep{Navaratnam},
uptake of vaccinations \citep{Shariff}, and longer term outcomes
from COVID-19 \citep{Chevinsky}.

Because of the special features of Canada's health
care system, administrative health databases are created to
administer Canadian provincial and territorial health insurance plans; however,
many restrictions and regulations apply to their access and use.
In addition, these administrative health databases are developed and maintained
to serve non-research purposes. Those features yield challenges to analyses of
administrative health data in general. This paper presents strategies
for dealing with the particular challenges
confronted in our analyses of the MHED data. The ideas and methods
are in fact applicable to many other administrative data analyses.

Each data extraction of the two MHED datasets was administered with
calendar time. A scientifically meaningful time origin is usually
subject-specific. We focus on the use of age as the individual time
in the paper. Also, eligible subjects were limited to those aged less than 18 at the time of their ED visit and the data extraction provided only the age in integer years. Those aspects result in the first major challenge
that the information availability varies from individual to
individual. Specifically, the two MHED datasets include the
subjects' information over individual-specific intervals. That is,
the MHED records of a subject are subject to double-censoring; the
left-censoring and right-censoring times vary from subject to
subject. The structure of doubly-censored recurrent event data
\citep[e.g.][]{zhang1996linear, cai2003local} indicates a particular
demand of loosely-structured models to be employed in data analyses
such as the marginal models of recurrent events
\citep[e.g.][]{lawless1995some, lin2000semiparametric}. Since the
range of age differences in the MHED records within individual
subjects and among them is rather wide, we consider the extended Cox
regression model with age-varying coefficients for the counting
process of an individual's cumulative MHED visits over age. Using
the indicator of the late group (i.e. the 2010 group) as one of the covariates, the model
permits exploring age-dependent differences between the two decades
and other risk/exposure (covariate) effects on the MHED visit
frequency. Analyses under the model include all the corresponding
stratified analyses with the data from the two decades. We follow
\citet{hu2016marginal} to adapt the local partial likelihood
procedure of \citet{cai2003local} and \citet{tian2005cox} for
estimating the model parameters with the recurrent event data.

The second major challenge confronted when analyzing the MHED data
is the inherent truncation issue
of administrative health data. Administrative health data usually provide information only
on people who seek health services due to a
particular disease or condition, which is a subgroup of the general population. To provide administrators and policy-makers with an overall
view of the disease burden and trends of using health services,
interests are usually in the population as a whole on its measures such
as rates of using health services. Examinations of only
health services users for a particular disease or condition during a limited time period  form
a truncated sample of the whole population and
examinations of their data only lead to biased conclusions in general.
Most well-established approaches in analysis of truncated data
focus on the situations where observations on the time to an event are
subject to length-biased sampling; see, for example,
\citet{wang1996hazards}, \citet{shen2009analyzing}, and \citet{qin2011maximum}.
There is relatively less work concerning zero-truncated recurrent event data.

In our application, individuals in Alberta who are younger than 18 years old in the two decades, no matter if they are
included in the MHED cohort or not, may be framed as a random sample of the general population.
By the features of Canadian medical insurance system with population-based administrative databases, eligible subjects who are not included
in the pediatric MHED cohort can be safely assumed not to have MHED visits during the two time periods.
Augmented by that fact, each of the two MHED datasets can be viewed as a collection of doubly-censored
recurrent event records where the covariates of all the subjects not in the MHED cohort are missing.
There is considerably rich literature available on statistical
analysis of right-censored recurrent event data \citep{cook2007statistical}.
Most well-established statistical tools are not directly applicable, however,
to analysis of the MHED data when targeting the general population
since covariates of the individuals not in the cohort are unavailable.
\citet{hu1996estimation} considered a similar situation where the awareness of a study unit and the associated
observation period is unavailable
 until the study unit experiences at least one event of interest.
They proposed to handle the truncated data using information on the size of the total sample including study units not having events over the observation period
and  considered robust nonparametric estimators of the rate and mean function
for the process of recurrent events. We adapt this idea to tackle the truncation issue. Particularly, we aim to overcome the challenge of unknown
covariates, which is posed by the zero-truncated data.

Some approaches have recently been
proposed to utilize aggregated data or summary statistics such as disease incidence
rate and survival probability to correct bias when analyzing a smaller-scale
sample from the target population; see, for example, \citet{zheng2021risk}, \citet{gao2021noniterative}, and \citet{li2022distributed}.
Those publications consider the situations where either the covariate effects
are constant over time or effects are assumed to be the same within the source cohort and the target population.
We are interested in exploring time/age-varying effects of the covariates on the aforementioned MHED visits.
We analyze the MHED data, the recurrent event data, in a combination of publicly available population-based aggregated data to provide insights into the MHED visit patterns of
the general  population in addition to targeting on the pediatric MHED cohort,
the subpopulation whose MHED visit records are extracted and available to us.

More practical challenges were encountered when  the MHED data were analyzed.
Privacy laws prevented patient birthdates to be extracted from the AHCIP
during the early time, and thus the 2002--2010 data do not include the
birthdates of the subjects. This data feature partly prompted \citet{hu2016marginal} to develop a marginal
regression model to handle recurrent events with coarsened censoring times,
which resulted because the start date for an individual (i.e., birthdate)
was not known exactly but known to be contained in an interval
(i.e., based on the age at ED visit). We adapt their approach
in the analyses of this paper. Further, the two MHED datasets were extracted
at different times, and the data extractions did not allow unique personal
identifiers. Thus, the subjects with MHED visits in both the early and late datasets cannot be identified
and linked. Despite being aware of the fact,
we start with assuming that all of the individuals in the MHED data are independent
with each other in the analyses. We then conducted a simulation study
to examine the consequences if the assumption is violated.

The contributions of this paper are twofold. We present innovative
strategies for processing doubly-censored recurrent event data,
and overcoming the truncation induced by the data collection. In addition,
through exploring the pediatric MHED visit records, we provide new insights into how the
mental health of children and youth changes over time.
We organize the rest of this paper as follows. The notation and modeling
are introduced together with discussions on useful model specifications in
Section~\ref{sec:Framework}. Section~\ref{sec:PMHEDVCohort} presents
analyses of the aforementioned MHED data from 2002--2010
and 2010--2017 concerning the pediatric MHED cohort.
Section~\ref{sec:GeneralPopulation} shows strategies and analyses of the MHED data supplemented by aggregated population
information concerning the general population.
In Section~\ref{sec:Simulation}, we report
simulation studies conducted to numerically investigate
two potential issues in the data analyses of Sections~\ref{sec:PMHEDVCohort}
and \ref{sec:GeneralPopulation}.
Some final remarks are given in Section~\ref{sec:Discussion}.

\section{Notation and Modeling}\label{sec:Framework}
\indent This section describes the formulation of the MHED data and the modeling
considered in the data analyses.

Denote by $\mathcal{O}$ the whole sample of children and youths
who were under 18 years old and residents of Alberta, Canada
during 2002--2017; $\mathcal{O}_1$,
the sub-sample that only includes the subjects who made MHED visits
during the time, i.e., the MHED cohort. That yields $\mathcal{O}
=\mathcal{O}_1 \bigcup \mathcal{O}_0$, where $\mathcal{O}_0$ is
the sub-sample of the eligible subjects who had no MHED visit
during the time. Moreover, denote the early and late groups
(i.e., the 2000 decade and the 2010 decade groups)
by $\mathcal{O}_E$ and $\mathcal{O}_L$ and, consequently,
the MHED cohorts are $\mathcal{O}_{E1}$ and $\mathcal{O}_{L1}$.
Thus $\mathcal{O} =\mathcal{O}_E \bigcup \mathcal{O}_L$ and
$\mathcal{O}_1 =\mathcal{O}_{E1} \bigcup \mathcal{O}_{L1}$.

Let $N(a)$ represent the number of MHED visits made by a subject
up to age $a$, and $\mathcal{H}(a)=\{N(t): 0\leq t <a\}$
be the history information prior to age $a$.
We use $Z$ to denote the covariates of interest, which in this
paper are the two categorical variables \texttt{Sex} and \texttt{Region}.
Further, let $X$ be the indicator of  whether or not a subject belongs to
the late subgroup $\mathcal{O}_L$, and
$[W^{E}_L, W^{E}_R]$ and $[W^{L}_L, W^{L}_R]$ be
the data extraction windows (in calendar time)
of the early and late MHED data, respectively.
The left and right censoring times (in age) are then, respectively,
$C^E_L=\max(0, W^{E}_L-B)$ and $C^E_R=\min(18, W^{E}_R-B)$
for a subject from the early subgroup $\mathcal{O}_E$ with birthdate $B$;
$C^L_L=\max(0, W^{L}_L-B)$ and $C^L_R=\min(18, W^{L}_R-B)$,
for $\mathcal{O}_L$. The above variable names with subscript $i$ are the
 corresponding values associated with subject $i$.

Given that the extraction windows are known, the MHED data can be expressed as follows:
\begin{equation}\label{eq:obsPMHEDVdata}
\mbox{\sc Data}_1=\mbox{\sc Data}_{E1} \bigcup \mbox{\sc Data}_{L1}
= \Big[\bigcup_{i\in \mathcal{O}_{E1}} \mbox{\sc Data}_{E1,i}\Big]
\bigcup \Big[\bigcup_{i\in \mathcal{O}_{L1}} \mbox{\sc Data}_{L1,i}\Big],
\end{equation}
where $\mbox{\sc Data}_{E1,i}
=\Big\{N_i(a)-N_i(C_{Li}): ~ a\in(C^E_{Li}, C^E_{Ri}]\Big\}\bigcup\big\{Z_i\big\}$
and $\mbox{\sc Data}_{L1,i}= \Big\{N_i(a)-N_i(C_{Li}):
~ a\in(C^L_{Li}, C^L_{Ri}]\Big\}\bigcup\big\{B_i, Z_i\big\}$ are
the available data of subject $i$ from the early cohort $\mathcal{O}_{E1}$
and the late cohort $\mathcal{O}_{L1}$, respectively.
Note that $C^E_{Li}$ and $C^E_{Ri}$ are unavailable for
$i\in \mathcal{O}_{E1}$ due to the missing birthdate $B_i$ in our application.

In this paper, we consider statistical inference on the target population,
denoted by $\mathcal{P}^*=\mathcal{P}^*_{E} \bigcup \mathcal{P}^*_{L}$
in the following two settings:
the MHED cohort of the two decades $\mathcal{O}_1=\mathcal{O}_{E1}
\bigcup \mathcal{O}_{L1}$ is a random sample of $\mathcal{P}^*=\mathcal{P}_1$, or
the whole group of interest
$\mathcal{O}=\mathcal{O}_{E} \bigcup \mathcal{O}_{L}$
is a random sample of $\mathcal{P}^*=\mathcal{P}=\mathcal{P}_1\cup\mathcal{P}_0$.
In the rest of this paper, $\mathcal{P}^*$ together with
its subpopulations $\mathcal{P}_{E}^*$ and $\mathcal{P}_{L}^*$
is specified in the context.

We assume that, for age $a \in (0,18)$ and $i\in\mathcal{P}^*$,
\begin{equation}\label{eq:modelGeneral}
\mbox{E}\big[dN_i(a)|Z_i,X_i\big]=\exp\big\{\alpha(a) X_i
+ \beta(a)\rq{}Z_i +\gamma(a)\rq{} X_iZ_i \big\}d\Lambda_{0}(a),
\end{equation}
where $\Lambda_0(a)=\int_0^a \lambda_0(u)du$ with $\lambda_0(\cdot)$ the
baseline rate function. This model is an extension of the proportional rates/means model discussed by,
for example, \citet{pepe1993some},
\citet{lawless1995some}, and \citet{lin2000semiparametric} in their marginal
analysis of recurrent events.

%\subsection{Model specifications}\label{subsec:ModelSpecifications}

The model includes many useful specifications. For example,
it is the Poisson process model with the conditional intensity function
in the same form as (\ref{eq:modelGeneral}) if assuming the
counting process $\big\{N(a): a \geq 0\big\}$ is Poisson.
The Poisson process model reduces to the Andersen-Gill model when
all the regression coefficients are time-independent \citep{andersen1982cox}.
Table~\ref{tab:modelSpecialCase}  displays relevant
semiparametric marginal models derived from
(\ref{eq:modelGeneral}).

%%%%%%%%%%%% Formerly Table S1 %%%%%%%%%%%%%%%%%%%%

\begin{table}[htbp]
\centering
\caption{Relevant model specifications.}
\label{tab:modelSpecialCase}
\begin{tabular}{lcccp{10cm}}
\hline
\hline
&\multicolumn{3}{c}{Coefficient of}&\\
\cline{2-4}
Model     & $X$ & $Z$ & $XZ$ &\multicolumn{1}{c}{Model Feature} \\
\hline
 CCC &     $\alpha$ & $\beta$ & $\gamma$
& proportional means model with parallel baselines\\
 VCC &     $\alpha(\cdot)$ & $\beta$ & $\gamma$
& stratified proportional means model\\
 CVC &     $\alpha$ & $\beta(\cdot)$ & $\gamma$
& parallel baselines with constant difference in time-independent covariate effect between two groups \\
 CCV &     $\alpha$ & $\beta$ & $\gamma(\cdot)$
& parallel baselines with time-varying difference in time-independent covariate effect between two groups\\
VVC &     $\alpha(\cdot)$ & $\beta(\cdot)$ & $\gamma$
& stratified model with constant difference in time-dependent covariate effect between two groups \\
 VCV &     $\alpha(\cdot)$ & $\beta$ & $\gamma(\cdot)$
& stratified model with time-varying difference in time-independent covariate effect between two groups \\
 CVV &     $\alpha$ & $\beta(\cdot)$ & $\gamma(\cdot)$
& parallel baselines with time-varying difference in time-dependent covariate effect between two groups \\
 VVV &     $\alpha(\cdot)$ & $\beta(\cdot)$ & $\gamma(\cdot)$ & Model (\ref{eq:modelGeneral}): stratified model with time-varying difference in time-dependent covariate effect between two groups \\
\hline\hline
\end{tabular}
\end{table}

We note that the model is equivalent to the following stratified model. With
$\beta_E(a)=\beta(a)$,
$d\Lambda_{0E}(a)= d\Lambda_{0}(a)$,
$\beta_L(a)=\beta(a)+\gamma(a)$, and
$d\Lambda_{0L}(a)= \exp\big\{\alpha(a)\big\} d\Lambda_{0}(a)$,
\[
\mbox{E}\big[dN_i(a)|Z_i,X_i\big]=\left\{\begin{array}{lr}
\exp\big\{\beta_E(a)\rq{}Z_i\big\}d\Lambda_{0E}(a), & \mbox{for } i\in \mathcal{P}^*_E \\
\exp\big\{\beta_L(a)\rq{}Z_i\big\} d\Lambda_{0L}(a), & \mbox{for } i\in \mathcal{P}^*_L. \end{array}\right.
\]
This representation provides a meaningful interpretation of the regression coefficients of
(\ref{eq:modelGeneral}) in our application. More discussion is given
in the following sections. When interests are in knowing about
each of the subpopulations, one can conduct a stratified analysis
without specifying how the two subpopulations are correlated.
%\newpage

%\begin{sidewaystable}

%\newpage
\section{Analyses Concerning the Pediatric MHED Cohort}\label{sec:PMHEDVCohort}
\indent This section focuses on the MHED cohort.
The MHED data from the two decades are analyzed with the target population
$\mathcal{P}^*=\mathcal{P}_1$  including $\mathcal{O}_1=\mathcal{O}_{E1}\bigcup \mathcal{O}_{L1}$ as a random sample, where $\mathcal{O}_1$ is
defined in Section~\ref{sec:Framework}
as all the subjects in the MHED cohort from the early and late decades.
We begin with a descriptive analysis of the MHED data. A comparison of the MHED visits between the two decades is conducted
through analyses of the MHED data under the model in (\ref{eq:modelGeneral})
and its specifications given in Table~\ref{tab:modelSpecialCase}.

\subsection{Two Collections of MHED Visit Records}
The dataset from the earlier time includes $36,310$ MHED visits
made by $24,922$ subjects; the later time, $58,166$ from $33,299$.
About $75\%$ of the children and youths (hereafter just referred to as children)
had one ED visit during the early time period whereas $67\%$
had one ED visit during the late time period;
see Table~S1 for more
summary statistics of the data.

We plot the number of visits and the number of subjects in the MHED cohort over age
according to risk factors \texttt{Sex} (Figure S1) and \texttt{Region}
(Figure S2). Most visits were made by subjects aged
between 13 and 17 years old ($86.1\%$ in the early time period
and $84.0\%$ in the late time period). There were apparently more visits
in the late subcohort compared to the early one,
especially for subjects aged $>10$ years. Moreover, Figure~S1b
shows that, in either of the two decades, females had more MHED visits
than males as their ages increase.
Interesting findings about different regions can be seen from
the histograms of MHED visits at different ages in Figure~S2a and from the subject numbers in the cohort shown in Figure~S2b.
Alberta is divided into five geographic regions for the delivery of health services (North, Edmonton, Central, Calgary, South), with Edmonton and Calgary
constituting the major metropolitan areas and comprising about two-thirds of the population.
During both of the time periods, regions Calgary and Edmonton had fewer MHED
visits than all the other regions together, referred to as ``other regions"
in the rest of the paper. These two major metropolitan regions also had less subjects
in the cohort. In addition, we notice
an increased frequency in Calgary during the late time period.

The privacy regulation during the early time did not allow
extraction of birthdates and thus there is no birthdate information
available in the 2002--2010 pediatric MHED visit data. The age at
each MHED visit was recorded in integer years and thus each subject's
birthyear is available. \citet{hu2016marginal} propose to handle the
resulting data, recurrent events with coarsened censoring times, by
assuming each missing birthdate follows the uniform distribution
over the subject's birthyear. \citet{rosychuk2020handling} and
\citet{pietrosanu2021handling} examine this strategy numerically.
They have shown that the approach for handling coarsened age
information caused by the missing birthdates performed well. To
check for the uniform assumption, we plot the histogram of the
available birthdates in the 2010--2017 pediatric MHED visit dataset
in Figure~S3 of the Supplementary Material. It
indicates the assumption is acceptable in practice.

\subsection{Comparisons of MHED Visits by Children in Two Decades}\label{sec:compare2}

\subsubsection{Estimation of model parameters}
\indent We assume that all of the subjects in $\mathcal{O}_1$, the MHED cohort,
are independent from each other. We estimate
the age-varying covariate effects under the model given in
(\ref{eq:modelGeneral}) by
adapting the estimation procedure proposed by \citet{hu2016marginal}.

We denote the indicator of subject $i$ at age $a$ within the subject-specific observation window by
$Y_i(a)$, which is $I\big\{a\in\big(C^E_{Li},C^E_{Ri}\big]\big\}$
or $I\big\{a\in\big(C^L_{Li},C^L_{Ri}\big]\big\}$ for subject $i$
from the early or late subcohort, respectively.
That indicator can be expressed as $Y(a|B_i)=I\big\{
\max(0,W_L-B_i)<a\leq \min(18, W_R) \big\}$ with $W_L,W_R$ to be
$W_L^E,W_R^E$ for subject $i$ from the early subgroup, and
to be $W_L^L,W_R^L$ for subject $i$ from the late subgroup.
Provided all $B_i$ are available, the following set of estimating functions
lead to a consistent estimator of
$\theta(\cdot)=\big(\alpha(\cdot),\beta(\cdot)^{'},\gamma(\cdot)^{'}\big)^{'}$,
the age-varying coefficients to the covariates $V=(X,Z^{'},XZ^{'})^{'}$.

We choose constants $0<\tau_L, \tau_R<18$ such that
the left and right censoring times satisfy
$P(C_{L}<\tau_L)>0$ and $P(C_{R}>\tau_R)>0$
to avoid the boundary problem in the local likelihood estimation.
For a fixed $a\in[\tau_L,\tau_R]$, approximate $\theta(u)$ with $u\in(0,18)$
by the Taylor expansion until the first order:
$\theta(a)+\dot{\theta}(a)(u-a)$ with $\dot{\theta}(a)=d\theta(a)/da$.
Let $\phi(a)=\left(\theta(a)^{'},\dot{\theta}(a)^{'}\right)^{'}$
and
$V_i^{*}(u,a)=\left(V_i^{'},(u-a)V_i^{'}\right)^{'}$. Using a kernel function
$K(\cdot)$ and substituting $\theta(u)$ with
its linear approximation yields
\begin{equation} \label{eq:estmFnBcoef}
U(\phi(a);a\big|\mathbf{B})
= \int_{0}^{18} K_h(u-a) \sum_{i\in \mathcal{O}_1} Y_i(u)
\left\{V_i^*(u,a) - \bar{V}^*(\phi(a);u,a)\right\} dN_i(u),
~ a \in[\tau_L,\tau_R],
\end{equation}
where
\begin{equation}\label{eq:middle1}
S^{(q)}(\phi(a);u,a)=\sum_{i\in \mathcal{O}_1} Y_i(u)
\big[V_i^*(u,a)\big]^{\otimes q}
\exp\{\phi^{'}(a)V_i^*(u,a)\}
\end{equation}
for $A^{\otimes q}=1, A, AA^{'}$ with $q=0,1,2$, respectively,
and $\bar{V}^*(\phi(a);u,a)$ is\\
 $S^{(1)}(\phi(a);u,a)\big / S^{(0)}(\phi(a);u,a)$.

The first component vector of the solution to
$U(\phi(a);a\big|\mathbf{B})=0$,
denoted by $\hat{\theta}(a), a\in [\tau_L,\tau_R]$, estimates $\theta(\cdot)
=\big(\alpha(\cdot),\beta(\cdot)^{'},\gamma(\cdot)^{'}\big)^{'}$.
The estimation procedure is referred to as {\it Approach A} in
this paper. When the counting process $\big\{N(a): a>0\big\}$
is Poisson, the set of estimating functions (\ref{eq:estmFnBcoef})
is the corresponding local linear partial score function of $\phi(\cdot)$,
and $\hat{\theta}(\cdot)$ is then
the local linear maximum partial likelihood estimator
of $\big(\alpha(\cdot),\beta(\cdot)^{'},\gamma(\cdot)^{'}\big)^{'}$.
The following propositions establish the consistency and weak convergence
of $\hat{\theta}(\cdot)$ as the sample size goes to infinity.
\begin{proposition}
Under conditions (C1)-(C6) stated in Appendix, with $n_1=|\mathcal{O}_1|$ and the bandwidth
$h=O(n_1^{-\nu})$ where $1/2<\nu<1$, the estimator $\hat{\theta}(a)$
is a strongly consistent estimator of $\theta_0(a)$ with fixed age
$a$, and $\sqrt{hn_1}(\hat{\theta}(a)-\theta_0(a))$ converges in
distribution to a multivariate normal random variable with mean zero
and covariance matrix $AV(\theta_0(a))$ given by
${\Pi}({\theta_0}(a))^{-1}{\Sigma}(\theta_0(a)){\Pi}(\theta_0(a))^{-1}$,
where
$\Pi({\theta_0}(a))=\{\upsilon^{(2)}({\theta_0}(a))-\upsilon^{(1)}({\theta_0}(a))^{\otimes
2}/\upsilon^{(0)}({\theta_0}(a))\}\lambda_0(a)$ with
$\upsilon^{(q)}({\theta_0}(a))=\mbox{E}[Y_i(a)V_i^{\otimes
q}e^{\theta_0(a)\rq{}V_i}]$ for $q=0,1,2$, and
${\Sigma}(\theta_0(a))$ is the asymptotic variance of the first
component vector of $\sqrt{hn_1^{-1}}U(\phi_0(a);a|\mathbf{B})$.
\end{proposition}
Proof of Proposition 1 follows the arguments of \citet{hu2016marginal}.
A consistent variance estimator for $\hat{\theta}(a)$ is
then the sub-matrix of
$\hat{\Pi}(\hat{\phi}(a);a|\mathbf{B})^{-1}\hat{\Sigma}(\hat{\phi}(a);a|\mathbf{B})\hat{\Pi}(\hat{\phi}(a);a|\mathbf{B})^{-1}$
corresponding to the estimator, where
\[
\hat{\Pi}({\phi}(a);a|\mathbf{B})
=\int_{0}^{18} K_h(u-a)\sum_{i\in\mathcal{O}_1}Y_i(u)\left[\frac{S^{(1)}({\phi}(a);u,a)}{S^{(0)}({\phi}(a);u,a)}-\big\{\frac{S^{(1)}({\phi}(a);u,a)}{S^{(0)}({\phi}(a);u,a)}\big\}^{\otimes 2}\right]dN_i(u),
\]
and $\hat{\Sigma}(\hat{\phi}(a);a|\mathbf{B})$ can be estimated by $\sum_{i\in\mathcal{O}_1}\Big(Q_i(\hat{\phi}(a);a)-\bar{Q}(\hat{\phi}(a);a)\Big)^{\otimes 2}$ with
\[
Q_i({\phi}(a);a)=\int_{0}^{18}\sqrt{h} K_h(u-a)Y_i(u)[V_i-S^{(1)}({\phi}(a);u,a)\big/S^{(0)}({\phi}(a);u,a)]dN_i(u)
\]
and $\bar{Q}({\phi}(a);a)$ is the sample average of $Q_i({\phi}(a);a)$ for $i \in \mathcal{O}_1$.

The birthdates of the subjects in the early subcohort $\mathcal{O}_{E1}$
are unavailable. Thus the estimating functions in (\ref{eq:estmFnBcoef})
are not directly applicable with the resulting coasened censoring times and
event records in years of age.
We adopt the approach to the problem in \citet{hu2016marginal}
to obtain the following applicable estimating functions of $\theta(\cdot)$.
Specifically, assuming birthdates follow the uniform distribution throughout
a year, conditional on the available data $\mbox{\sc Data}_{E1,i}$ the birthdate of subject $i\in \mathcal{O}_{E1}$
follows the uniform distribution over the interval
$\big(W^E_L-18, W^E_R\big] \bigcap \Big\{
\bigcap_{j=1}^{N_i} \big(T_{ij}-\big\lfloor A_{ij} \big\rfloor -1,
    T_{ij}- \big\lfloor A_{ij} \big\rfloor \big] \Big\}$
with $T_{ij}$ and $\big\lfloor A_{ij} \big\rfloor$ the recorded calendar time and age
in year at subject $i$'s $j$th recorded MHED visit, respectively.
Denote the conditional distribution of $B_i$ by $G_i(\cdot)$ for $i \in\mathcal{O}_{E1}$.
Let $\tilde{Y}_i(a)= \int Y_E(a\big|b) dG_i(b)$.
Plugging $\tilde{Y}_i(\cdot)$ for $i \in \mathcal{O}_{E1}$
in the estimating functions (\ref{eq:estmFnBcoef})
leads to a set of feasible estimating functions with the MHED data.
One may approximate
$\tilde{Y}_i(a)$ in the new estimating functions by
$\tilde{Y}_{iK}(a) =\sum_{k=1}^K Y_E(a\big|b_{ik})\big/K$
using a random sample of $b_{ik}, k=1,\ldots,K$ from $G_i(\cdot)$
with sufficiently large $K$. We refer that procedure to as
Hu-Rosychuk Approach B.

Moreover, adapting the Breslow estimator \citep{breslow1972discussion},
an estimator of the baseline $\Lambda_{0}(\cdot)$ can be
$\hat{\Lambda}_{0}(a)=\hat{\Lambda}_{0}(a;\hat{\theta}(\cdot))$, where
\begin{equation}\label{eq:baselineEstm}
\hat{\Lambda}_{0}(a;\theta(\cdot))
        = \int_{\tau_L}^a\frac{\sum_{i\in \mathcal{O}_{E1}} \tilde{Y}_i(u) dN_i(u)
+\sum_{i\in \mathcal{O}_{L1}} Y_i(u) dN_i(u)}{
        \sum_{l\in \mathcal{O}_{E1}} \tilde{Y}_l(u)\exp\{\theta(u)^{'}V_l\}
+\sum_{l\in \mathcal{O}_{L1}} Y_l(u)\exp\{\theta(u)^{'}V_l\} }.
\end{equation}
We plug in \eqref{eq:baselineEstm} with $\theta(u)=\theta(\tau_L)$ if $a\in(0,\tau_L)$ and $\theta(u)=\theta(\tau_R)$ if $a\in(\tau_R,18)$.

In the Supplementary Material, Table S2
presents two sets of estimates and standard errors with the
2010--2017 MHED data for the age-independent effects under Model
CCC: one applies {\it Approach A} using the available
birthdates, and the other adapts the approach of
\citet{hu2016marginal} without using the birthdate information,
referred to as {\it Approach B} in this paper. The corresponding estimates in the two
sets are very close to each other. This result indicates {\it Approach B}
is plausible in our application. We applied
it in all of the analyses of
2002--2010 MHED data presented in this paper.

The above estimation procedure can be straightforwardly adapted to
the situations under the relevant special cases of
model~(\ref{eq:modelGeneral}) listed in Table~\ref{tab:modelSpecialCase}.
We implemented the estimation procedures with the MHED data using codes written in Rcpp~\citep{dirk}. We used two months
as a time unit and set the bandwidth to be $h=6$ units. The Epanechnikov
kernel function, $K(u)=3(1-u^2)/4,-1\leq u \leq 1$ is used to evaluate the regression coefficients.

\subsubsection{Analysis outcomes}
Table~\ref{table:AGmodelAll} presents the estimates and
the associated standard error estimates of the age-independent regression coefficients
under the models listed in Table~\ref{tab:modelSpecialCase}.
With all of the models assuming the marginal difference between
the two groups is age-independent (the Models CEE listed in
Table~\ref{tab:modelSpecialCase}, where E stands for either C or V) in
the presence of the covariates, the estimates indicate the increase
in ED visits during 2010--2017 is significant as $\alpha$ in all of the models
is significantly positive from $0$ based on the analyses.
Based on the estimates of the regression coefficients for \texttt{Sex} and \texttt{Region}, we
see that the ED visit frequency associated with male subjects is overall significantly
lower compared to female subjects; Edmonton has higher visit frequencies than Calgary and Other regions.
The ED visit patterns of the two decades appear not to differ from each other
much in their associations with either
male vs.\ female or among regions when assuming the associations are
age-independent throughout the age of over $(0,18)$ except in their associations with male vs.\ female (i.e., Models ECC). We also report the AIC values to compare different models. The models with age-varying effects have smaller AIC than the purely age-independent model (i.e., Model CCC), suggesting modeling with age-varying coefficients is more appropriate.
\begin{table}[!h]
\caption{Estimates of age-independent covariate effects
and standard errors under models listed in Table~\ref{tab:modelSpecialCase},
concerning the MHED cohort.}
\centering
%\hspace{-.5cm}%{-2cm}
%\addtolength{\leftskip} {-2cm}
%\addtolength{\rightskip}{-2cm}
\label{table:AGmodelAll}
%\label{table:AGmodelPart2d}
\scalebox{.75}{
\begin{tabular}{@{}lcr@{ }rcr@{ }rcr@{ }rcr@{ }rcr@{ }rcr@{ }rcr@{ }rcr@{ }r@{}}
\hline
\hline
%\multirow{3}{*}{Covariate} &\multicolumn{14}{c}{Models$^\ast$}\\
Models$^\ast$
&& \multicolumn{2}{c}{CCC} &&\multicolumn{2}{c}{VCC}&&\multicolumn{2}{c}{CVC}
&&\multicolumn{2}{c}{CCV} &&\multicolumn{2}{c}{VVC }&&\multicolumn{2}{c}{VCV}
&&\multicolumn{2}{c}{CVV} &&\multicolumn{2}{c}{VVV} \\
 \cline{3-25}
Covariate&&  Est$^\dagger$   & SE$^\dagger$    && Est   & SE
&& Est   & SE &&  Est   & SE    && Est   & SE
&& Est   & SE &&  Est   & SE    && Est   & SE
\\ \hline
\multicolumn{25}{l}{Decade (reference $=$  Early)}\\
$\quad$ Late&&\bf .282&.020 && --$^\ddagger$& --&& \bf.263&.016&&\bf.281&.014 && --& --&& --& --&&\bf.270&.011&& --& --\\
\multicolumn{25}{l}{Sex (reference $=$  Female)}\\
$\quad$ Male  &&  \bf -.049& .016 &&\bf-.049 & .015&& --& --&&\bf-.035&.011&& --& --&&\bf-.035 &.011&& --& --&& --& --\\
\multicolumn{25}{l}{Region (reference $=$  Other)}\\
$\quad$ Edmonton && \bf.050 & .021 &&\bf .056 & .019   && --& --&&\bf .052& .015&& --& --&&\bf.056&.015&& --& --&& --& -- \\
$\quad$ Calgary  && .003 & .019&&  .007& .018&& --& --&&.001&.013&& --& --&&.003&.013&& --& --&& --& --  \\
Decade$\times$Sex&&\bf -.058&.023&& \bf-.055&.020&&-.022&.016&& --& --&&-.020&.014&& --& --&& --& --&& --& --  \\
Decade$\times$Edmonton&&.003&.029&&.001&.025&&.007&.020&& --& --&&.007&.018&& --& --&& --& --&& --& --\\
Decade$\times$Calgary&&.015&.026&&.015&.022&&.014&.018&& --& --&&.005&.015&& --&--&& --& --&& --& --\\
\hline
\multicolumn{25}{l}{Ratio of Measures of Goodness-of-Fit (using model CCC as the reference)}\\
\cline{3-25}
log-likelihood&&
     \multicolumn{2}{c}{1$^{\dagger\dagger}$} &&\multicolumn{2}{c}{9.841e-3} && \multicolumn{2}{c}{9.842e-3} && \multicolumn{2}{c}{9.843e-3} && \multicolumn{2}{c}{9.832e-3} && \multicolumn{2}{c}{9.832e-3} && \multicolumn{2}{c}{9.841e-3}&& \multicolumn{2}{c}{9.832e-3} \\
     AIC$^\S$&& \multicolumn{2}{c}{1$^{\ddagger\ddagger}$} &&  \multicolumn{2}{c}{9.844e-3} &&  \multicolumn{2}{c}{9.844e-3} &&  \multicolumn{2}{c}{9.845e-3} &&  \multicolumn{2}{c}{9.833e-3} &&  \multicolumn{2}{c}{9.834e-3} &&  \multicolumn{2}{c}{9.843e-3} &&  \multicolumn{2}{c}{9.833e-3} \\
     \hline
     \hline
\multicolumn{25}{l}{$\ast$as defined in
Table~\ref{tab:modelSpecialCase}}\\
\multicolumn{25}{l}{$\dagger$Est: estimate of the age-independent regression coefficient; SE:
estimate of the standard error of the parameter estimator}\\
\multicolumn{25}{l}{$\ddagger$Estimates of age-varying regression coefficients and associated standard error estimates are plotted in the Supplementary Material.}\\
\multicolumn{25}{l}{$\S$AIC$=2\nu-2\Big(l_n\big(\hat{\theta}(\cdot)|\mathbf{B}\big)\Big)$, where $l_n\big(\hat{\theta}(\cdot)|\mathbf{B}\big)$ is the log-likelihood and $\nu$ is the effective number of parameters with }\\
\multicolumn{25}{l}{$\quad\nu=tr[\Big(l\rq{}\rq{}_n\big(\hat{\theta}(\cdot)|\mathbf{B}\big)\Big)^{-1}l\rq{}\rq{}_n\big(\hat{\theta}(\cdot)|\mathbf{B}\big)]$
following \citet{li2008variable}. }\\
\multicolumn{25}{l}{$\dagger\dagger$ The log-likelihood for model CCC is -103,954,536.}\\
\multicolumn{25}{l}{$\ddagger\ddagger$ The AIC for model CCC is 207,909,086.}
\end{tabular}}
\end{table}

Figure S4 presents the local linear estimates of $\alpha(\cdot)$ along with the approximated $95\%$ point-wise confidence
intervals with each of Models VEE of Table~\ref{tab:modelSpecialCase}, which have age-varying coefficients of the decade indicator $X$ and either age-independent or
age-varying coefficients of $Z$ and $XZ$. For comparison,
added in each plot is the estimate of $\alpha$ with the corresponding Model CEE, which assumes the coefficient for the decade indicator is age-independent.
All of the estimates of $\alpha(\cdot)$ indicate an overall increase of the late decade in ED visits throughout the age of 0--18
after adjusting for the factors of \texttt{Sex} and \texttt{Region}. This finding agrees to what we observed from the analysis
assuming the difference between the
two decade groups is constant. It shows how the two decade groups differ from each other over ages.
Specifically, ED visits made at ages 13 to 16 notably increase in the late decade.
The estimated cumulative baselines under Model CCC (the proportional means model) and Model VVV (model \eqref{eq:modelGeneral})
are plotted in Figure S5. As expected, the two estimated cumulative baselines are close.

\begin{figure}[!h]
\includegraphics[scale=0.31]{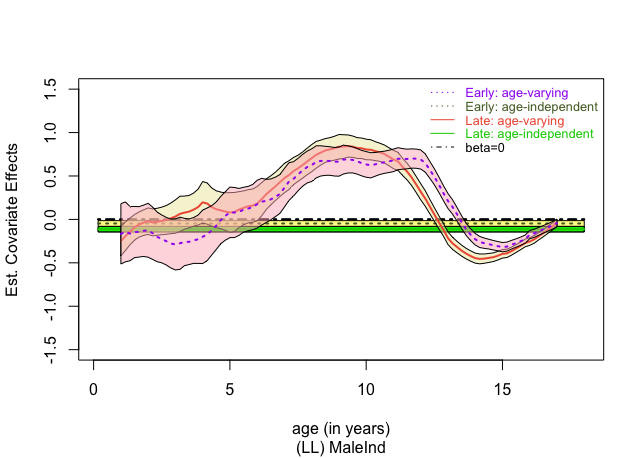}
\includegraphics[scale=0.31]{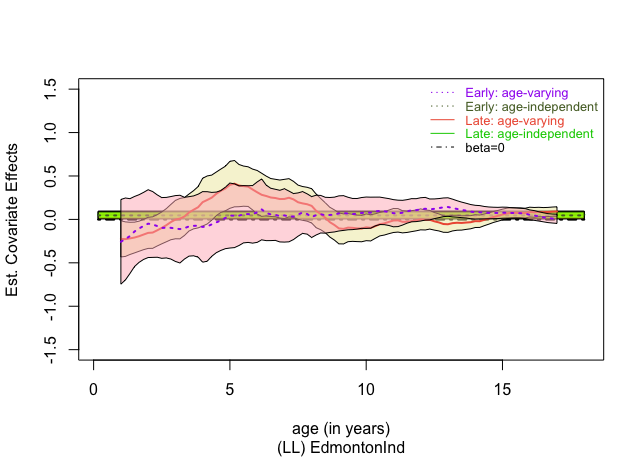}
\includegraphics[scale=0.31]{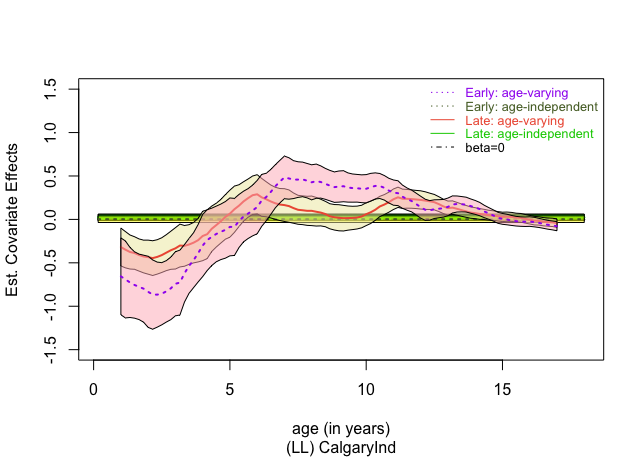}
        \caption{Estimates for the covariate effects for the early and late groups concerning the MHED cohort.}
      \label{fig:estbED}
\end{figure}

The estimated functions $\beta(\cdot)$ and
$\beta(\cdot)+\gamma(\cdot)$ with Model VVV together with
the estimated constant $\beta$ with Model CCC are shown in
Figure \ref{fig:estbED}. The estimated age-varying effects,
$\hat{\beta}(\cdot)$ and
$\hat{\beta}(\cdot)+\hat{\gamma}(\cdot)$, reveal how the ED visits
associated with male vs.\ female and different regions vary over
subject age $0$ to $18$ years in early and late decades, respectively. The curves in Figure \ref{fig:estbED}
indicate that the associations in the two decade groups are
similar except for the ones associated with \texttt{Sex}. The
difference between male vs.\ female in ED visits varies over age and
the curves for both early and late periods are different from constant. This discrepancy suggests
that the difference between male vs.\ female in ED visits is not
age-independent for either of the two decade groups. We also observe
the dotted curve for the late subgroup declines at a younger age.
That indicates an increase in ED visits for girls
starts at a younger age in the late time period. We notice
similar findings from the local constant estimates of the regression
coefficients, which are presented in
Figures~S7--S9 of the
Supplementary Material.

Since the unique identifiers of the cohorts' subjects are not disclosed, in the analysis reported above we assumed that the two
decade groups are independent from each other. The two subcohorts likely overlap: there are subjects who belong to both decade groups.
That assumption can be inappropriate and may result in under-estimated variance of
the proposed estimator. A simulation study, reported in Section~\ref{sec:Simulation}, was conducted
to examine the consequences of the assumption. It indicates the inference under the assumption can be quite robust when the overlap of the two
groups is small relative to the sizes of the groups themselves.

\section{Analyses Concerning the General Population}\label{sec:GeneralPopulation}

The aforementioned data provide information only
on the individuals in the pediatric MHED cohort, who were Alberta residents, under 18 years of age and experienced
at least one MHED visit during 2002--2017. Therefore, the analysis reported in Section~\ref{sec:PMHEDVCohort} cannot reveal the frequency and associations with the factors/exposures of
the ED visits concerning the whole province.
We thus conducted alternative analyses of the MHED data
following the approach proposed by \citet{hu1996estimation} to adjust for the truncation.
The key idea is to make use of the fact that eligible subjects who are not included
in the MHED cohort must not have had any MHED visit during the two time periods.
This assumption is plausible due to the single-payer government medical insurance
system in Canada and that population-based administrative databases are used in the administration of health services. The MHED data augmented by that information can then be framed into
a collection of doubly-censored recurrent event data where all of the covariate
data of the subjects not included in the MHED cohort are missing.
The following presents an analysis of the MHED data integrated with publicly available
population census information.
\subsection{Alberta Census Information}\label{sec:ABcensus}
The Alberta population information available to us
is in the form of all of the counts of Alberta residents in each year
during 2002--2017 belonging to each of the Alberta subpopulations defined
by \texttt{Sex} (male vs.\ female), \texttt{Region} (Edmonton, Calgary, Other),
and age subgroups $(0,1),[1,2),\cdots,[17,18)$ years old.

Table~S3 in the Supplementary Material presents the numbers of Alberta residents younger than 18 years old during 2002--2017.
Compared with the numbers of male and female subjects who had recorded
MHED visits in Table~S1,
the counts of males and females are
more equally distributed across Alberta: there are slightly
more male than female Alberta residents
while there are more females in the MHED cohort.
Figure S11a presents the numbers of males and females
of different age groups in the early and the late decades;
Figure S11b, residents in Edmonton, Calgary,
and Other regions.
Calgary has the largest population size in recent years, with many more residents in the pre-school age group compared to Edmonton and
the Other regions. However, the number of subjects with records of ED visits
in Calgary is less than in the Other region; see Figure S2b.
 More ED visits in other regions can be due to limited access to mental health care services in rural areas and families may be more likely to seek
 help at the ED.

\subsection{Comparison of MHED Visits Made by Children in Two Decades}\label{sec:comparison2dec}

\subsubsection{Estimation of model parameters} \label{sec:estmodparams}
We now take the target population $\mathcal{P}^*=\mathcal{P}$, which includes $\mathcal{O}=\mathcal{O}_1 \bigcup \mathcal{O}_0$, all of the Alberta
residents under age 18 throughout 2002--2017
(i.e., born over the time period of $(W_L^E-18,W_R^L]$) as a random sample.

We assume all of the subjects in $\mathcal{O}$ are independent from each other.
Extending the approach of \citet{hu1996estimation},
we estimate the age-varying covariate effects and the baseline function
under model (\ref{eq:modelGeneral}) or its variants listed in
Table~\ref{tab:modelSpecialCase} based on the MHED data of subjects
in the cohort $\mathcal{O}_1$: $\mbox{\sc Data}_1$ (\ref{eq:obsPMHEDVdata}).
Specifically, we first integrate the MHED data $\mbox{\sc Data}_1$
with the fact that individuals in $\mathcal{O}_0$ experienced
zero MHED visits throughout $[W_L^E,W_R^L]$:
\begin{equation}\label{eq:infoP0}
\mbox{\sc Data}_0=\bigcup_{i\in \mathcal{O}_0}\big\{N_i(a)-N_i(C_{Li})\equiv 0: a \in (C_{Li},C_{Ri}]\big\},
\end{equation}
where $C_{Li}=\max(0,W_L^E-B_i)$ and $C_{Ri}=\min(18,W_R^L-B_i)$ with
missing $B_i$. The aforementioned available sizes of the subpopulations
are used to further augment the MHED cohort data $\mbox{\sc Data}_1$
as given in (\ref{eq:obsPMHEDVdata}).

Let $V^*(X,Z;u,a)= \big(V^{'},(u-a)V^{'}\big)^{'}$
for $a \in [\tau_L,\tau_R]$ and $u\in(0,18)$ with
the covariates $V=(X,Z^{'},XZ^{'})^{'}$ introduced in Section~\ref{sec:PMHEDVCohort}. Here
$X$ is the categorical variable \texttt{Decade} with two levels, early
and late decade for time periods 2002--2010 and 2010--2017,
and $Z$ includes component \texttt{Sex} with female and male categories and
component \texttt{Region} with its three levels for the aforementioned
three Alberta regions: Edmonton, Calgary, and the other areas (Other).
Further, let $M(d,g,r,\big\lfloor u\big\rfloor)$ be
the number of individuals aged $\big\lfloor u\big\rfloor$ years in
the subpopulations defined by \texttt{Decade} $d$, \texttt{Sex} $g$ and \texttt{Region} $r$.
When applying the estimating functions in (\ref{eq:estmFnBcoef}) to the
target population $\mathcal{P}^*=\mathcal{P}$, we approximate
$U(\phi(a);a\big|\mathbf{B})$ using
\begin{equation} \label{eq:estmFnBcoefGP}
\tilde{U}(\phi(a);a\big|\mathbf{B})
= \int_{0}^{18} K_h(u-a) \sum_{i\in \mathcal{O}_1} Y_i(u)
\left\{V_i^*(u,a) - \tilde{\bar{V}}^*(\phi(a);u,a)\right\} dN_i(u),
%~\mbox{for} ~
a \in[\tau_L,\tau_R]
\end{equation}
if $B_i$'s are available for $i\in \mathcal{P}_1$,
where $\tilde{\bar{V}}^*(\phi(a);u,a)$ is the approximation to
$\bar{V}^*(\phi(a);u,a) =S^{(1)}(\phi(a);u,a)\big / S^{(0)}(\phi(a);u,a)$
of (\ref{eq:estmFnBcoef}) with the denominator and numerator replaced respectively by
\begin{equation}\label{eq:middle2}
\tilde{S}^{(j)}(\phi(a);u,a)=\sum_{d}\sum_{g}\sum_{r}
M(d,g,r,\big\lfloor u\big\rfloor)
(V^{*})^{j} \exp\{\phi^{'}(a)V^*)\}\big|_{V^*=V^*(d,g,r;u,a)}
\end{equation}
for $j=0,1$.
The approximation may work very well when the sizes of
the subpoulations within each age year group are similar
and the ages of the subjects within their groups
are close to evenly distributed.

Denote the solution to $\tilde{U}(\phi(a);a\big|\mathbf{B})=0$
by $\tilde{\phi}(a)$ for $a \in[\tau_L,\tau_R]$. We then use the first
component vector of $\tilde{\phi}(a)$, denoted by $\tilde{\theta}(a)$, as
the estimator for $\theta(a)$. The consistency and weak convergence
of $\tilde{\theta}(\cdot)$ are stated in the following proposition.
\begin{proposition}
Under conditions (C1)--(C6) stated in Appendix and provided that $\tilde{S}^{(q)}(\phi(a);u,a)$ satisfy the condition (AC), the estimator $\tilde{\theta}(a)$ is a strongly consistent estimator of $\theta_0(a)$ with fixed age $a$,
 and \\
 $\sqrt{hn}(\tilde{\theta}(a)-\theta_0(a))\overset{d}\rightarrow MVN\big(0,{\Pi}({\theta_0}(a))^{-1}{\Sigma}(\theta_0(a)){\Pi}(\theta_0(a))^{-1}\big)$,
where $n=|\mathcal{O}|$ and the bandwidth $h=O(n^{-\nu})$ with $1/2<\nu<1$.
\end{proposition}
The asymptotic derivation follows arguments in \citet{hu2016marginal}.
A proof of this proposition is outlined together with
the required conditions in the  Appendix.
By Proposition 2, we can estimate the variance of $\tilde{\theta}(a)$
by the corresponding component matrix of $\tilde{\Pi}(\tilde{\phi}(a);a|\mathbf{B})^{-1}\tilde{\Sigma}(\tilde{\phi}(a);a|\mathbf{B})\tilde{\Pi}(\hat{\phi}(a);a|\mathbf{B})^{-1}$, where
\[
\tilde{\Pi}({\phi}(a);a|\mathbf{B})=\int_{0}^{18}\!\!\! \!\!\! K_h(u-a)\sum_{i\in\mathcal{O}_1}Y_i(u)\left[\frac{\tilde{S}^{(1)}({\phi}(a);u,a)}{\tilde{S}^{(0)}({\phi}(a);u,a)}-\big\{\frac{\tilde{S}^{(1)}({\phi}(a);u,a)}{\tilde{S}^{(0)}({\phi}(a);u,a)}\big\}^{\otimes 2}\right]dN_i(u),
\]
and $\tilde{\Sigma}(\tilde{\phi}(a);a|\mathbf{B})$ can be estimated by $\sum_{i\in\mathcal{O}_1}\Big(\tilde{Q}_i(\tilde{\phi}(a);a)-\bar{\tilde{Q}}(\tilde{\phi}(a);a)\Big)^{\otimes 2}$ with\\
$\tilde{Q}_i({\phi}(a);a)=\int_{0}^{18}\sqrt{h} K_h(u-a)Y_i(u)[V_i-\tilde{\bar{V}}^*(\phi(a);u,a)]dN_i(u)$ and $\bar{\tilde{Q}}({\phi}(a);a)$ is
 the sample average of $\tilde{Q}_i({\phi}(a);a)$ for $i \in \mathcal{O}_1$.

Applying {\it Approach B} for dealing with missing
birthdates, we can then obtain an estimator
for $\theta(\cdot)$ with the MHED data. The baseline $\Lambda_0(\cdot)$
can then be estimated by adapting (\ref{eq:baselineEstm}) to
the current setting. In addition,
the approximation of $S^{(q)}(\phi(a);u,a)$ given
in (\ref{eq:middle2}) assumes the count of the individuals aged $u$
in the subpopulations defined by \texttt{Decade}, \texttt{Sex} and \texttt{Region}
is the same as it is throughout the age year. That assumption is appropriate for our application. A simulation study, reported
in Section~\ref{sec:Simulation}, was conducted to examine performance of the
estimators of
$\theta(\cdot)=\big(\alpha(\cdot),\beta(\cdot),\gamma(\cdot)\big)^{'}$
and $\Lambda_0(\cdot)$.

\subsubsection{Analysis outcomes}

Table \ref{tab:AGmodelAll2}
summarizes estimates for all of the age-independent regression coefficients with the models listed in Table~\ref{tab:modelSpecialCase}.
It shows a significant increase of ED visits in the late decade overall with adjustments for the exposures. Overall, the visits associated with males
appeared significantly less compared to females. The difference was further increased in the late decade.
The number of children from the other regions who experienced MHED visits was significantly higher than from either Edmonton or Calgary overall. Moreover,
the number of ED visits in the other regions was significantly more than in Edmonton, but less than in Calgary.
The corresponding local linear estimates for the age-varying regression coefficients are provided in
Section B of the Supplementary Material. The findings from the age-varying estimates are consistent
with the findings based on the models with constant exposure effects but more informative.
We plot the estimated cumulative baseline rates under Model CCC and Model VVV in Figure S17
and they are also close to each other.
\begin{table}[!htbp]
\caption{Estimates of age-independent covariate effects
and standard errors under models listed in Table~\ref{tab:modelSpecialCase},
concerning the general population.}
\centering
%\addtolength{\leftskip} {-2cm}
%\addtolength{\rightskip}{-2cm}
\label{tab:AGmodelAll2}
\scalebox{.72}{
\begin{tabular}{lcr@{ }rcr@{ }rcr@{ }rcr@{ }rcr@{ }rcr@{ }rcr@{ }rcr@{ }rcr@{ }r}
\hline
\hline
%\multirow{3}{*}{Covariate} &\multicolumn{14}{c}{Models$^\ast$}\\
Models
&& \multicolumn{2}{c}{CCC} &&\multicolumn{2}{c}{VCC}&&\multicolumn{2}{c}{CVC}
&&\multicolumn{2}{c}{CCV} &&\multicolumn{2}{c}{VVC }&&\multicolumn{2}{c}{VCV}
&&\multicolumn{2}{c}{CVV}&&\multicolumn{2}{c}{VVV}  \\
 \cline{3-25}
Covariate
&&  Est   & SE    && Est   & SE && Est   & SE && Est   & SE
&& Est   & SE && Est   & SE && Est   & SE && Est   & SE
\\ \hline
\multicolumn{25}{l}{Decade (reference $=$  Early)}\\
$\quad$ Late&&\bf  .605 & .020 && --& --&& \bf.599 & .016&&\bf.604 & .011 && --& --&& --& --&&\bf.590 & .011 && --& --\\
\multicolumn{25}{l}{Sex (reference $=$  Female)}\\
$\quad$ Male   &&  \bf -.414 & .016 &&\bf-.401 & .015&& --& --&&\bf-.410 & .013&& --& --&&\bf-.380 & .011&& --& -- && --& --\\
\multicolumn{25}{l}{{Region} (reference $=$  Other)}\\
$\quad$  Edmonton && \bf-.227 & .021 &&\bf -.206 & .019   && --& --&&\bf -.225 & .015&& --& --&&\bf-.211 & .015&& --& --  && --& --\\
$\quad$  Calgary  &&\bf-.386 & .019&&  \bf-.372 & .018&& --& --&&\bf-.410 & .013&& --& --&&\bf-.396 & .013&& --& --   && --& --\\
Decade$\times$Sex&&\bf -.124 & .023&& \bf-.141 & .020&& \bf-.097 & .016&& --& --&&\bf-.103 & .014&& --& --&& --& --   && --& --\\
Decade$\times$Edmonton&& \bf-.080 & .029&& \bf-.104 & .025&&\bf-.072 & .020&& --& --&&\bf -.089 & .019&& --& --&& --& -- && --& --\\
Decade$\times$Calgary&&\bf.196 & .026&& \bf.167 & .022&&\bf.171 & .018&& --& --&&\bf.145 & .015&& --& -- && --& -- && --& --\\
\hline
\multicolumn{25}{l}{Ratio of Measures on Goodness-of-Fit (using model CCC as the reference)}\\
\cline{3-25}
log-likelihood&&
     \multicolumn{2}{c}{1$^{\dagger}$} && \multicolumn{2}{c}{9.852e-3} && \multicolumn{2}{c}{9.856e-3} && \multicolumn{2}{c}{9.841e-3} && \multicolumn{2}{c}{9.839e-3} && \multicolumn{2}{c}{9.820e-3} && \multicolumn{2}{c}{9.851e-3} && \multicolumn{2}{c}{9.813e-3}  \\
AIC&& \multicolumn{2}{c}{  1$^{\ddagger\ddagger}$} && \multicolumn{2}{c}{9.853e-3} && \multicolumn{2}{c}{9.857e-3} && \multicolumn{2}{c}{9.842e-3} && \multicolumn{2}{c}{9.840e-3} && \multicolumn{2}{c}{9.820e-3} && \multicolumn{2}{c}{9.850e-3} && \multicolumn{2}{c}{9.819e-3} \\
     \hline
     \hline
%\multicolumn{25}{l}{$\ast$Models as defined in
%Table~\ref{tab:modelSpecialCase}}\\
%\multicolumn{25}{l}{$\dagger$Est: estimate of the parameter; SE:
%estimate of the standard error of the parameter estimator}\\
%\multicolumn{25}{l}{$\ddagger$Age-varying regression parameter}\\
%\multicolumn{25}{l}{$\S$AIC$=2\nu-2\Big(l_n\big(\hat{\theta}(\cdot)|\mathbf{B}\big)\Big)$, where $l_n\big(\hat{\theta}(\cdot)|\mathbf{B}\big)$ is the log-likelihood and $\nu$ is the effective number of parameters with }\\
%\multicolumn{25}{l}{$\quad\nu=tr[\Big(l\rq{}\rq{}_n\big(\hat{\theta}(\cdot)|\mathbf{B}\big)\Big)^{-1}l\rq{}\rq{}_n\big(\hat{\theta}(\cdot)|\mathbf{B}\big)]$. }\\
\multicolumn{25}{l}{$\dagger$ The log-likelihood for model CCC is -133,280,640.}\\
\multicolumn{25}{l}{$\ddagger\ddagger$ The AIC for model CCC is 266,561,294.}
\end{tabular}}
\end{table}

\begin{figure}[!h]
\includegraphics[scale=0.31]{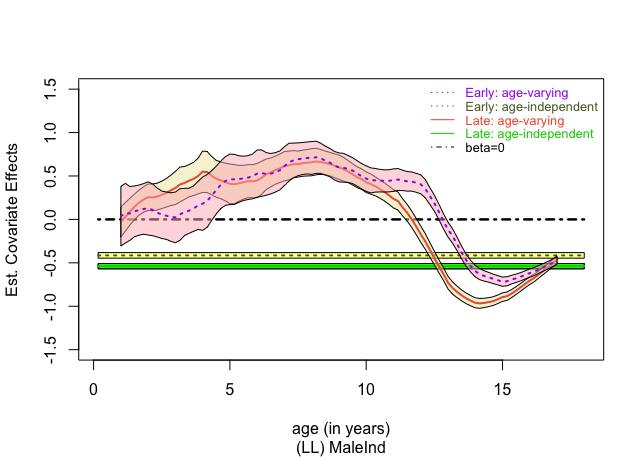}
\includegraphics[scale=0.31]{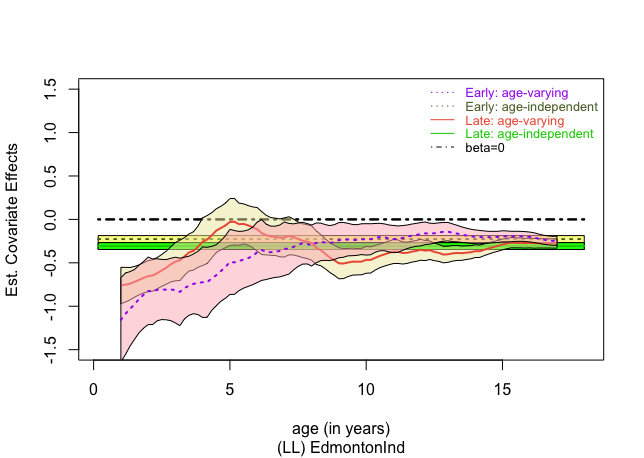}
\includegraphics[scale=0.31]{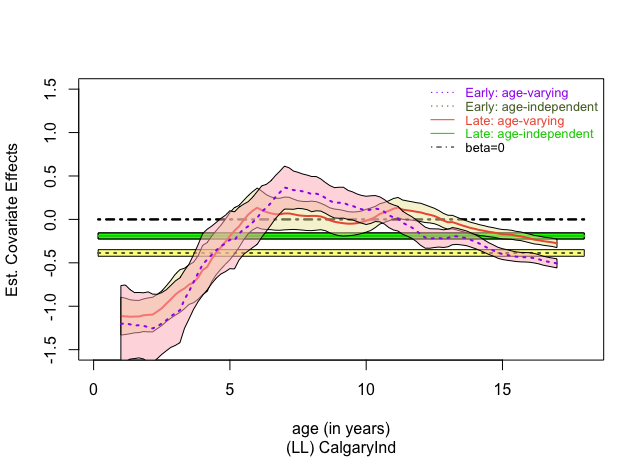}
        \caption{Estimates for the covariate effects for the early and late groups concerning the general population.}
      \label{fig:estbpop}
\end{figure}

Figure \ref{fig:estbpop} shows estimates for $\beta(\cdot)$ and $\beta(\cdot)+\gamma(\cdot)$ with Model VVV, which are the estimated age-varying associations of the ED visits with the
exposures in
the early and late decades, respectively. For comparison, the estimates of the age-independent coefficients with Model CCC are added into each of the plots. 
Interestingly we observe more substantially significant exposure effects on the ED visits when concerning the general population even though the dynamic patterns of MHED visits frequencies are similar to the ones in
Section~\ref{sec:PMHEDVCohort}, concerning the ED cohort. For example, the difference in the exposure effects between subcohorts from two decades become more significant in this
analysis concerning the general population.

\section{Simulation Studies}\label{sec:Simulation}
\indent We conducted simulation studies to examine the finite-sample
performance of our approach and to verify the findings of the MHED
data analyses. Two settings were considered in the simulation: {\sc
Setting 1} mimicked the real data example, generating recurrent
events associated with two eras; {\sc Setting 2} simulated a
situation where subjects were divided into two generations based on
their birthdates and our approach is applied to analyze the
generated data. {\sc Setting 2} allows us to explore the performance
of our approach with a misspecified model. In each setting, we
conducted analyses of the generated data concerning with the target
population $\mathcal{P^*}$ being $\mathcal{P}_1$, the population of
subjects with at least one event during a given period, and
$\mathcal{P}$, the whole population of all individuals no matter if
included in $\mathcal{P}_1$. When $\mathcal{P^*}=\mathcal{P}_1$, we
checked how analysis outcomes are affected when the independence
assumption for the two groups does not hold. When the target
population is $\mathcal{P^*}=\mathcal{P}=\mathcal{P}_1\bigcup
\mathcal{P}_0$, we examined how well the approximation in the
approach performs using the population-level census information.
\subsection{Setting 1}
We simulated data as a random sample $\mathcal{O}$ from
$\mathcal{P}$ with $n=50,000$ subjects. Two groups were defined by the two extraction time windows, $[W^{E}_L, W^{E}_R]$ and $[W^{L}_L, W^{L}_R]$ (in calendar time).
Here we took $W^{E}_R$ as one year before $W^{L}_L$.
For subject $i$, the occurrences of recurrent events before the subject\rq{}s $18^{th}$ birthday are generated as follows.
%\vspace{-1cm}
%
% \newenvironment{Ventry}[1]%
%  {\begin{list}{}{\renewcommand{\makelabel}[1]{##1 \hfil}%
%       \settowidth{\labelwidth}{#1}%
%       \setlength{\leftmargin}{\labelwidth+1.45\labelsep}}}%
%       {\end{list}}
\floatname{algorithm}{Simulation Setting}
%\begin{singlespace}
 \begin{algorithm}
 \caption{}
    \begin{algorithmic}
\State Step 1. Define $X_i(a)=\mbox{I}(A_i<a)$, the indicator of $a$ older than $A_i$,
the age of subject $i$ at $W^{L}_L$.
\State Step 2. Generate event occurrences associated with
subject $i$ at age $a$ with
\[
E[dN_i(a)|Z_i,X_i(a)]=\exp\{\alpha X_i(a)+\beta Z_i+\gamma Z_iX_i(a)\}d\Lambda_0(a),\]
where $Z_i\sim B(1,0.6), \Lambda_0(a)=\int_{0}^a \lambda_{0}(u)du$.
\end{algorithmic}
 \end{algorithm}
%\end{singlespace}

Recall the definitions of the subpopulations corresponding to the two groups,
$\mathcal{O}_E$ includes subjects born within $[W^{E}_L-18,W^E_R]$
and $\mathcal{O}_L$ includes subjects born within $[W^{L}_L-18,W^L_R]$.
Therefore, those subjects with $A_i\leq 0$ and $X_i(\cdot)=1$ can only
belong to $\mathcal{O}_E$; those with $A_i\geq 18$ and $X_i(\cdot)=0$,
$\mathcal{O}_L$. Subject $i$ with $A_i\in (0,18)$ may
belong to both $\mathcal{O}_E$ and $\mathcal{O}_L$.
With the assumed model in { Step 2}, we can generate different patterns
of events in $\mathcal{O}_E$ and $\mathcal{O}_L$.
When $\alpha=0$ and $\gamma=0$, the patterns of events for subjects
in $\mathcal{O}_E$ and $\mathcal{O}_L$ remain the same.
We chose $W^{E}_L=$\lq\lq{}2002-04-01\rq\rq{},
$W^{E}_R=$\lq\lq{}2010-03-31\rq\rq{},
$W^{L}_L=$\lq\lq{}2010-04-01\rq\rq{}, and
$W^{L}_R=$\lq\lq{}2017-03-31\rq\rq{}, and assumed the birthdates
of subjects were uniformly distributed between \lq\lq{}1984-04-01\rq\rq{}
and \lq\lq{}2017-03-31\rq\rq{}.
Two simulation cases were considered:
{\sc Case 1} $\alpha(\cdot)=0, \gamma(\cdot)=0$, and {\sc Case 2} $\alpha(\cdot)=.3, \gamma(\cdot)=.15$,
where $\lambda_0(\cdot)=.002$ and $\beta(\cdot)=.7$ in both cases.
The analyses with different target populations were carried out as follows.

\subsubsection{Concerning $\mathcal{P}_1$}
With the simulated data, the sample from the target population $\mathcal{P}_1$
was created by selecting subjects who made visits during 2002-04-01
to 2017-03-31. To follow the data extraction mechanism in the practical
situation, we also created subpopulations for the early and late time periods,
$\mathcal{O}_{E1}$ and $\mathcal{O}_{L1}$, by selecting subjects who
made visits during 2002-04-01 to 2010-03-31 and 2010-04-01 to 2017-03-31,
respectively. We compared the following approaches to conduct the analysis.
%\vspace{-1cm}
%\begin{singlespace}

\begin{itemize}
\item[](A.1.1) By assuming the individuals in $\mathcal{O}_{E1}$ and $\mathcal{O}_{L1}$ are independent, conduct the analysis with stratified modelling presented in Section~\ref{sec:Framework}.
\item[](A.1.2) By assuming the individuals in the early and late subgroups are independent, construct the combined collection $\mathcal{O}_{E1}\cup\mathcal{O}_{L1}$  and perform the analysis with model \eqref{eq:modelGeneral}, where $X_i=I(\text{subject $i$ }\in \mathcal{O}_{L1})$.
\item[](A.1.3) Perform the analysis for subjects from $\mathcal{O}_{1}$
as a random sample from $\mathcal{P}_1$ with
\[
E[dN_i(a)|Z_i, X_i(a)]=\exp\{\alpha(a) X_i(a)+\beta(a) Z_i+\gamma(a) Z_iX_i(a)\}d\Lambda_0(a)
\]
for $i\in \mathcal{P}_{1}$.
In Case 1 with $\alpha(\cdot)=0,\gamma(\cdot)=0$, we conducted the
analysis of the generated data under the model
$E[dN_i(a)|Z_i, X_i(a)]=\exp\{\beta(a) Z_i\}d\Lambda_0(a)$ for
$i\in \mathcal{P}_{1}$.
\end{itemize}
%\end{singlespace}

With the MHED datasets, the approach (A.1.3) is not feasible
and but (A.1.1) and (A.1.2) are since subjects in the early and late subgroups do not
have the unique identification numbers to allow $\mathcal{O}_{E1}$
and $\mathcal{O}_{L1}$ to be linked.
We implemented the approach (A.1.3) with the simulated data
to use its outcomes as a reference (a gold standard) to make comparisons
with the outcomes from (A.1.1) and (A.1.2), which assume subjects
in $\mathcal{O}_{E1}$ and $\mathcal{O}_{L1}$ are independent as we did in
the real data analysis.

We first started by assuming age-independent coefficients in each analysis.
Table S5 presents the sample means (SMean), the sample
standard deviations (SSE), and the sample means of the estimated standard errors (ESE) of the estimates for (A.1.1), (A.1.2) and (A.1.3) based on 300 repetitions.
The estimates of $\lambda_0$ in the table were obtained by averaging
out $\hat{\lambda}_0(t)$ over time. As expected, the sample means of estimates
are not close to the true values. This result is because subjects in $\mathcal{P}_1$
constitute a biased subpopulation of $\mathcal{P}$, from which the data were generated. The sample means of estimated standard errors also deviate
from the empirical standard errors of the estimates since the true model
structure for the recurrent event data in $\mathcal{P}_1$ may be rather
different from the one we used to simulate the data. On the other hand,
we observe that the SMean of $\hat{\beta}$ in (A.1.2) is similar to
the one of $\hat{\beta}_{E}$ in (A.1.1), which verifies the consistency
between the general model \eqref{eq:modelGeneral} and the stratified model.
In Case 1 where the number of overlapped subjects in $\mathcal{O}_{E1}$ and $\mathcal{O}_{L1}$ is relatively small,
it is further noted that both $\hat{\beta}_{E}$ from (A.1.1) and $\hat{\beta}$ from (A.1.2) approach to estimates from (A.1.3)
in which the independence assumption between $\mathcal{O}_{E1}$ and $\mathcal{O}_{L1}$ does not hold.

The estimates of the cumulative baseline rate function are presented in Figure S18 in the
Supplementary Material. The SMeans of $\hat{\Lambda}_{E1}(\cdot)$ with $\mathcal{O}_{E1}$ and  $\hat{\Lambda}_{0}(\cdot)$
with $\mathcal{O}_{E1}\cup\mathcal{O}_{L1}$ are close and both of them
are smaller than $\hat{\Lambda}_{L1}(\cdot)$ for $\mathcal{O}_{L1}$.
On the other hand, the analysis with subjects from $\mathcal{O}_1$ results
in the smallest estimates of the cumulative baseline rate function.
When analyzing data from subjects in $\mathcal{O}_1$, the data extraction
time window is $[W^{E}_E,W^L_L]$. This leads to a wider censoring interval
and results in a larger risk set compared to the ones from (A.1.1) and (A.1.2)
that assumed subjects from $\mathcal{O}_{E1}$ and $\mathcal{O}_{L1}$
are independent. As there are more subjects in the risk set with the
analysis of $\mathcal{O}_1$, the estimated $\Lambda_0(\cdot)$ is smaller.
Further, we assumed age-dependent coefficients in the model for each analysis and presented estimates of $\beta(\cdot)$ with simulation Case 1
in Figure S19a of the Supplementary Material. All of the estimates are very close and they are approximately constant over age.
This finding is consistent with what we have observed from the age-independent
 estimates.

\subsubsection{Concerning $\mathcal{P}$}
We next considered the following approaches to conduct analysis with the target population being the whole
 population $\mathcal{P}$.
%\vspace{-1cm}
%\begin{singlespace}
\begin{itemize}
\item[](B.1.1) By assuming the individuals in $\mathcal{O}_{E}$ and $\mathcal{O}_{L}$ are independent, conduct the analysis with stratified modelling.
\item[](B.1.2) By assuming the individuals in the early and late subgroups are independent, consider the stratified modelling to analyze $\mathcal{O}_{E1}$ and $\mathcal{O}_{L1}$ together with the supplementary information $\{M(d,z,\big\lfloor u\big\rfloor), \text{for $d=E,L, z=0,1, u\in (0, 18)$}\}$.
\item[](B.1.3) By assuming the individuals in the early and late subgroups are independent, construct the combined collection $\mathcal{O}_{E}\cup\mathcal{O}_{L}$  and
perform the analysis with model \eqref{eq:modelGeneral}, where $X_i=I(\text{subject $i$ }\in \mathcal{O}_{L})$.
\item[](B.1.4) By assuming the individuals in the early and late subgroups are independent, construct the combined collection $\mathcal{O}_{E1}\cup\mathcal{O}_{L1}$  and use the procedure presented in Section~\ref{sec:estmodparams} to analyze the combined data $\mathcal{O}_{E1}\cup\mathcal{O}_{L1}$   supplemented by the summary information $\{M(d,z,\big\lfloor u\big\rfloor), \text{for $d=E,L, z=0,1, u\in (0, 18)$}\}$.
\item[](B.1.5) Analyze data from subjects in $\mathcal{O}$, which is a random sample of $\mathcal{P}$, with the true model
\[
E[dN_i(a)|Z_i, X_i(a)]=\exp\{\alpha X_i(a)+\beta Z_i+\gamma Z_iX_i(a)\}d\Lambda_0(a), i\in \mathcal{P}.
\]
\item[](B.1.6) Use information of subjects from $\mathcal{O}_1$ together with the supplementary information $\{M(d,z,\big\lfloor u\big\rfloor)$, for $d=E,L, z=0,1, u\in (0, 18)\}$ to conduct analysis with the true model.
\end{itemize}
%\end{singlespace}
The approaches (B.1.1), (B.1.3) and (B.1.5) used the true information for all of individuals in $\mathcal{O}_E$, $\mathcal{O}_L$, $\mathcal{O}_E\cup\mathcal{O}_L$, and $\mathcal{O} $
that can all be viewed as random samples of $\mathcal{P}$. The estimates from these analyses can serve as the reference to evaluate the approaches (B.1.2), (B.1.4), and (B.1.6),
which are the proposed approaches to conduct analysis  concerning the target population $\mathcal{P}$. Furthermore, (B.1.1), (B.1.2), (B.1.3) and (B.1.4) followed the MHED example that the identifiers for
subjects are unknown. They used an age-independent indicator to divide subjects into
early and late subgroups rather than using the true $X(a)$. They also assumed subjects from the
early and late subgroups
are independent. Therefore, it is worth comparing them with (B.1.5) and (B.1.6) to evaluate the performance of feasible approaches with the MHED data.

\begin{figure}[htbp]
 \begin{subfigure}[t]{3in}%{.45\textwidth}
\includegraphics[scale=.25]
{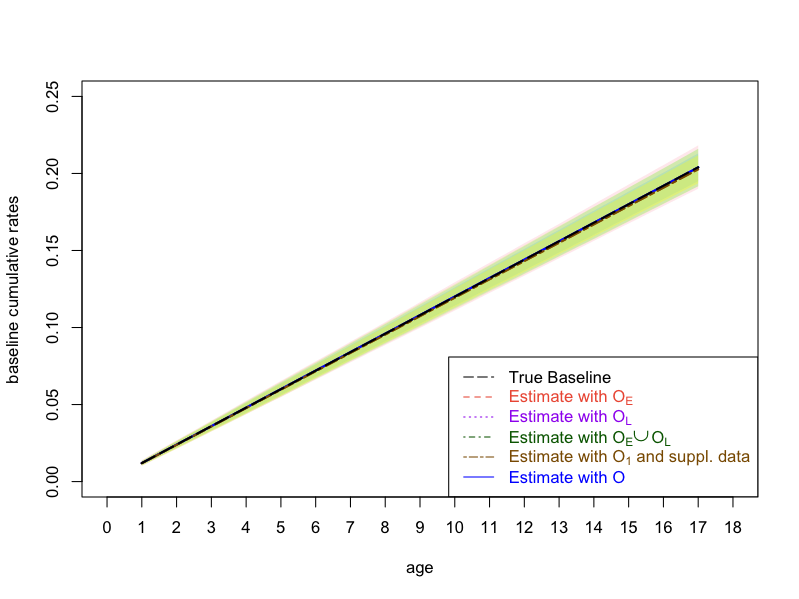}
\caption{\footnotesize Case 1}
   \label{fig:Lambda002B1}
 \end{subfigure}
 \hspace{.5cm}
 \begin{subfigure}[t]{3in}%{.45\textwidth}
\includegraphics[scale=.25]
{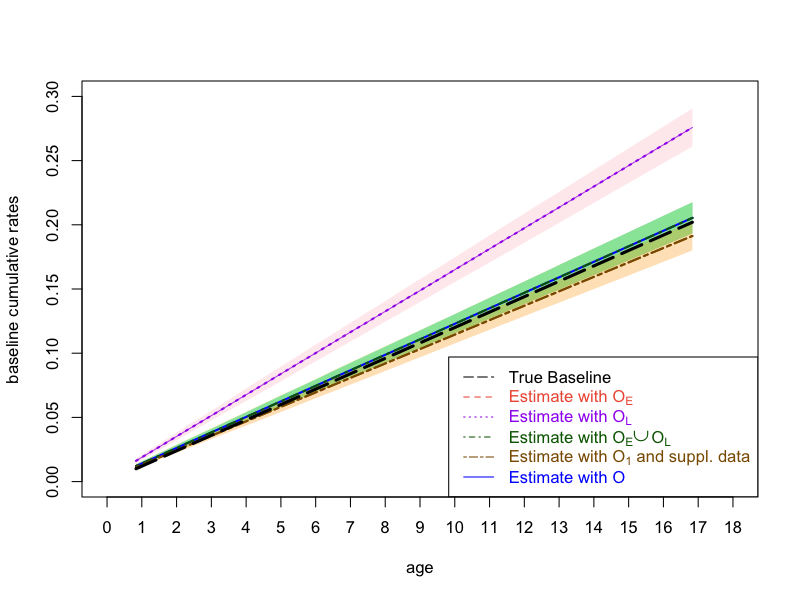}
\caption{\footnotesize Case 2}
   \label{fig:Lambda002B2}
 \end{subfigure}
       \caption{Simulation Setting 1: estimates for $\Lambda_0(\cdot)$ concerning $\mathcal{P}$.}
      \label{fig:Lambda002B}
\end{figure}

We present outcomes from simulation Case 2 (i.e., $\alpha=0.3, \beta=0.7, \gamma=0.15$) in Table
\ref{tab:simB12}. The SMeans of estimates from analyses (B.1.2), (B.1.4) and (B.1.6) are
close to the true values. In addition, those SMeans, SSEs and ESEs of estimates from (B.1.2), (B.1.4) and (B.1.6)
are also comparable with (B.1.1), (B.1.3) and (B.1.5) that used the
individual-level information. These findings indicate the consistency of the approach that uses the
supplemental data and suggest using supplemental data can recover
the efficiency loss due to the truncated information. We also note that the ESEs are similar to
those from (A.1.1)--(A.1.3) concerning  $\mathcal{P}_1$. This pattern is in close agreement
with the comparisons of standard errors from analyses presented in Sections~\ref{sec:ABcensus} and
\ref{sec:comparison2dec}. Moreover, (B.1.1), (B.1.2), (B.1.3) and (B.1.4) also led to unbiased estimates,
suggesting approaches that assumed independence between the early and late subgroups can perform well with
the MHED data. We observe similar findings from simulation outcomes in Table S6 with Case 1.
\begin{table}[!h]
\caption{Simulation Setting 1 Case 2: $\alpha=.3,\beta=.7,\gamma=.15$,  concerning  $\mathcal{P}$.}
\label{tab:simB12}
\centering
\begin{tabular}{llllllllll}
  \hline
&  \multicolumn{4}{c}{$\mathcal{O}_E,\mathcal{O}_L$ (B.1.1)}&&
 \multicolumn{4}{c}{$\mathcal{O}_{E1},\mathcal{O}_{L1}$ with Suppl. Data (B.1.2)}\\
%&  \multicolumn{4}{c}{(B.1.1)}&&
% \multicolumn{4}{c}{(B.1.2)}\\
& $\hat{\beta}_E$ & $\hat{\beta}_L$&$\hat{\lambda}_{0,E}$ & $\hat{\lambda}_{0,L}$&&
  $\tilde{\beta}_E$ & $\tilde{\beta}_L$&$\tilde{\lambda}_{0,E}$ & $\tilde{\lambda}_{0,L}$\\
 \cline{2-5}\cline{7-10}
  SMean &.6990 & .8486 &.0020 & .0027 && .6967 & .8488&  .0020 & .0020 \\
    SSE & .0338 & .0314 &.0001 & .0001&& .0340 & .0316&.0001 & .0001 \\
   ESE$^{(a)}$  &.0375 & .0344 && &&.0378 & .0346&& \\
  ESE$^{(b)}$ & .0357 & .0324 &&&& .0360 & .0326&& \\
  \hline
 \hline
  &  \multicolumn{4}{c}{$\mathcal{O}_E\cup\mathcal{O}_L$ (B.1.3)}&
&    \multicolumn{4}{c}{$\mathcal{O}_{E1}\cup\mathcal{O}_{L1}$ with Suppl. Data (B.1.4)}\\
%&  \multicolumn{4}{c}{(B.1.3)}&&
% \multicolumn{4}{c}{(B.1.4)}\\
    & $\hat{\alpha}$ &$\hat{\beta}$&$\hat{\gamma}$ &$\hat{\lambda}_0$&&$\tilde{\alpha}$ &$\tilde{\beta}$&$\tilde{\gamma}$&$\tilde{\lambda}_0$ \\
   \cline{2-5}\cline{7-10}
 SMean &.2949 & .6990 &  .1497&.0026&&  .2969 & .6967 & .1521&.0020 \\
     SSE &.0391 & .0338 & .0442 &.0001&& .0395 & .0340 & .0448&.0001 \\
   ESE$^{(a)}$&  .0440 & .0375 & .0509 &&& .0442 & .0378 & .0512& \\
   ESE$^{(b)}$& .0421 & .0357 & .0482 &&& .0424 & .0360 & .0485 &\\
   \hline
 \hline
 & \multicolumn{4}{c}{$\mathcal{O}$ (B.1.5)}&&\multicolumn{4}{c}{$\mathcal{O}_1$ and Suppl. data (B.1.6)}\\
% &  \multicolumn{4}{c}{(B.1.5)}&&
% \multicolumn{4}{c}{(B.1.6)}\\
  & $\hat{\alpha}$ &$\hat{\beta}$&$\hat{\gamma}$ &$\hat{\lambda}_0$&&$\tilde{\alpha}$ &$\tilde{\beta}$&$\tilde{\gamma}$&$\tilde{\lambda}_0$ \\
 \cline{2-5}\cline{7-10}
SMean&.2948 & .6991 & .1495&.0020 && .2964 & .6992 & .1497&.0019 \\
 SSE  &.0391 & .0337 & .0442&.0001 && .0393 & .0337 & .0445&.0001 \\
  ESE$^{(a)}$  &.0427 & .0375&  .0491&& & .0428 & .0375 & .0492& \\
 ESE$^{(b)}$ & .0421 & .0357 & .0482 &&& .0422 & .0357 & .0483& \\
    \hline
\end{tabular}
\end{table}
%\clearpage
The estimates of cumulative baseline functions are shown in Figure \ref{fig:Lambda002B}. Estimates from all of the analyses are close to the true value of $\Lambda_0(\cdot)$, which corroborates the observations from Table \ref{tab:simB12}.

\subsection{Setting 2}
In this section, we report simulation results with Setting 2 that generated data under a misspecified model.
We defined $X^*$ as an indicator that $X^*=I(B\geq \text{\lq\lq{}2001-03-31\rq\rq{}})$. In this way, we created two generations of subjects in
$\mathcal{O}$, which is a random sample from $\mathcal{P}$. The occurrences of events for
subject $i$ at age $a$ are then simulated from the model
\[
E[dN_i(a)|Z_i,X^*_i]=\exp\{X^*_i\alpha+Z_i\beta+X^*_iZ_i\gamma\}d\Lambda_0(a), a\in (0,18), i\in
\mathcal{O},
\]
where $\Lambda_0(a)=\int_{0}^a \lambda_{0}(u)du$ and the true values
of $(\lambda_0, \alpha, \beta, \gamma)$ were chosen as (0.002, 0.6,
0.7, 0.35).

Analyses were conducted in similar fashion to those in Setting 1. We denoted these approaches by (A.2.1), (A.2.2) and (A.2.3) for analyses concerning
$\mathcal{P}_1$ and (B.2.1), (B.2.2), (B.2.3), (B.2.4), (B.2.5) and (B.2.6) for analyses concerning $\mathcal{P}$. For (A.2.3), (B.2.5) and (B.2.6), we used the true indicator $X^*$ instead of using $X$ that distinguishes $\mathcal{P}_{E}$ and $\mathcal{P}_L$.

Table \ref{tab:simB22} presents
age-independent estimates for analyses (B.2.1)--(B.2.6) concerning $\mathcal{P}$. Estimates from (B.2.2), (B.2.4) and (B.2.6) are comparable with those
from (B.2.1), (B.2.3) and (B.2.5), indicating the proposed strategy synthesizing supplemental data still performs well with this simulation setting.
Despite of the misspecified indicator used in (B.2.1), (B.2.2), (B.2.3) and (B.2.4), these analyses can identify the significant effect of $Z$ and
detect the different patterns of recurrent events between the two generations. These observations also match with the age-varying estimates for
$\alpha(\cdot),\beta(\cdot)$ and $\gamma(\cdot)$ shown in Figure S20. Estimates concerning with $\mathcal{P}_1$ are presented in Table S7. Although $\mathcal{P}_1$ is a biased cohort of $\mathcal{P}$, we can identify the significant effect of
covariate $Z$ from all three analyses (A.2.1), (A.2.2) and (A.2.3). One can still conclude the effect of $Z$ on the occurrences of recurrent events has become much stronger for the late generation based on results from analyses (A.2.1) and (A.2.2),
even though these two analyses used a misspecified indicator variable and under-estimated the $\alpha$ and $\gamma$.

We also can clearly identify the difference between $\mathcal{P}_{E1}$ and $\mathcal{P}_{L1}$ (see Figure S21a) and the
difference between $\mathcal{P}_{E}$ and $\mathcal{P}_{L}$ (see Figure S21b). These findings suggest that the reported analysis of the MHED data can
still be meaningful and interpretable even if the mechanism to divide two generations is subject to misspecification.

\begin{table}[htbp]
\caption{Simulation Setting 2: $\alpha=.6,\beta=.7,\gamma=.35$, concerning with the general population  $\mathcal{P}$.}
\label{tab:simB22}
\centering
\begin{tabular}{llllllllll}
  \hline
&  \multicolumn{4}{c}{$\mathcal{O}_E,\mathcal{O}_L$}&&
 \multicolumn{4}{c}{$\mathcal{O}_{E1},\mathcal{O}_{L1}$ with Suppl. Data}\\
&  \multicolumn{4}{c}{(B.2.1)}&&
 \multicolumn{4}{c}{(B.2.2)}\\
& $\hat{\beta}_E$ & $\hat{\beta}_L$&$\hat{\lambda}_{0,E}$ & $\hat{\lambda}_{0,L}$&&
  $\tilde{\beta}_E$ & $\tilde{\beta}_L$&$\tilde{\lambda}_{0,E}$ & $\tilde{\lambda}_{0,L}$\\
 \cline{2-5}\cline{7-10}
  SMean & .8611 & .9951&.0024&.0031& & .8601 & .9947&.0024&.0031 \\
    SSE & .0319 & .0291 &.0001 & .0001&& .0321 & .0290&.0001 & .0001 \\
   ESE$^{(a)}$  &.0338 & .0319 && &&.0340 & .0321&& \\
  ESE$^{(b)}$ & .0316 & .0295 &&&& .0318 & .0297&& \\
  \hline
 \hline
  &  \multicolumn{4}{c}{$\mathcal{O}_E\cup\mathcal{O}_L$}&
&    \multicolumn{4}{c}{$\mathcal{O}_{E1}\cup\mathcal{P}_{L1}$ with Suppl. Data}\\
&  \multicolumn{4}{c}{(B.2.3)}&&
 \multicolumn{4}{c}{(B.2.4)}\\
    & $\hat{\alpha}$ &$\hat{\beta}$&$\hat{\gamma}$ &$\hat{\lambda}_0$&&$\tilde{\alpha}$ &$\tilde{\beta}$&$\tilde{\gamma}$&$\tilde{\lambda}_0$ \\
   \cline{2-5}\cline{7-10}
 SMean &.2447 & .8611 & .1340 &.0024&& .2434 & .8602 & .1346&.0024 \\
     SSE &.0363 & .0320 & .0427 &.0001&& .0363 & .0322 & .0425&.0001 \\
   ESE$^{(a)}$& .0406 & .0338 & .0465 &&& .0408 & .0340 & .0467& \\
   ESE$^{(b)}$&.0385 & .0316 & .0433&& & .0387 & .0318 & .0435& \\
   \hline
 \hline
 & \multicolumn{4}{c}{$\mathcal{O}$}&&\multicolumn{4}{c}{$\mathcal{O}_1$ and Suppl. data}\\
 &  \multicolumn{4}{c}{(B.2.5)}&&
 \multicolumn{4}{c}{(B.2.6)}\\
  & $\hat{\alpha}$ &$\hat{\beta}$&$\hat{\gamma}$ &$\hat{\lambda}_0$&&$\tilde{\alpha}$ &$\tilde{\beta}$&$\tilde{\gamma}$&$\tilde{\lambda}_0$ \\
 \cline{2-5}\cline{7-10}
SMean&.5960 & .6972 & .3548 &.0020&& .5852 & .6972 & .3549&.0019 \\
 SSE  &.0439 & .0368 & .0458&.0001 && .0424 & .0368 & .0455&.0001 \\
  ESE$^{(a)}$  &.0464 & .0392 & .0499&& & .0456 & .0392 & .0500 \\
 ESE$^{(b)}$ &.0429 & .0363 & .0452 &&&.0426 & .0363 & .0453 \\

    \hline
\end{tabular}
\end{table}

\section{Final Remarks}\label{sec:Discussion}

In this article, we compared frequencies of ED visits for mental health reasons made by children during two time periods (decades). We examined the temporal patterns of covariate effects with recurrent events data analysis.
We were able to provide both innovative strategies for processing doubly-censored recurrent event data in the presence of the
truncation induced by the data collection and
new insights into the evolution of ED visits for mental health reasons over
time.

Rates of mental health related ED visits are higher for older compared to younger children. We considered various model specifications to explore both age-independent and age-varying covariate effects.
The age-varying effects were estimated via Kernel-weighted estimating functions. We first evaluated the age-independent and age-varying differences of covariate effects on the
MHED visits over time with the target population being the cohort of children who had ED visits (ED cohort). While access to administrative data allowed for analyses that were population-based, the separate
data extractions in different time periods prevented linkage of subjects in both extractions. This
feature gave rise to a challenge to the comparison of trends in recurrent events between two
periods and motivated our proposed approach. We expect that comparisons between periods will
be of even greater interest as researchers seek to understand and quantify effects of
the COVID-19 pandemic by comparing data before and during the pandemic, or before and after the pandemic.

We next provided an approach to draw inference with a broader population that takes all
Alberta children and youths as a random sample. Access to an administrative health data source in the single-payer health system for
an entire jurisdiction enabled the use of census data to determine the number of children who
did not visit for mental health related ED care to complement the data on children who did visit.
The proposed approach leverages summary information from census data and recovers the efficiency
due to the loss of information. Numerical studies show that the covariate effects are not the same in
the different populations and thus caution is required when applying models to populations beyond the cohorts
that were used for building these models.

There are several directions for further investigations.
It is also of great practical interest to extend the estimation procedure to domain selection,
i.e., to identify time domains in which the covariate effects remains constant and varying. We may also consider other health conditions that may lead to ED visits
 and extend analyses with multi-type recurrent event data
 to accommodate ED visits with multiple causes.
In our motivating example, the proposed approach to tackle the zero-truncation issue is applied to regression analysis with discrete covariate variables. It is worth investigating how to extend it to incorporate continuous covariate variables.

\section*{Appendix}
In this section, we first introduce the regularity conditions that are assumed for both Propositions 1 and 2 together with an additional requirement for approximations ${\tilde{S}^{(q)}(\phi(a);u,a)}$ for $q=0,1,2$ that are used in estimation with the general population. We then outline the proofs for Proposition 2.

The following are the conditions assumed to derive asymptotic
properties in Propositions 1 and 2. We adapt the conditions given in
Hu and Rosychuk (2016).

\begin{itemize}
\item[]{\bf (C1)} $\{N_i(\cdot), Z_i, X_i, B_i\}$ for $i=1,\cdots,n$ are independent and identically distributed;
\item[]{\bf (C2)} $P(C_{Li}<\tau_L)>0,P(C_{Ri}>\tau_R)>0$ for $i=1,\cdots,n$, where $0<\tau_L<\tau_R<18$ are predetermined constants;
\item[]{\bf (C3)} $N_i(18)$ for $i=1,\cdots,n$ are bounded by a constant;
\item[]{\bf (C4)} $\lambda_0(\cdot)>0$ and continuous and $\theta(\cdot)=(\alpha(\cdot)\rq{},\beta(\cdot)\rq{},\gamma(\cdot)\rq{})\rq{}$ have continuous second derivative over $(0,18)$;
\item[]{\bf (C5)} The kernel function $K(\cdot)$ is a bounded and symmetric density with a bounded support;
\item[]{\bf (C6)} The matrix $\{\upsilon^{(2)}({\theta_0}(a))-\upsilon^{(1)}({\theta_0}(a))^{\otimes 2}/\upsilon^{(0)}({\theta_0}(a))\}\lambda_0(a)$ is positive semi-definite for $a\in(0,18)$.
\item[]{\bf (AC)} Let $n$ be the number of individuals in the general population $\mathcal{P}$. The approximation of $n^{-1}{S}^{(q)}(\phi(a);u,a)$ converges a.s. to
${s}^{(q)}(\phi(a);u,a)$, where ${s}^{(q)}(\phi(a);u,a)=\mbox{E}[Y_i(u)V^*_i(u,a)^{\otimes q}\exp\{\phi(a)\rq{}V^*_i(u,a)\}]$ for $q=0,1,2$.
\end{itemize}
{\sc PROOF OF PROPOSITION 2.} Note that when a time unit is defined
as one year, $M(d,g,r,\big\lfloor u\big\rfloor)$, the number of
individuals aged $\big\lfloor u\big\rfloor$ years is equivalent to
$\sum_{i\in\mathcal{P}(d,g,r)}Y_i(u)$ where $\mathcal{P}(d,g,r)$
represents the subpopulation defined by decade $d$, sex $g$  and
region $r$. Therefore,
$\tilde{S}^{(q)}(\phi(a);u,a)={S}^{(q)}(\phi(a);u,a)$. Since
$n^{-1}{S}^{(q)}(\phi(a);u,a)$ converges a.s.\ to
${s}^{(q)}(\phi(a);u,a)$ by the strong law of large numbers,
$\tilde{S}^{(q)}(\phi(a);u,a)$ satisfy the condition (AC) when using
one year as a time unit. As we also assume that the number of
individuals aged $u$ in the subpopulation $\mathcal{P}(d,g,r)$
remains the same throughout the age year $\big\lfloor u\big\rfloor$,
the condition (AC) can be satisfied with other choices of time unit.

Provided that $n^{-1}\tilde{S}^{(q)}(\phi(a);u,a)\overset{a.s.}\rightarrow{s}^{(q)}(\phi(a);u,a)$, we have
$\tilde{\bar{V}}^*(\phi(a);u,a)$, which is $\tilde{S}^{(1)}(\phi(a);u,a)\big/\tilde{S}^{(0)}(\phi(a);u,a)$, converges to ${s}^{(1)}(\phi(a);u,a)\big/{s}^{(0)}(\phi(a);u,a)$.
Following Hu and Rosychuk (2016), we can show the estimating function $n^{-1}\tilde{U}(\phi(a);a\big|\mathbf{B})$ converges a.s.\ to 0 and establishes the point-wise consistency and weak convergence of $\tilde{\theta}(\cdot)$.

%%%%%%%%%%%%%%%%%%%%%%%%%%%%%%%%%%%%%%%%%%%%%%
%% Support information, if any,             %%
%% should be provided in the                %%
%% Acknowledgements section.                %%
%%%%%%%%%%%%%%%%%%%%%%%%%%%%%%%%%%%%%%%%%%%%%%
\section*{Acknowledgments}
The authors acknowledge Alberta Health Services Data Integration and Management
Repository (DIMR) for the use of MHED data. We thank Professor Jerry Lawless for helpful comments and suggestions.

%%%%%%%%%%%%%%%%%%%%%%%%%%%%%%%%%%%%%%%%%%%%%%
%% Funding information, if any,             %%
%% should be provided in the                %%
%% funding section.                         %%
%%%%%%%%%%%%%%%%%%%%%%%%%%%%%%%%%%%%%%%%%%%%%%
\section*{Funding}
An operating
grant from the Canadian Institutes of Health Research (CIHR) supported
data extraction. This work was supported by individual Discovery Grants from the
Natural Sciences and Engineering Research Council of Canada (NSERC)
held by XJ Hu and RJ Rosychuk, an NSERC Discovery Accelerator Supplement held by XJ Hu, and a postdoctoral fellowship sponsored by Fred Hutchinson Cancer Center.

%%%%%%%%%%%%%%%%%%%%%%%%%%%%%%%%%%%%%%%%%%%%%%
%% Supplementary Material, including data   %%
%% sets and code, should be provided in     %%
%% {supplement} environment with title      %%
%% and short description. It cannot be      %%
%% available exclusively as external link.  %%
%% All Supplementary Material must be       %%
%% available to the reader on Project       %%
%% Euclid with the published article.       %%
%%%%%%%%%%%%%%%%%%%%%%%%%%%%%%%%%%%%%%%%%%%%%%
%NOT SURE WHAT SHOULD BE IN SUPPLEMENTARY VS APPENDIX BUT BECAUSE WE HAD SUPPLEMENTARY BEFORE, DOING THE SAME THING HERE
%THEY DO SAY THAT THEY DON'T WANT REFERENCES USING \ref FOR SUPPLEMENTARY, BUT I HAVEN'T CHANGED THAT YET. WILL CHANGE IF ACCEPTED.

\section*{Supplementary material}
\subsubsection*{A Additional Tables and Figures for Section 3}
\noindent{Additional tables and figures related to Section 3.}

\subsubsection*{B Additional Tables and Figures for Section 4}
\noindent{Additional tables and figures related to Section 4.}

\subsubsection*{C Additional Tables and Figures for Simulation Study}
\noindent{Additional tables and figures related to the Simulation Study.}

%%%%%%%%%%%%%%%%%%%%%%%%%%%%%%%%%%%%%%%%%%%%%%%%%%%%%%%%%%%%%
%%                  The Bibliography                       %%
%%                                                         %%
%%  imsart-nameyear.bst  will be used to                   %%
%%  create a .BBL file for submission.                     %%
%%                                                         %%
%%  Note that the displayed Bibliography will not          %%
%%  necessarily be rendered by Latex exactly as specified  %%
%%  in the online Instructions for Authors.                %%
%%                                                         %%
%%  MR numbers will be added by VTeX.                      %%
%%                                                         %%
%%  Use \cite{...} to Cite references in text.             %%
%%                                                         %%
%%%%%%%%%%%%%%%%%%%%%%%%%%%%%%%%%%%%%%%%%%%%%%%%%%%%%%%%%%%%%
% \usepackage{natbib}
% \setcitestyle{authoryear,open={((},close={))}} %Citation-related commands
\bibliographystyle{abbrvnat}

\bibliography{ref1}       % Bibliography file (usually '*.bib')

%% or include bibliography directly:
% \begin{thebibliography}{}
% \bibitem[\protect\citeauthoryear{???}{???}]{b1}
% \end{thebibliography}

\end{document}

\title{\bf Exploring Differences between Two Decades of Mental Health
Related Emergency Department Visits by Youth via Recurrent Events
Analyses}

\date{}
\maketitle

\doublespacing
\begin{center} 
{Yi Xiong\\
Department of Biostatistics, University at Buffalo, Buffalo, New York, USA,\\
%and \\
    X. Joan Hu
     \\Department of Statistics and Actuarial Science, Simon Fraser University, British Columbia, Canada,\\    
%        and \\
    Rhonda J.\ Rosychuk \\
Department of Pediatrics,  University of Alberta, Alberta, Canada}
\end{center}

\vspace{1cm}
\centerline{\bf\large Supplementary Material}

\newpage
\section{\bf Additional Tables and Figures of the Real Data Analysis in Section 3}

\indent Table~\ref{tab:charPMHEDCohort} presents summary statistics of MHED cohort data.

\indent Figure~\ref{fig:histSex1a} shows the number of MHED visits from different age groups during 2002--2010 and 2010--2017 made by males and females and Figure~\ref{fig:histSex1b} shows the number of male and female subjects in the MHED cohort.

\indent Figure~\ref{fig:histRegiona} shows the number of MHED visits made by subjects from different regions and Figure~\ref{fig:histRegionb} shows the number of MHED cohort subjects in each region.

\indent Table \ref{table:AGmodelPart2} shows two sets of estimates and standard errors with the 2010--2017 MHED data
for the age-independent effects. Approach A uses available birthdates for the MHED cohort and Approach B follows Hu and Rosychuk (2016) without using the available birthdates.

Figure \ref{fig:histbirth} presents the histogram of the birthdates of
subjects in 2010--2017 cohort.

Figure \ref{fig:estall} shows the local linear estimates for $\alpha(\cdot)$ with each of the Models VEE. For comparison,
added in each plot is the estimate of $\alpha$ with the corresponding Model CEE, which assumes the coefficient for the decade indicator is age-independent.

Figure \ref{fig:estcumED1} shows the estimates for the cumulative baseline rate functions concerning the ED cohort under Models VVV and CCC.

Figure \ref{fig:fig56diffbeta1}        shows the difference in $\beta_E(\cdot)$ and $\beta_L(\cdot)$ with model VVV.
The Hu-Rosychuk Approach B is applied here for both 2002--2010 data and 2010--2017 data to obtain estimates.

Figures \ref{fig:esta}--\ref{fig:estc} present local constant estimates for $\alpha(\cdot)$, $\beta(\cdot)$ and $\gamma(\cdot)$ concerning the MHED cohort.

Figure \ref{fig:estc2} presents local linear estimates for
$\beta(\cdot)+\gamma(\cdot)$ concerning the MHED cohort.

\newpage

\input{supp2023floatsB}

\clearpage

\section{\bf Additional Tables and Figures of the Real Data Analysis in Section 4}

\indent Table~\ref{tab:charGeneralPopu} presents the numbers of Alberta residents younger than 18 years old during 2002--2017.

Figure \ref{fig:riskpop} shows the numbers of Alberta residents over age for subcohorts from the two decades.

%   Figure \ref{fig:fig56diffbetapop1} shows the difference in $\beta_E(\cdot)$ and $\beta_L(\cdot)$ with model VVV concerning the general population.

Table \ref{tab:AGmodelAll2} presents estimates of age-independent regression coefficients concerning the general population under each of the models given in Table~\ref{tab:modelSpecialCase}.

Figure \ref{fig:estapop} shows local linear estimates for $\alpha(\cdot)$ concerning the general population under each of the models VEE. For comparison,
added in each plot is the estimate of $\alpha$ with the corresponding Model CEE, which assumes the coefficient for the decade indicator is age-independent.

Figures \ref{fig:estapopLC}--\ref{fig:estcpopLC} present local constant estimates for $\alpha(\cdot)$, $\beta(\cdot)$ and $\gamma(\cdot)$ concerning the general population.

Figure \ref{fig:estcpop2} presents local linear estimates for $\beta(\cdot)+\gamma(\cdot)$ concerning the general population.

Figure \ref{fig:estcumPop1} shows the estimated cumulative baseline rate  functions under Models CCC and VVV concerning the general population.
\input{supp2022floatsC}
\clearpage
\newpage

\section{\bf Additional Tables and Figures of Simulation Study}

\indent Table \ref{tab:simA11} presents the summary of estimated regression coefficients from analyses (A.1.1)--(A.1.3) with 300 runs of simulation under Setting 1, concerning the target population $\mathcal{P}_1$.

 Figure \ref{fig:Lambda002A} shows sample means of the estimates and the approximated 95$\%$ confidence bands of $\Lambda_0(\cdot)$ from analyses (A.1.1)--(A.1.3) concerning $\mathcal{P}_1$.

Figure \ref{fig:simBeta002LL} shows sample means of the local linear estimates for $\beta(\cdot)$ with 300 runs of simulation under Setting 1 Case 1 concerning $\mathcal{P}_1$ (Figure \ref{fig:Beta002A}) and $\mathcal{P}$ (Figure \ref{fig:Beta002B}).

Table \ref{tab:simB11} presents the summary of estimated regression coefficients from analyses (B.1.1)--(B.1.6) concerning the target population $\mathcal{P}$.

\indent Figure \ref{fig:simLLB2} shows sample means of the local linear estimates for $\beta(\cdot)$ with 300 runs of simulation under Setting 2 concerning $\mathcal{P}$.

\indent Table \ref{tab:simA22} presents the summary of estimated regression coefficients from analyses (A.2.1)--(A.2.3) with 300 runs of simulation under Setting 2, concerning the target population $\mathcal{P}_1$.

\indent Figures \ref{fig:simBaseline002AG2} show sample means of the estimates and the approximated 95$\%$ confidence bands of $\Lambda_0(\cdot)$ concerning $\mathcal{P}_1$ (see Figure \ref{fig:Lambda002A2}) and concerning $\mathcal{P}$ (see Figure \ref{fig:Lambda002B2}).
\input{supp2023floatsD}

\clearpage
\newpage

\input{supp2023floatsE}
\clearpage
\newpage

%% file: supp2023floatsB.tex
\begin{table}[htbp]
%\small
\caption{Characteristics of subjects in the MHED cohort during two
decades.}
\centering
%\addtolength{\leftskip} {-2cm}
%\addtolength{\rightskip}{-2cm}
\label{tab:charPMHEDCohort}
\scalebox{.9}{
\begin{tabular}{lrrrr}
\hline
\hline
%\multirow{2}{*}{Covariate} & \multicolumn{2}{c}{2002--2010} & \multicolumn{2}{c}{2010--2017}   \\
& \multicolumn{2}{c}{2002--2010} & \multicolumn{2}{c}{2010--2017}   \\
& \multicolumn{2}{c}{n=24,922} & \multicolumn{2}{c}{n=33,299}  \\ \hline
Sex & \multicolumn{4}{c}{}\\
 \phantom{xx}Female                       &14,150                & (56.8\%)                     & 19,210                           & (57.7\%)           \\
 \phantom{xx}Male                         &10,772                & (43.2\%)                     &14,089                            & (42.3\%)  \\
 \multicolumn{5}{l}{{Region}}\\
 \phantom{xx}Edmonton    & 7,013              & (28.1\%)                      & 8,365                 & (25.1\%)                               \\
  \phantom{xx}Calgary    &   7,167            &  (28.8\%)                     &   11,579                &  (34.8\%)        \\
  \phantom{xx}Other    & 10,742 & (43.1\%)                      & 13,355               & (4.2\%)                                       \\
Number of Recorded ED visits & \multicolumn{4}{c}{}\\
\phantom{xx}1           &  18,749           & (75.2\%)                      &22,328         & (67.1\%)                \\
\phantom{xx}2          &   3758 &  (15.1\%)                      & 5,728        & (17.2\%)          \\
\phantom{xx}3          &   1,307          &  (5.2\%)                     &  2,381       &  (7.2\%)      \\
\phantom{xx}$\ge$ 4    &   1108 &  (4.4\%)                     &  2,862     &  (8.6\%)                 \\
 \hline
 \hline
\end{tabular}}
\end{table}

\begin{figure}[h]
\begin{subfigure}{\textwidth}
%\hspace{0.6cm}
  \includegraphics[width=0.475\linewidth]{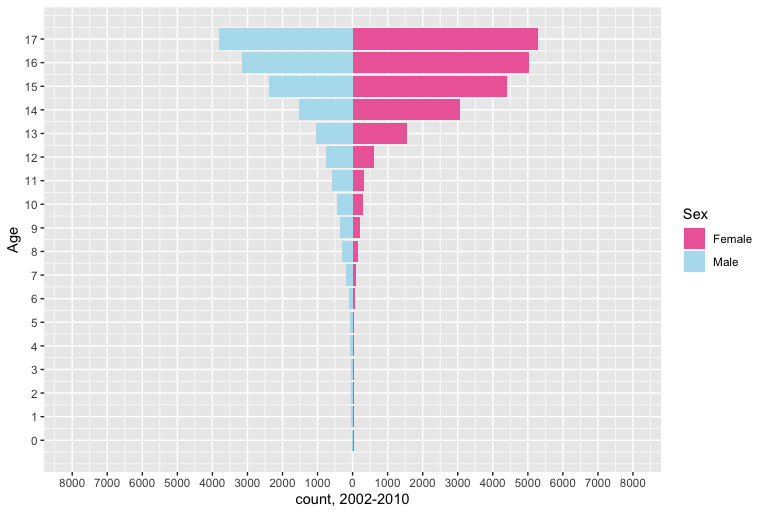}
  \hfill
  \includegraphics[width=0.475\linewidth]{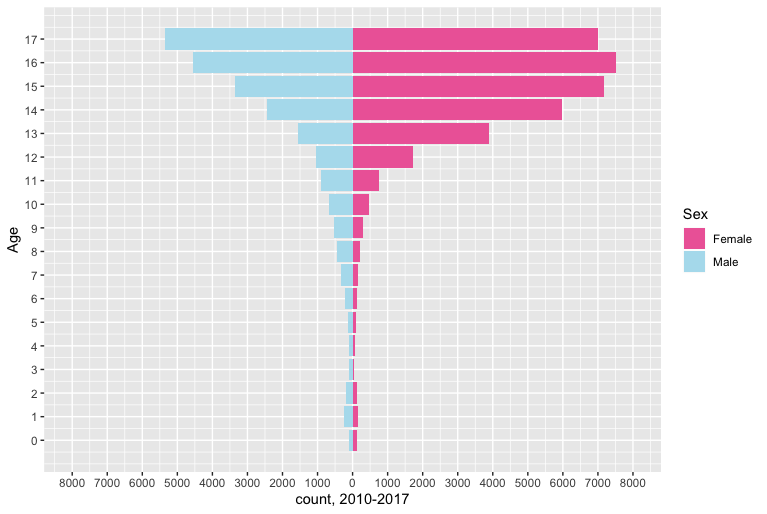}
%  \hspace{-3cm}
%     \parbox{3in}{
%     \vspace{-0.02in}
  \caption{Number of  MHED visits from males and females}
   \label{fig:histSex1a}
\end{subfigure}
\begin{subfigure}{\textwidth}
%\hspace{0.6cm}
\includegraphics[width=0.475\linewidth]{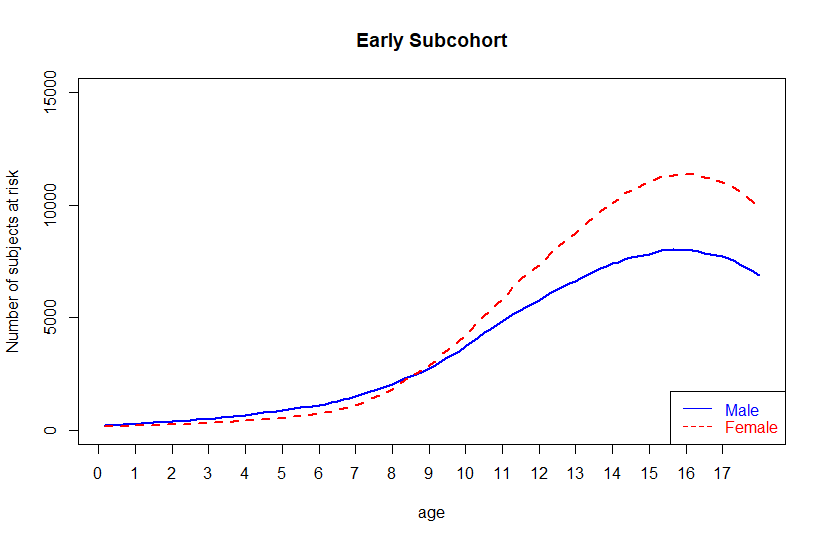}
\hfill
\includegraphics[width=0.475\linewidth]{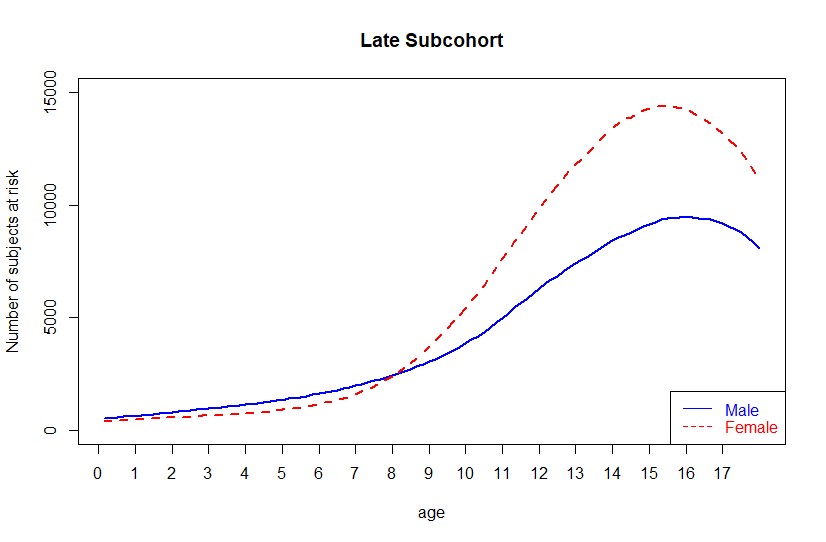}
 \caption{Number of male and female subjects in the MHED cohort}
   \label{fig:histSex1b}
\end{subfigure}
       \caption{Pediatric MHED visits from males and females  in the 2002--2010 and 2010--2017 subcohorts.}
      \label{fig:histage}
\end{figure}

\begin{figure}[t]
\begin{subfigure}{\textwidth}
 \includegraphics[width=0.475\linewidth]{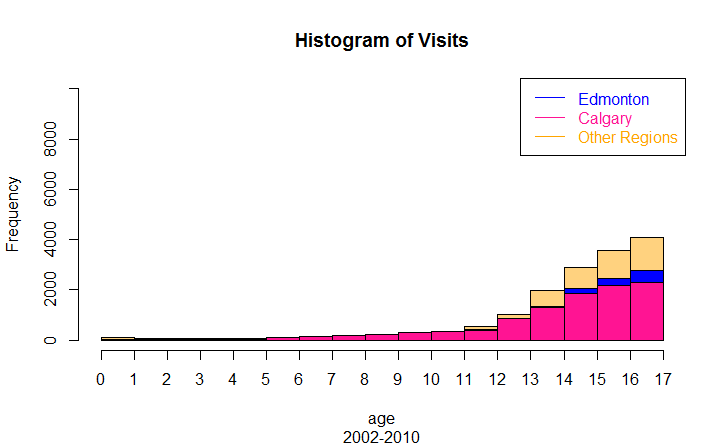}
  \hfill
  \includegraphics[width=0.475\linewidth]{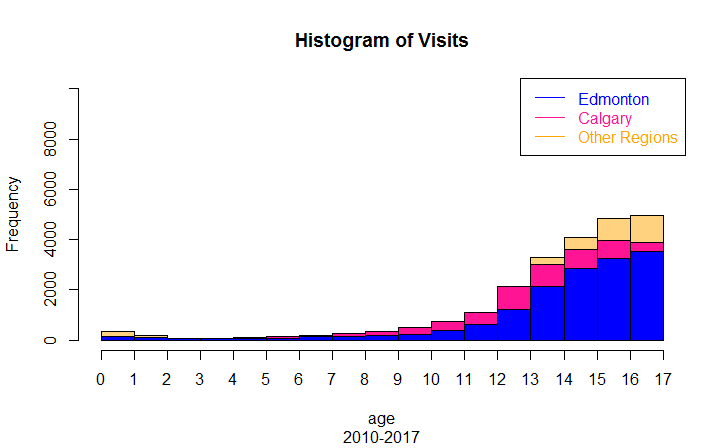}
%  \parbox{2in}{
%   \hspace{-7cm}
%  \vspace{-0.02in}
  \caption{Number of MHED visits for Edmonton, Calgary and other regions}
  \label{fig:histRegiona}
  \end{subfigure}
  \begin{subfigure}{\textwidth}
\includegraphics[width=0.475\linewidth]{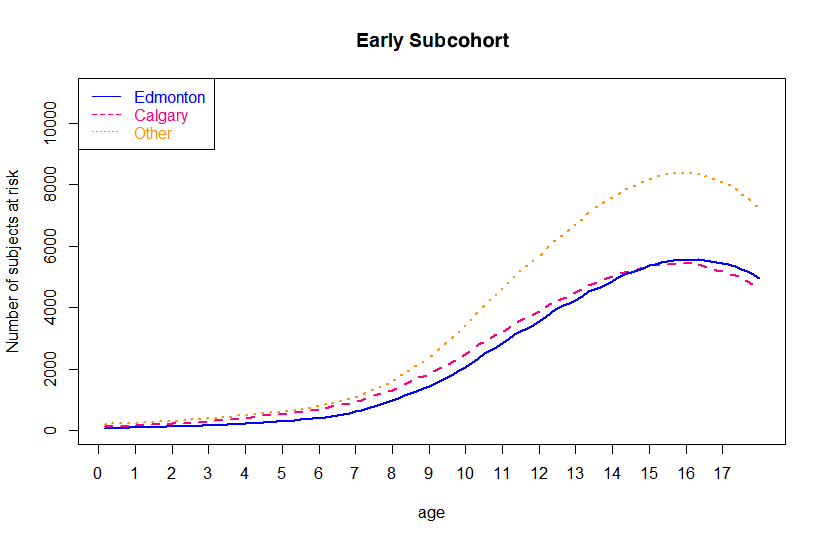}
\hfill
\includegraphics[width=0.475\linewidth]{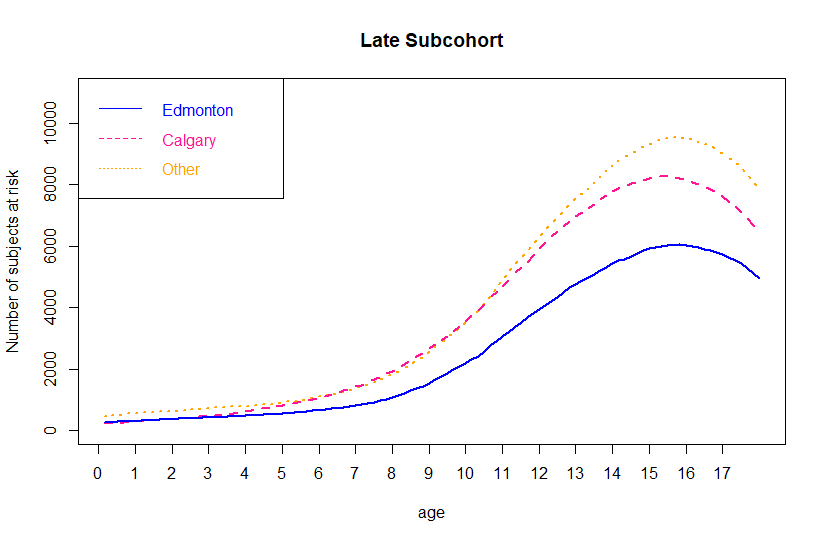}
       \caption{Number of subjects from different regions in the MHED cohort}
      \label{fig:histRegionb}
   \end{subfigure}
    \caption{Pediatric MHED visits from different regions in the 2002--2010 and 2010--2017 subcohorts.}
      \label{fig:histage2}
\end{figure}

\begin{table}[htbp]%[!htbp]
%\small
\caption{Coefficient estimates and standard error estimates under Model CCC from 2010--2017 data.}
%\addtolength{\leftskip} {-2cm}
%\addtolength{\rightskip}{-2cm}
\label{table:AGmodelPart2}
%\hspace{-1cm}
\scalebox{.8}{
\begin{tabular}{lccccccc}
\hline
\hline
Covariate &  \multicolumn{3}{c}{Hu-Rosychuk Approach B} && \multicolumn{3}{c}{Hu-Rosychuk Approach A}   \\
 \cline{2-4}\cline{6-8}
& $\tilde{\beta_L}$    & $SE(\tilde{\beta_L})$    &$SE_{FI}(\tilde{\beta_L})$    &&
                           ${\hat{\beta_L}}$    & $SE({\hat{\beta_L}})$& $SE_{FI}({\hat{\beta_L}})$  \\ \hline
                           Sex (reference $=$  Female) &&&&&&&\\
$\quad$Male
&\bf -.10570 & .01550 & .00856 &&\bf -.10441 & .01550 & .00856 \\
\multicolumn{8}{l}{Region (reference $=$  Other)}\\
$\quad$Edmonton
&\bf .05283 & .02036 & .01048 &&\bf .05225 & .02036 & .01048 \\
$\quad$Calgary
&.01784 & .01779 & .00966 &&  .01596 & .01779 & .00967 \\
\hline
\hline
\end{tabular}}
\end{table}

%%%%%%%%%%%%%%%%%%%%%%% Figure S3 %%%%%%%%%%%%%%%%%%%%%%%%%%%%%%%
\begin{figure}[htbp]%[b]
%\hspace{0.6cm}
\centering
\includegraphics[scale=0.5]{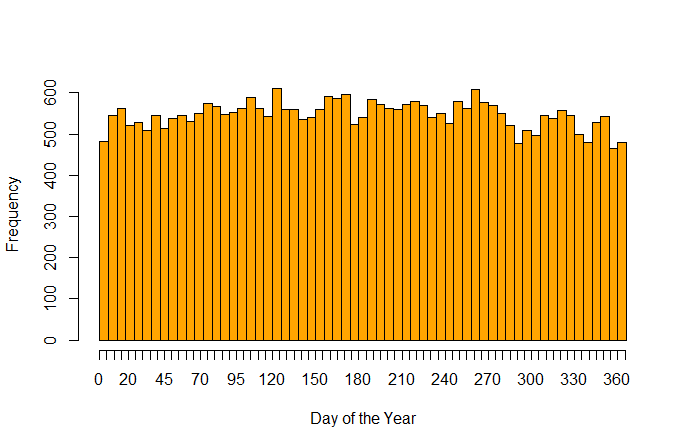}
       \caption{Histogram of birthdates of subjects in 2010--2017.}
      \label{fig:histbirth}
\end{figure}

\begin{figure}[h]
%\hspace{-1cm}
 \begin{subfigure}[t]{0.45\textwidth}
\includegraphics[scale=0.3]
{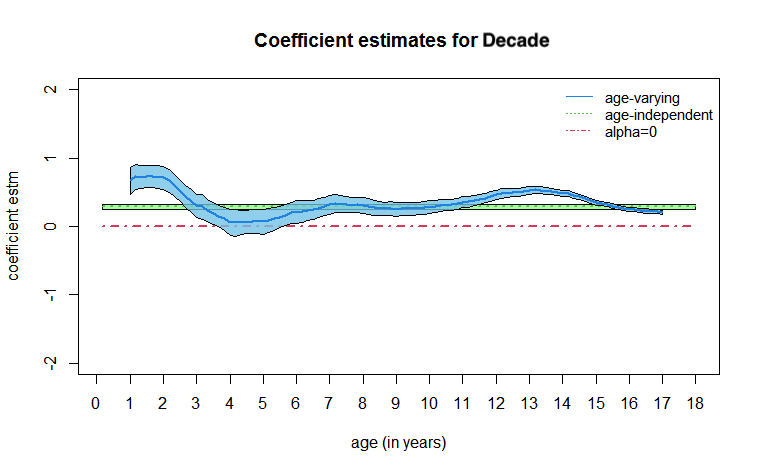}
\caption{\footnotesize Model VCC vs. Model CCC}
   \label{fig:M100a}
 \end{subfigure}
\hspace{.8cm}
 \begin{subfigure}[t]{0.45\textwidth}
\includegraphics[scale=0.3]
{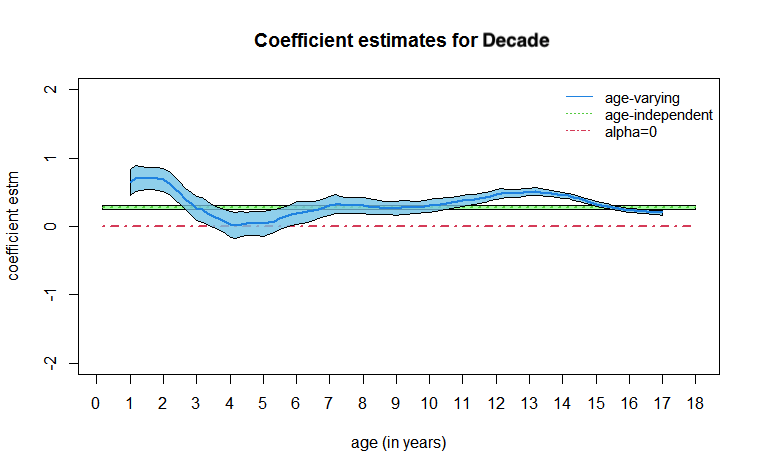}
\caption{\footnotesize Model VVC vs. Model CVC}
   \label{fig:M110a}
\end{subfigure}

 \begin{subfigure}[t]{0.45\textwidth}
\includegraphics[scale=0.3]
{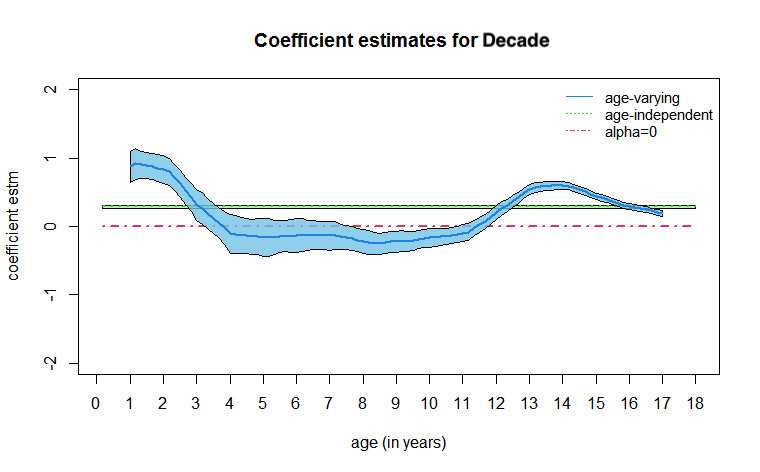}
\caption{\footnotesize Model VCV vs. Model CCV}
   \label{fig:M101a}
  \end{subfigure}
 \hspace{.8cm}
 \begin{subfigure}[t]{0.45\textwidth}
\includegraphics[scale=0.3]
{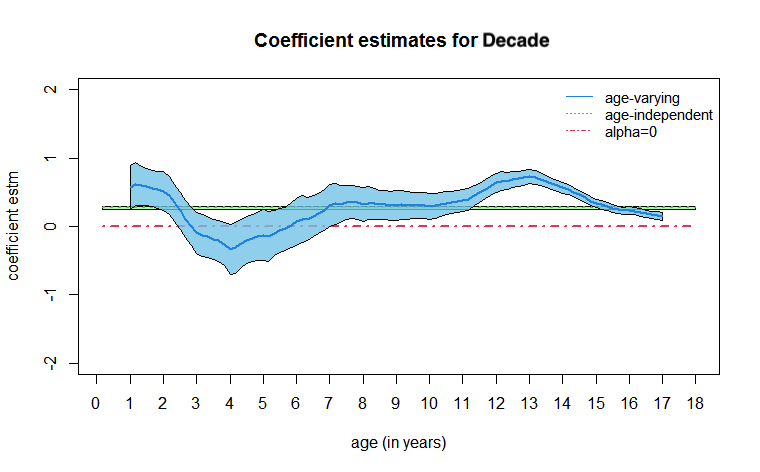}
\caption{\footnotesize Model VVV vs. Model CVV}
   \label{fig:M111a}
\end{subfigure}
\caption{Estimates of $\alpha(\cdot)$, i.e. the effects of \texttt{Decade} under different models: the solid curves are local linear estimates of the age-varying coefficients,
and the dotted lines are estimates from the model assuming $\alpha$ being age-independent.}
   \label{fig:estall}
\end{figure}

\begin{figure}[h]
\centering
 \includegraphics[scale=0.35]{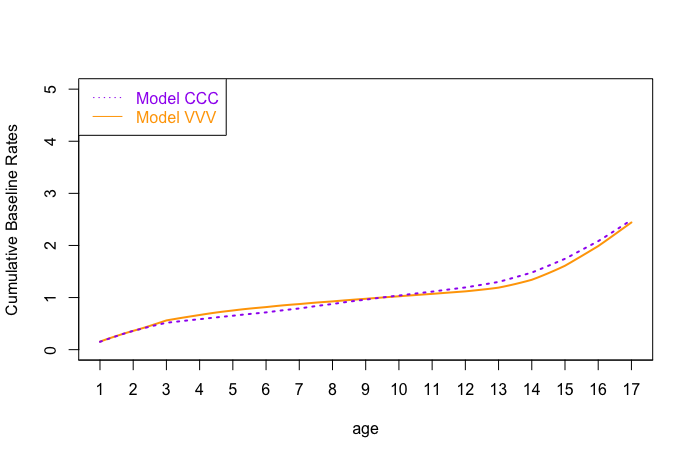}
 \caption{Estimates for $\Lambda_0(\cdot)$ under Models CCC and VVV concerning the general population}
         \label{fig:estcumED1}
\end{figure}

\begin{figure}[htbp]%[htbp]
\includegraphics[scale=0.36]{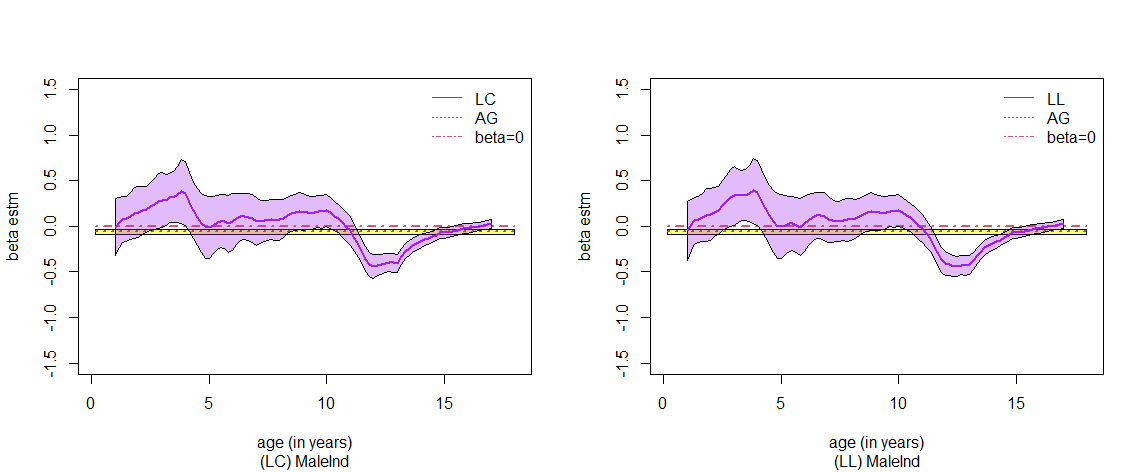}
\includegraphics[scale=0.36]{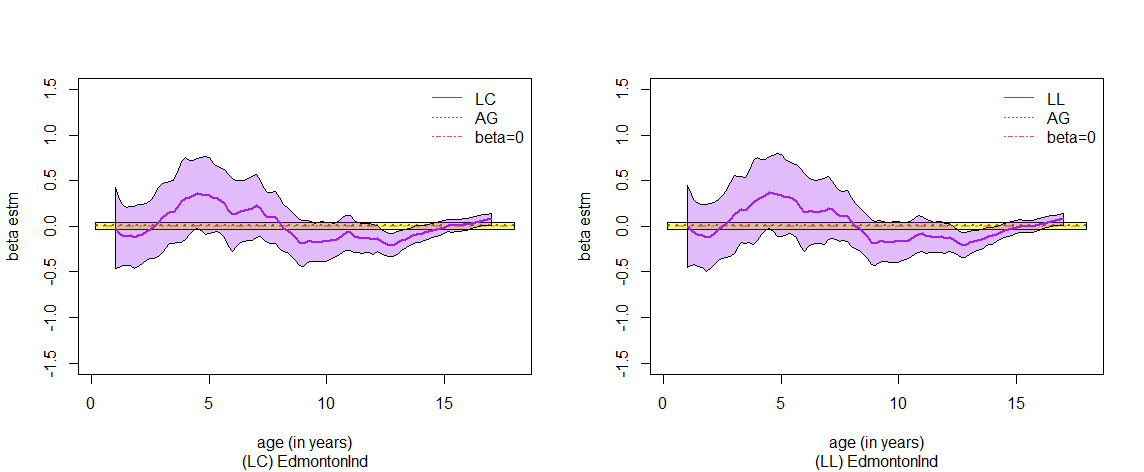}
\includegraphics[scale=0.36]{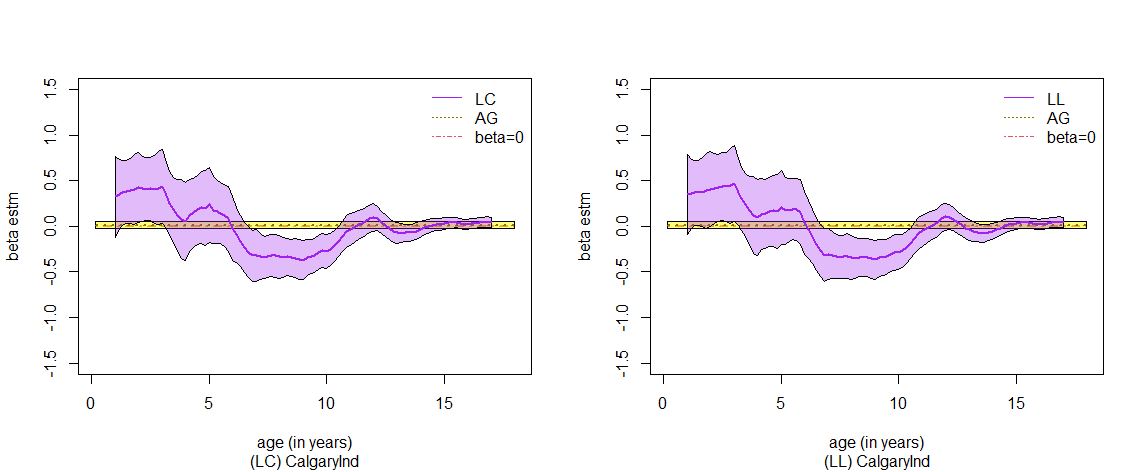}
        \caption{Estimates for the difference in covariate effects between the two time periods 2002--2010 and 2010--2017 concerning the MHED cohort. The left/right panels are the local constant/linear estimates.}
      \label{fig:fig56diffbeta1}
\end{figure}

\begin{figure}%[h]
%\hspace{-1cm}
 \begin{subfigure}[t]{0.45\textwidth}
\includegraphics[scale=0.3]
{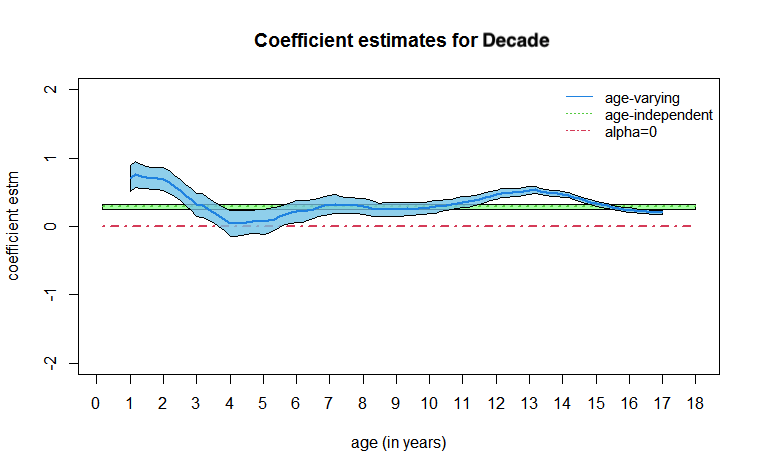}
\caption{\footnotesize Model VCC vs. Model CCC}
   \label{fig:M100a}
 \end{subfigure}
\hspace{.8cm}
 \begin{subfigure}[t]{0.45\textwidth}
\includegraphics[scale=0.3]
{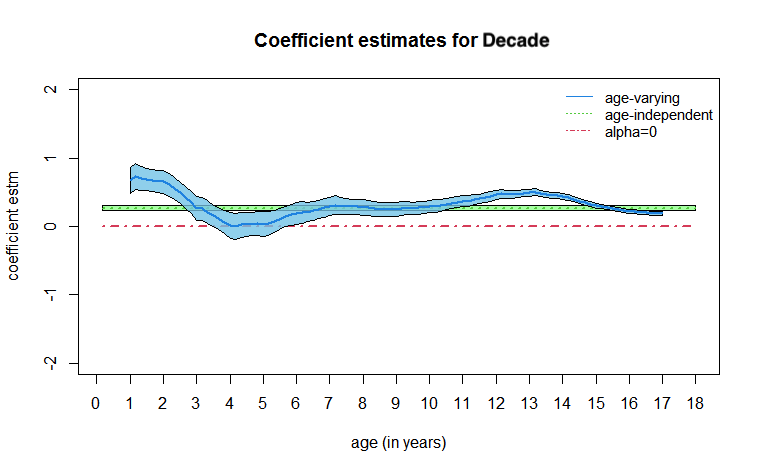}
\caption{\footnotesize Model VVC vs. Model CVC}
   \label{fig:M110a}
\end{subfigure}

 \begin{subfigure}[t]{0.45\textwidth}
\includegraphics[scale=0.3]
{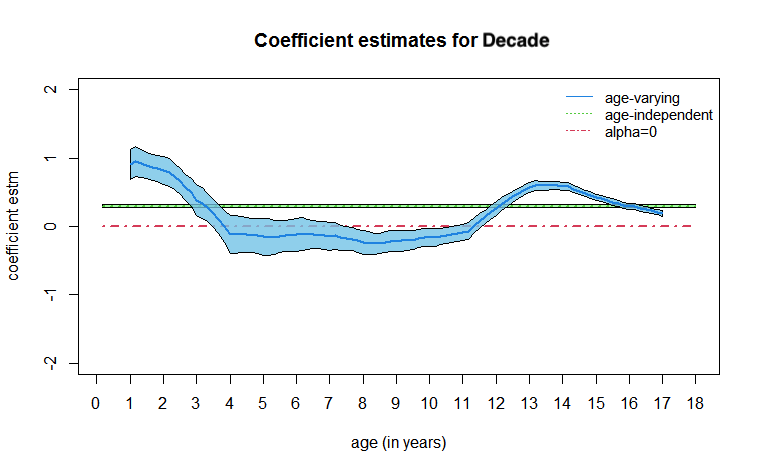}
\caption{\footnotesize Model VCV vs. Model CCV}
   \label{fig:M101a}
  \end{subfigure}
 \hspace{.8cm}
 \begin{subfigure}[t]{0.45\textwidth}
\includegraphics[scale=0.3]
{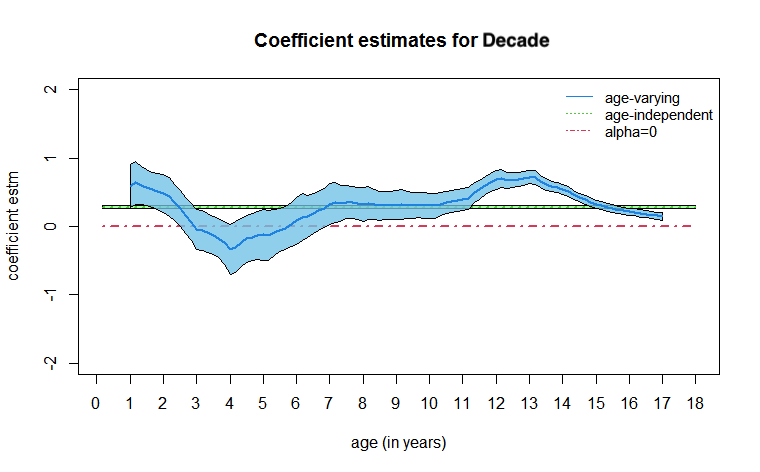}
\caption{\footnotesize Model VVV vs. Model CVV}
   \label{fig:M111a}
\end{subfigure}
\caption{Estimates of $\alpha(\cdot)$, i.e., the effects of \texttt{Decade} under different models: the solid curves are local constant estimates of the age-varying coefficients, and the dotted lines are estimates from the model assuming $\alpha$ being age-independent}
   \label{fig:esta}
\end{figure}

\begin{figure}[h]
 \begin{subfigure}[t]{\textwidth}
\includegraphics[width=0.3\linewidth]
{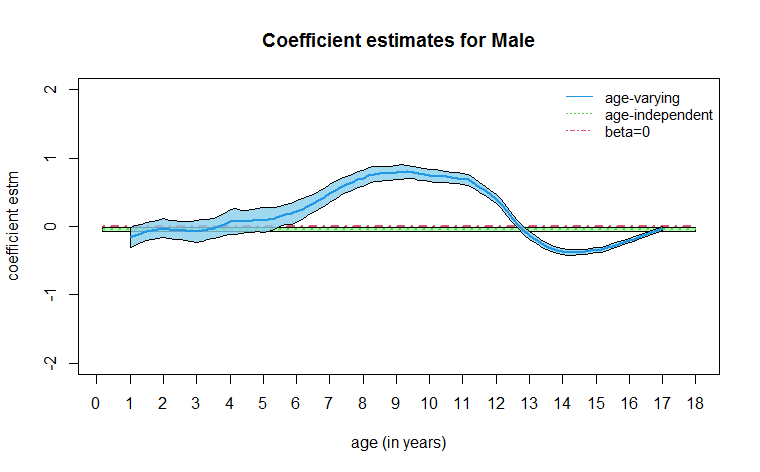}
\includegraphics[width=0.3\linewidth]
{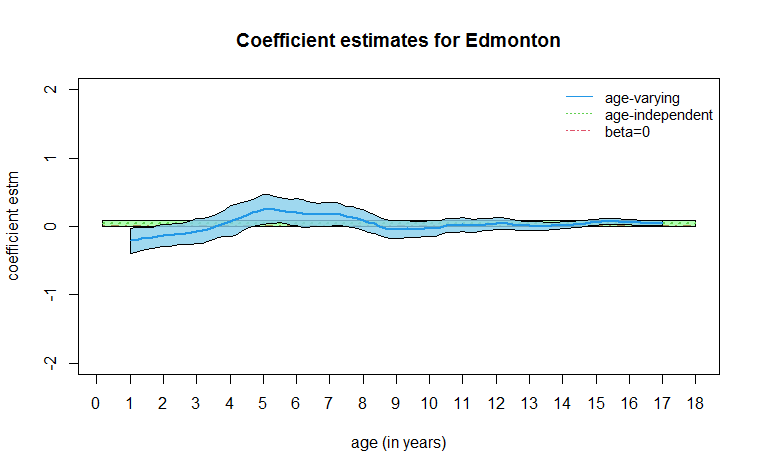}
\includegraphics[width=0.3\linewidth]
{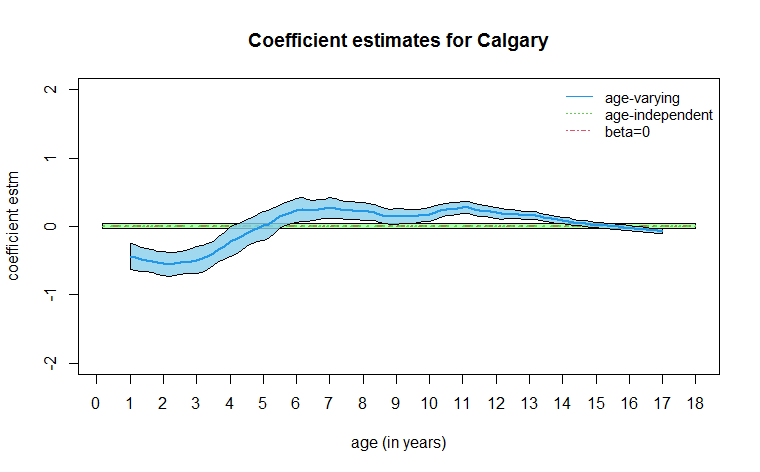}
\caption{\footnotesize Model CVC vs. Model CCC}
   \label{fig:M010}
 \end{subfigure}

  \begin{subfigure}[t]{\textwidth}
\includegraphics[width=0.3\linewidth]
{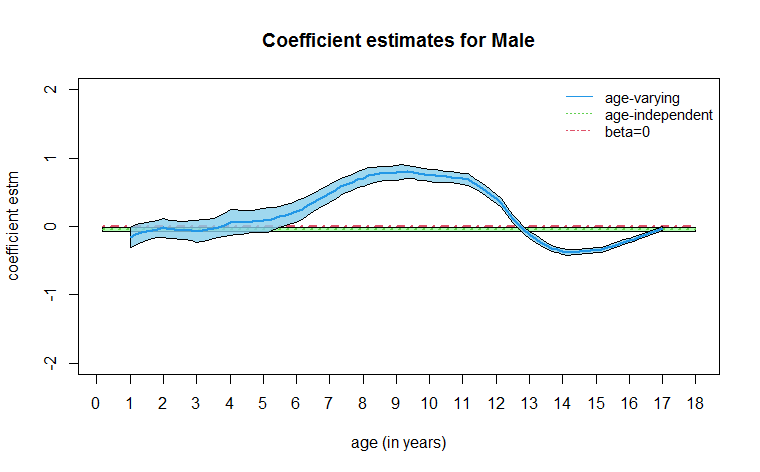}
%\hfill
\includegraphics[width=0.3\linewidth]
{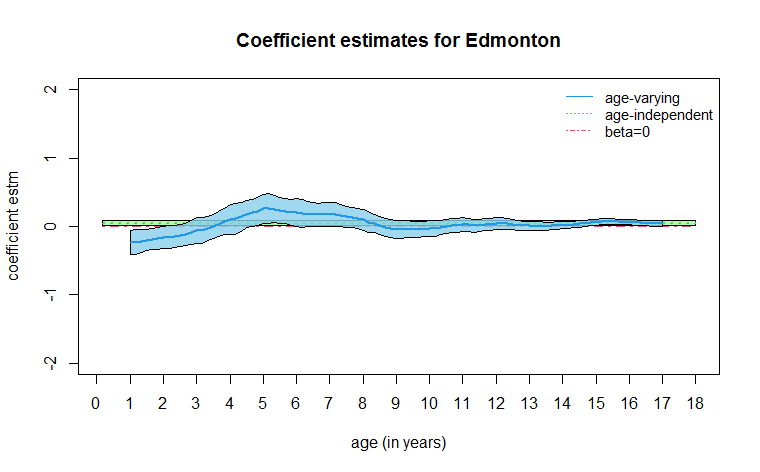}
%\hfill
\includegraphics[width=0.3\linewidth]
{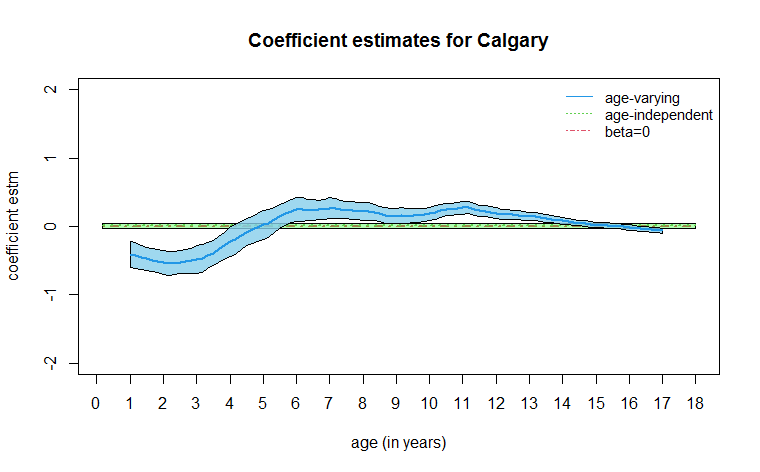}
\caption{\footnotesize Model VVC vs. Model VCC}
   \label{fig:M110}
 \end{subfigure}

  \begin{subfigure}[t]{\textwidth}
\includegraphics[width=0.3\linewidth]
{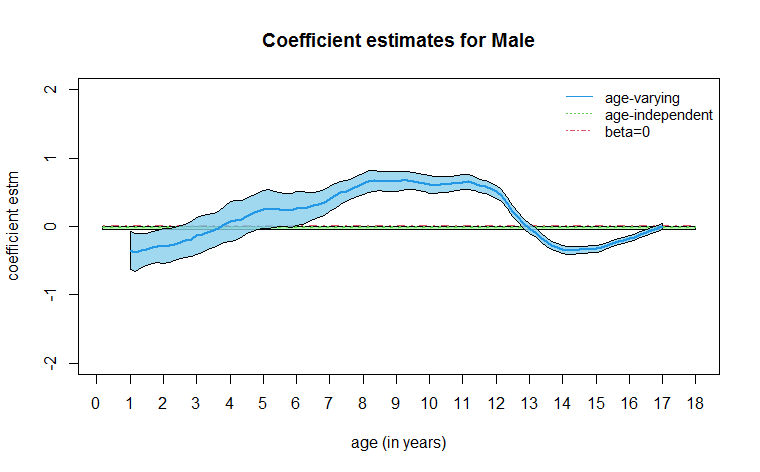}
%\hfill
\includegraphics[width=0.3\linewidth]
{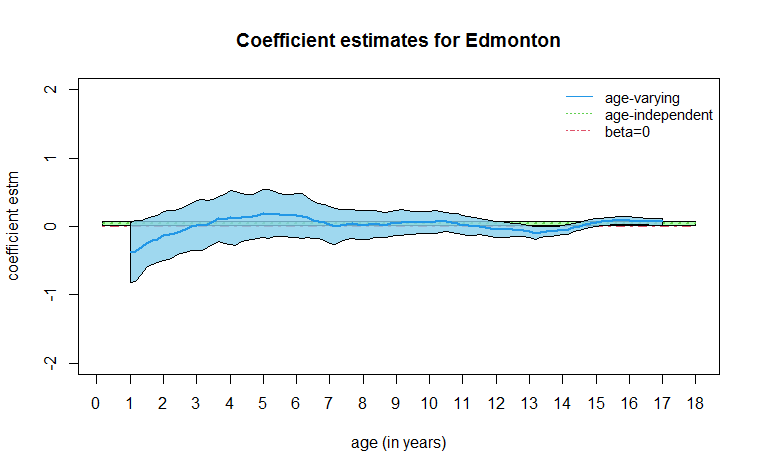}
%\hfill
\includegraphics[width=0.3\linewidth]
{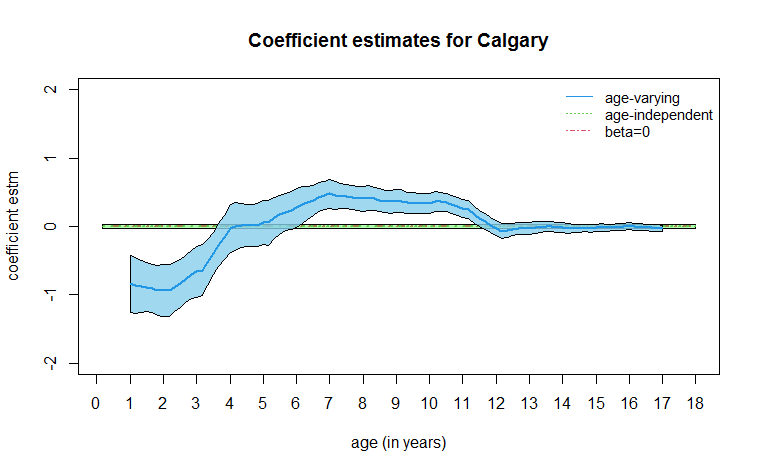}
\caption{\footnotesize Model CVV vs Model CCV}
   \label{fig:M011}
 \end{subfigure}
  \hspace{1cm}
  \begin{subfigure}[t]{\textwidth}
\includegraphics[width=0.3\linewidth]
{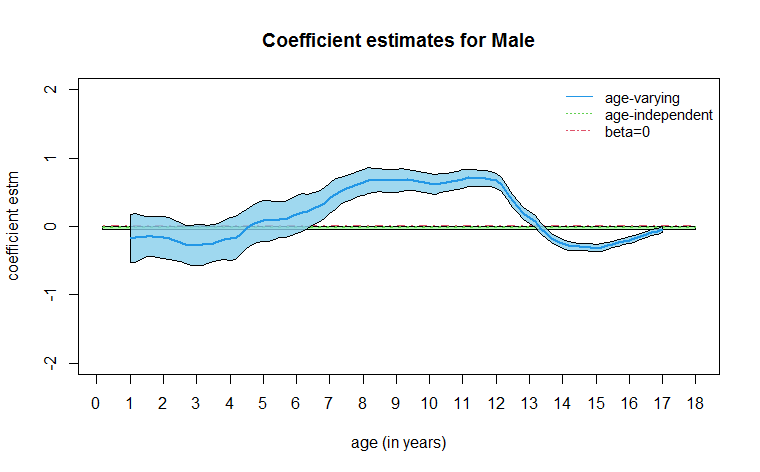}
%\hfill
\includegraphics[width=0.3\linewidth]
{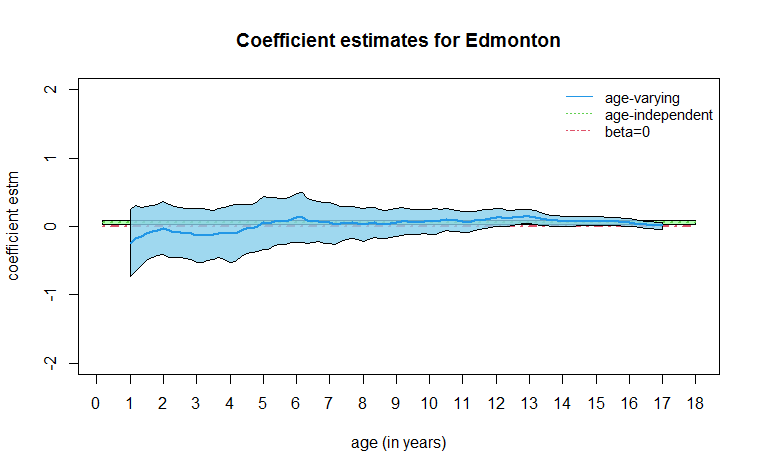}
%\hfill
\includegraphics[width=0.3\linewidth]
{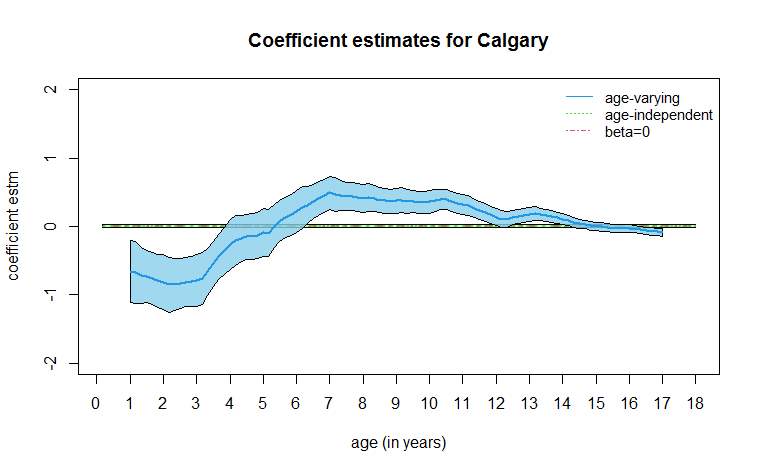}

\caption{\footnotesize Model VVV vs. Model VCV}
   \label{fig:M111}
 \end{subfigure}

\caption{Estimates of $\beta(\cdot)$ under different models: the solid curves are local constant estimates of the age-varying coefficients, and the dotted lines are
estimates from the model assuming $\beta$ being age-independent.}
   \label{fig:estb}
\end{figure}

\begin{figure}[h]

 \begin{subfigure}[t]{\textwidth}
\includegraphics[width=0.3\linewidth]
{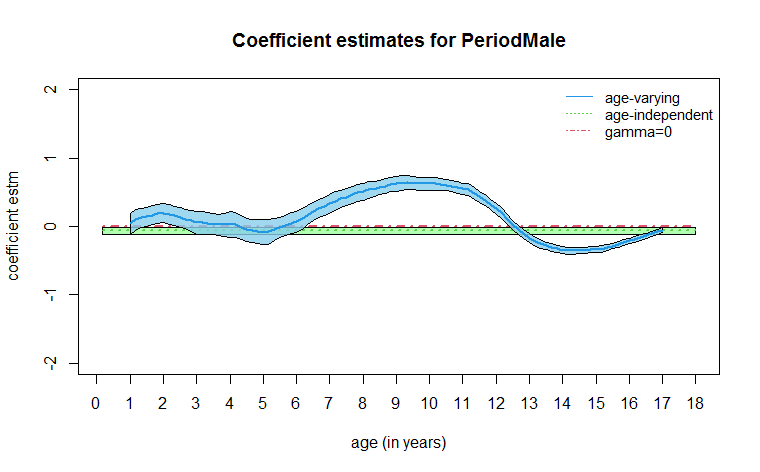}
\includegraphics[width=0.3\linewidth]
{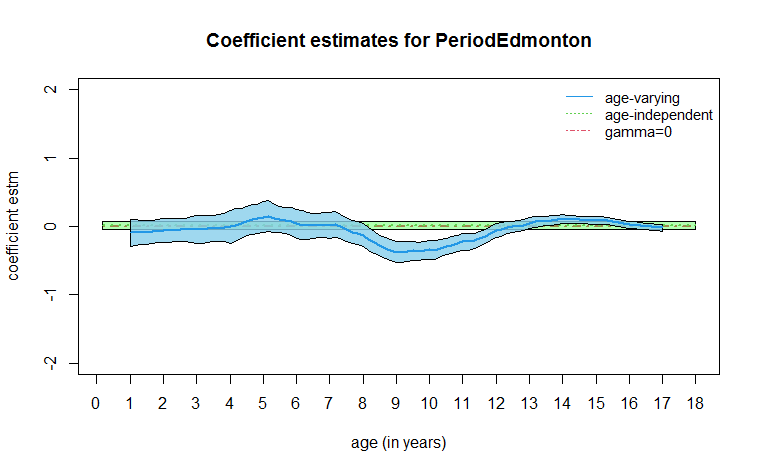}
\includegraphics[width=0.3\linewidth]
{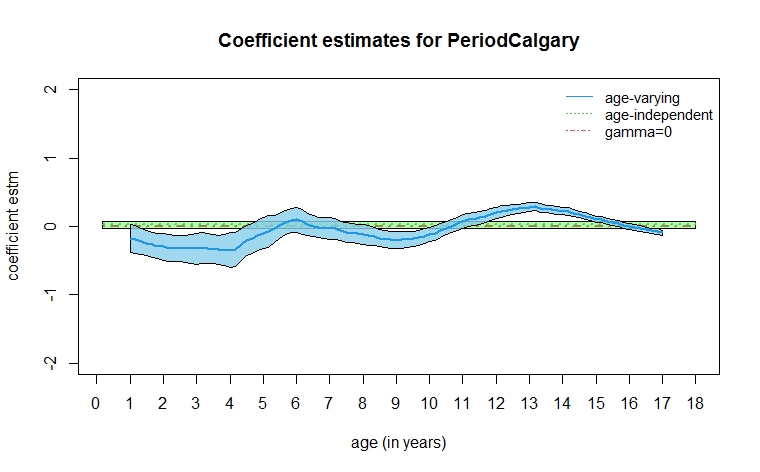}
\caption{\footnotesize Model CCV vs. Model CCC}
   \label{fig:M001c}
 \end{subfigure}

  \begin{subfigure}[t]{\textwidth}
\includegraphics[width=0.3\linewidth]
{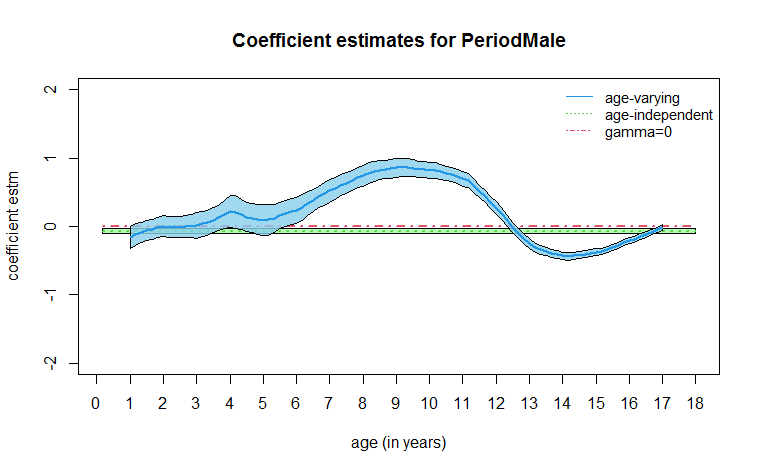}
\includegraphics[width=0.3\linewidth]
{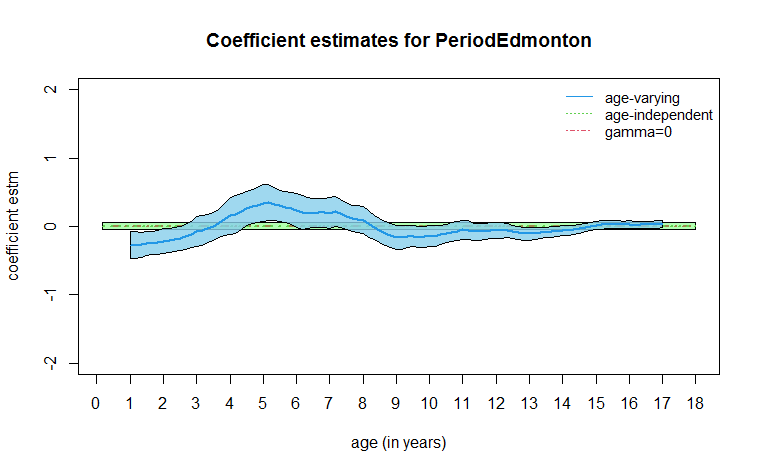}
\includegraphics[width=0.3\linewidth]
{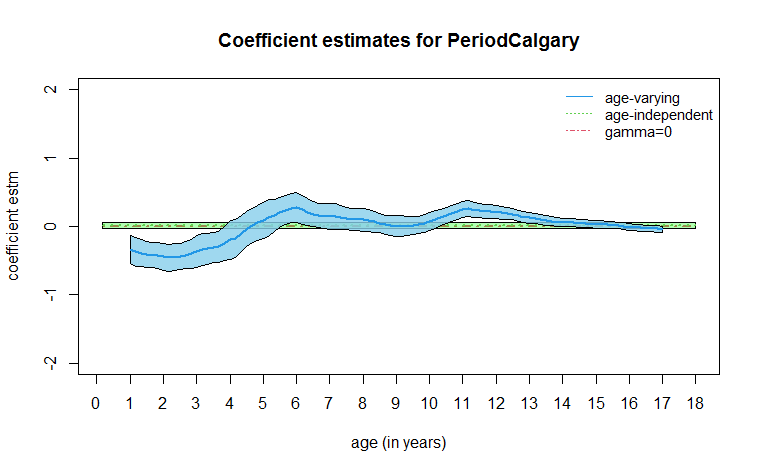}
\caption{\footnotesize Model VCV vs. Model VCC}
   \label{fig:M101c}
 \end{subfigure}

  \begin{subfigure}[t]{\textwidth}
\includegraphics[width=0.3\linewidth]
{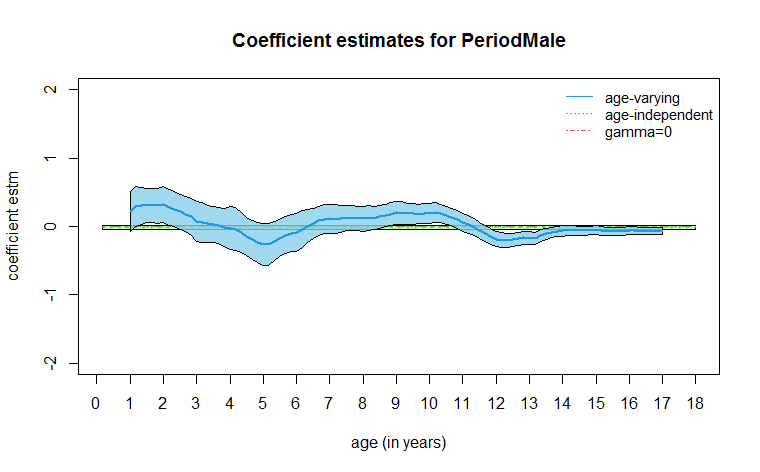}
\includegraphics[width=0.3\linewidth]
{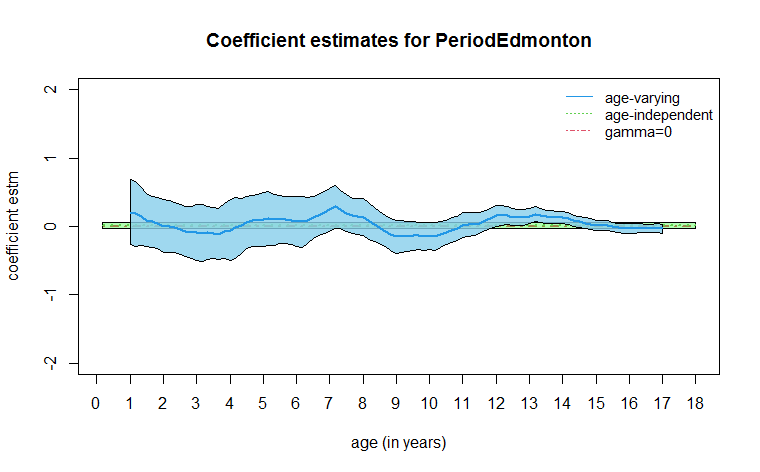}
\includegraphics[width=0.3\linewidth]
{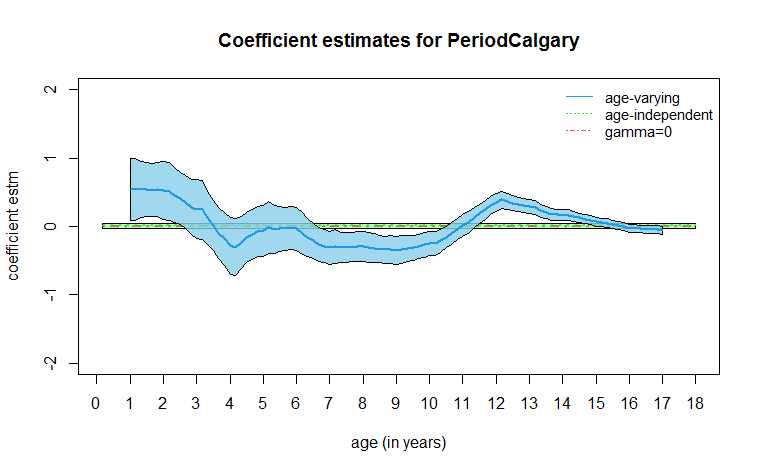}
\caption{\footnotesize Model CVV vs. Model CVC}
   \label{fig:M011c}
 \end{subfigure}

  \begin{subfigure}[t]{\textwidth}
\includegraphics[width=0.3\linewidth]
{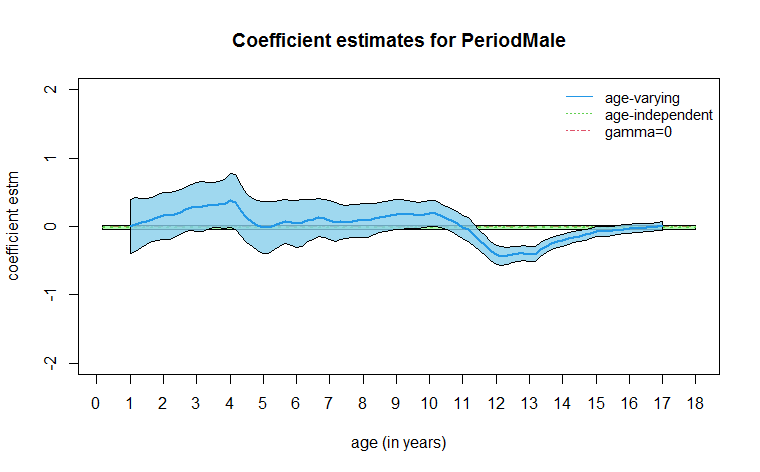}
\includegraphics[width=0.3\linewidth]
{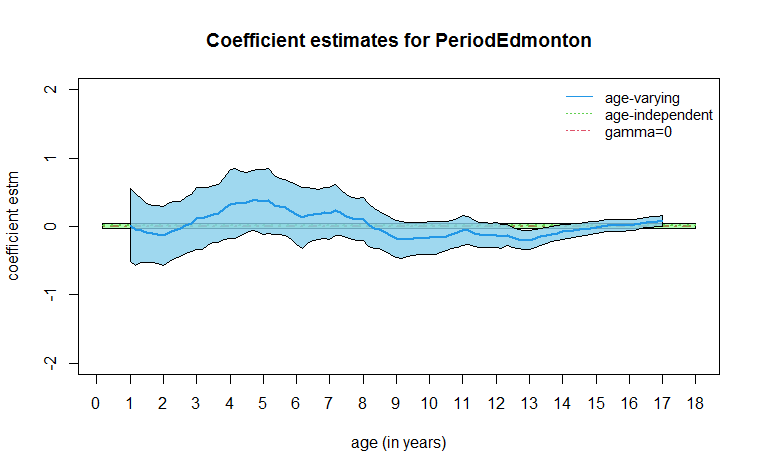}
\includegraphics[width=0.3\linewidth]
{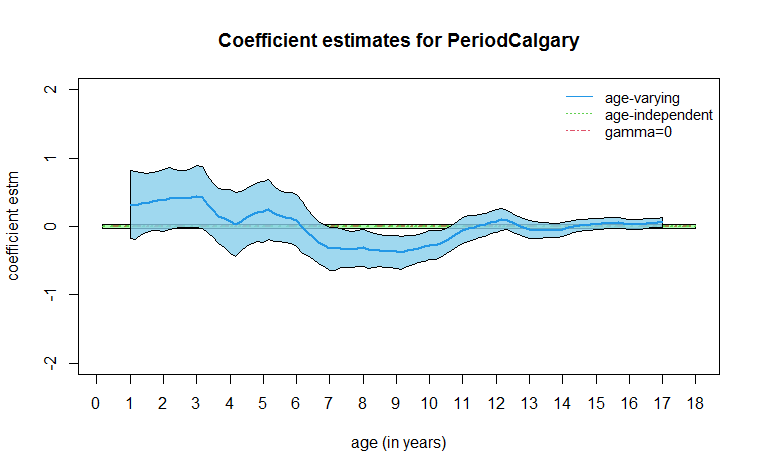}
\caption{\footnotesize Model VVV vs. Model VVC}
   \label{fig:M111c}
 \end{subfigure}
\caption{Estimates of $\gamma(\cdot)$ under different models: the solid curves are local constant estimates of the age-varying coefficients, and the dotted lines are estimates from the model assuming $\gamma$ being age-independent}
   \label{fig:estc}
\end{figure}

\begin{figure}[h]

 \begin{subfigure}[t]{\textwidth}
\includegraphics[width=0.3\linewidth]
{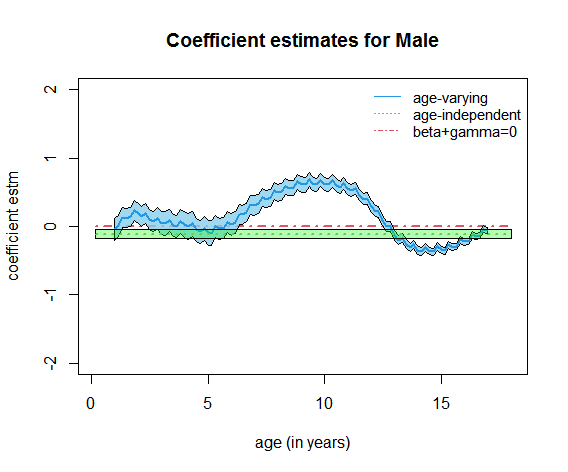}
\includegraphics[width=0.3\linewidth]
{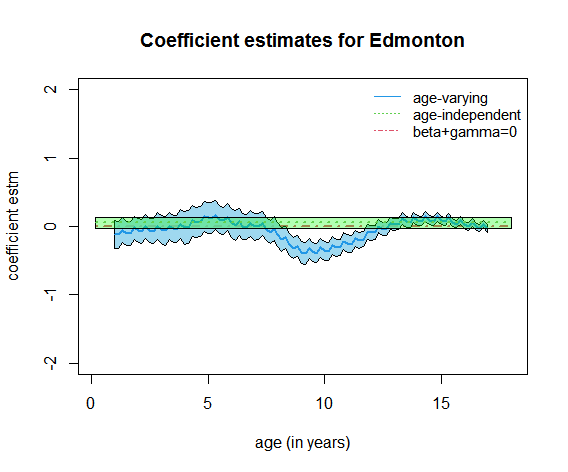}
\includegraphics[width=0.3\linewidth]
{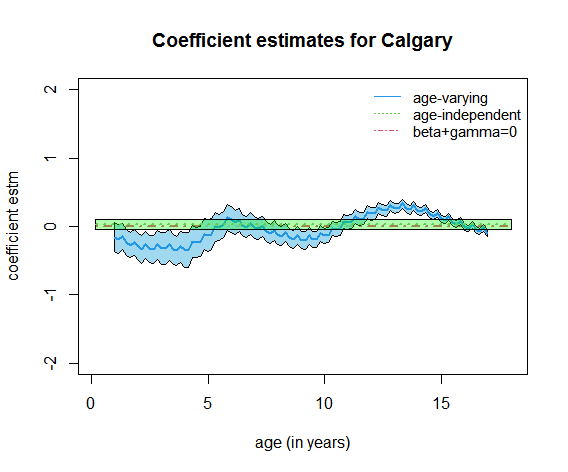}
\caption{\footnotesize Model CCV vs. Model CCC}
   \label{fig:M001c2}
 \end{subfigure}

  \begin{subfigure}[t]{\textwidth}
\includegraphics[width=0.3\linewidth]
{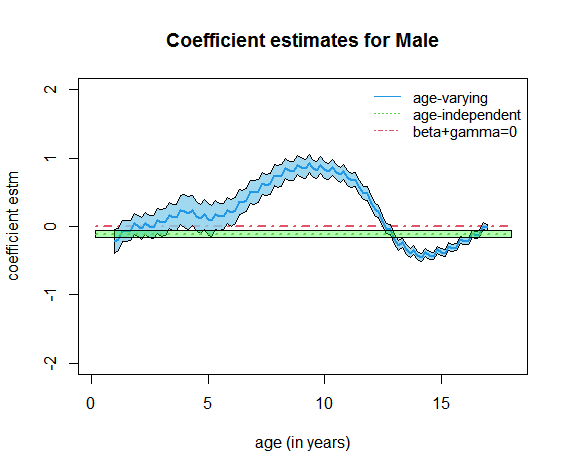}
\includegraphics[width=0.3\linewidth]
{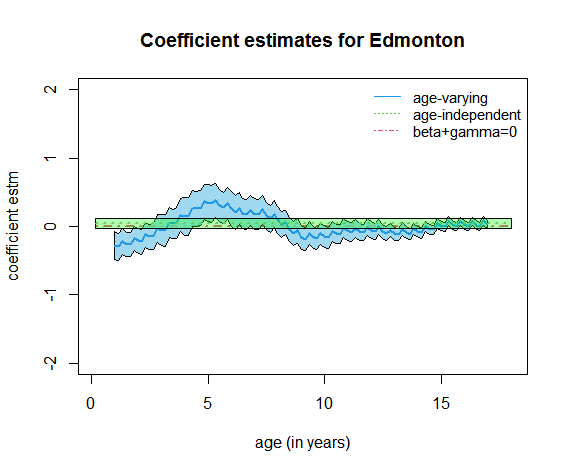}
\includegraphics[width=0.3\linewidth]
{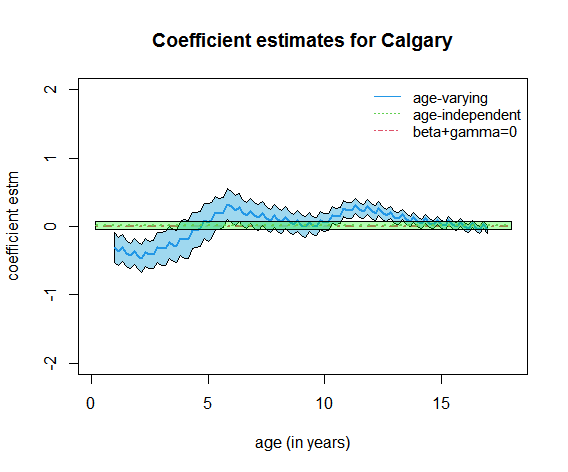}
\caption{\footnotesize Model VCV vs. Model VCC}
   \label{fig:M101c2}
 \end{subfigure}

  \begin{subfigure}[t]{\textwidth}
\includegraphics[width=0.3\linewidth]
{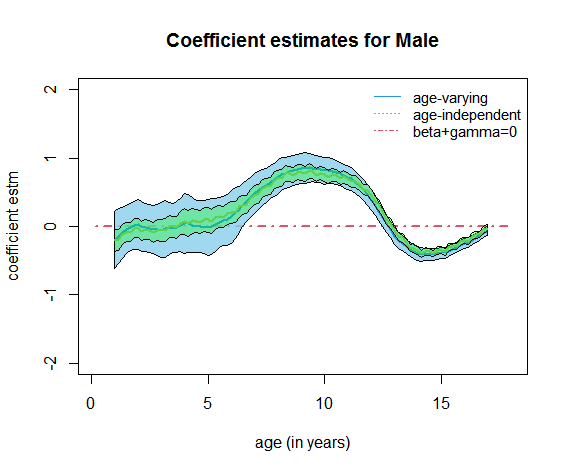}
\includegraphics[width=0.3\linewidth]
{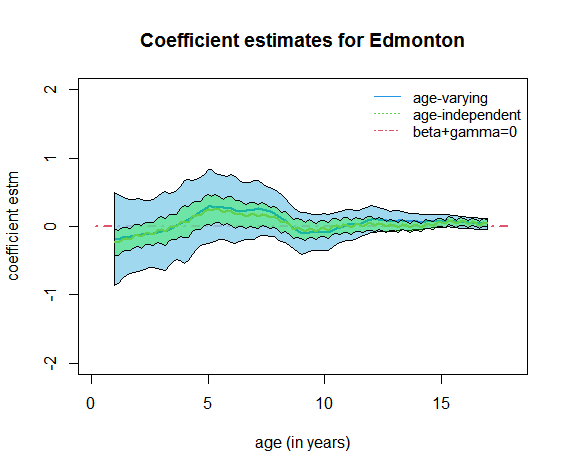}
\includegraphics[width=0.3\linewidth]
{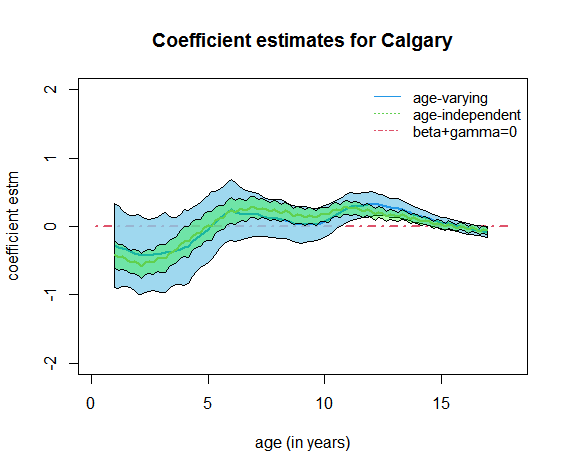}
\caption{\footnotesize Model CVV vs. Model CVC}
   \label{fig:M011c2}
 \end{subfigure}

  \begin{subfigure}[t]{\textwidth}
\includegraphics[width=0.3\linewidth]
{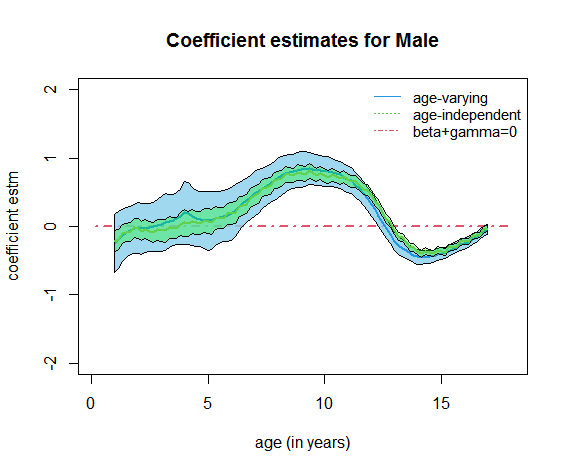}
\includegraphics[width=0.3\linewidth]
{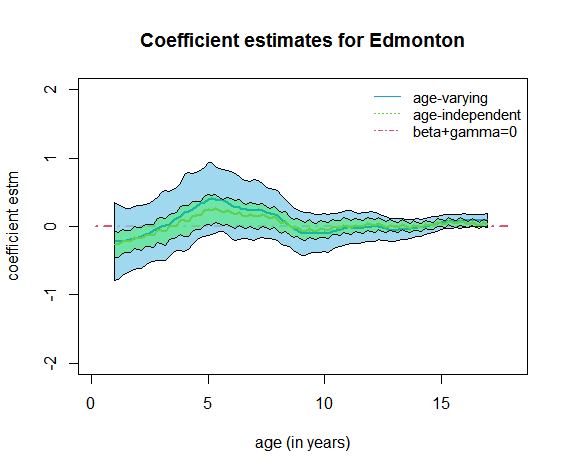}
\includegraphics[width=0.3\linewidth]
{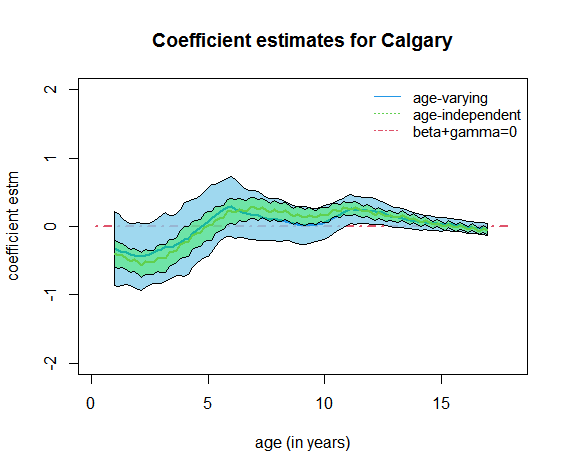}
\caption{\footnotesize Model VVV vs. Model VVC}
   \label{fig:M111c2}
 \end{subfigure}
\caption{Estimates of $\beta(\cdot)+\gamma(\cdot)$ under different models: the solid curves are obtained by assuming $\gamma(\cdot)$ being age-varying, and the dotted lines are estimates from the model assuming $\gamma$ being age-independent.}
   \label{fig:estc2}
\end{figure}

%% file: supp2023floatsD.tex
\begin{table}[htbp]
\caption{Simulation Setting 1: concerning $\mathcal{P}_1$}
\label{tab:simA11}
\hspace{-2cm}
\begin{tabular}{lllllllllllllll}
  \hline
&  \multicolumn{5}{c}{$\mathcal{O}_{E1},\mathcal{O}_{L1}$}& \multicolumn{5}{c}{$\mathcal{O}_{E1}\cup\mathcal{O}_{L1}$}&\multicolumn{4}{c}{$\mathcal{O}_1$}\\
&  \multicolumn{5}{c}{(A.1.1)}& \multicolumn{5}{c}{(A.1.2)}&\multicolumn{4}{c}{(A.1.3)}\\
\cline{2-5}\cline{7-10}\cline{12-15}
\multicolumn{13}{l}{\it Case 1: $, \alpha=0, \beta=.7,\gamma=0$}\\
\cline{1-4}
     & $\hat{\beta}_E$ & $\hat{\beta}_L$&$\hat{\lambda}_{0,E}$ & $\hat{\lambda}_{0,L}$& & $\hat{\alpha}$ &$\hat{\beta}$&$\hat{\gamma}$ &$\hat{\lambda}_0$ && $\hat{\alpha}$ &$\hat{\beta}$&$\hat{\gamma}$&$\hat{\lambda}_0$ \\
     \cline{2-5}\cline{7-10}\cline{12-15}
SMean & .0434 & .0384 &.0258&.0287&& .1040 & .0432 & -.0050& .0256 &&& .0697&&.0166 \\
 SSE &.0110 & .0113&.0002&.0003 && .0120 & .0104 & .0148&.0002 &&& .0111&&.0002 \\
  ESE$^{(a)}$ & .0378 & .0403&& && .0472 & .0376 & &.0549 &&& .0283&& \\
   ESE$^{(b)}$ &.0360 & .0386&& && .0455 & .0358 & &.0525&& & .0262&& \\
   \hline
   \hline
    \multicolumn{15}{l}{\it Case 2: $, \alpha=.3, \beta=.7,\gamma=.15$}\\
  \cline{1-4}
     & $\hat{\beta}_E$ & $\hat{\beta}_L$&$\hat{\lambda}_{0,E}$ & $\hat{\lambda}_{0,L}$& & $\hat{\alpha}$ &$\hat{\beta}$&$\hat{\gamma}$ &$\hat{\lambda}_0$ && $\hat{\alpha}$ &$\hat{\beta}$&$\hat{\gamma}$&$\hat{\lambda}_0$ \\
\cline{2-5}\cline{7-10}\cline{12-15}
SMean &.0420 & .0674 &.0260 & .0292&& .1169 & .0420 & .0254& .0260 && .2405 & .0230 & .1245& .0150 \\
 SSE &.0106 & .0114 &.0002 & .0003 && .0136 & .0107 & .0154& .0002   && .0340 & .0231 & .0382& .0003  \\
  ESE$^{(a)}$ &.0378 & .0346&& && .0442 & .0378 & .0512 &&& .0443 & .0375 & .0492&\\
   ESE$^{(b)}$ &.0360 & .0326&& && .0424 & .0360 & .0485& &&.0426 & .0357 & .0483& \\
   \hline
  \multicolumn{15}{l}{ESE$^{(a)}$: sample mean of the standard error estimates using the sandwich estimator}\\
    \multicolumn{15}{l}{ESE$^{(b)}$: sample mean of the standard error estimates using the Fisher\rq{}s Information}\\
\end{tabular}
\end{table}

\begin{figure}[htbp]
 \begin{subfigure}[t]{.45\textwidth}
\includegraphics[scale=.25]
{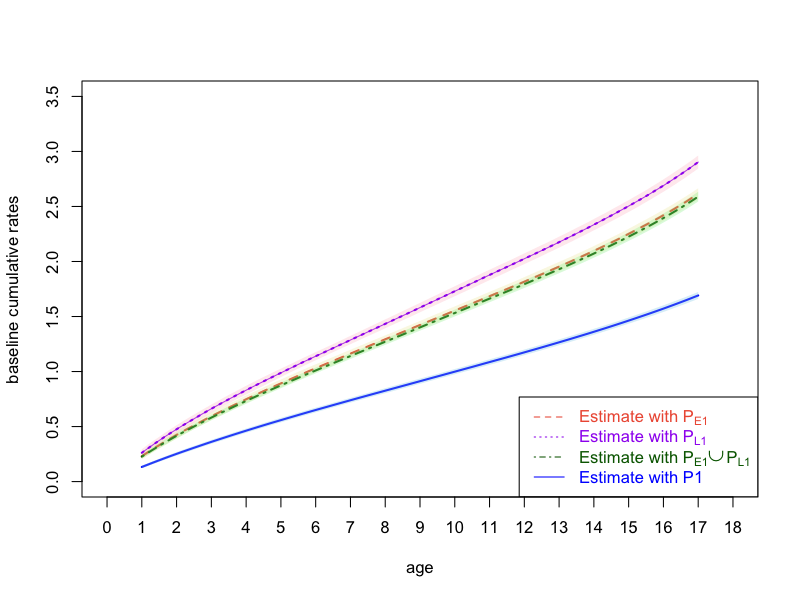}
\caption{\footnotesize Case 1}
   \label{fig:Lambda002A1}
 \end{subfigure}
 \hspace{.5cm}
 \begin{subfigure}[t]{.45\textwidth}
\includegraphics[scale=.25]
{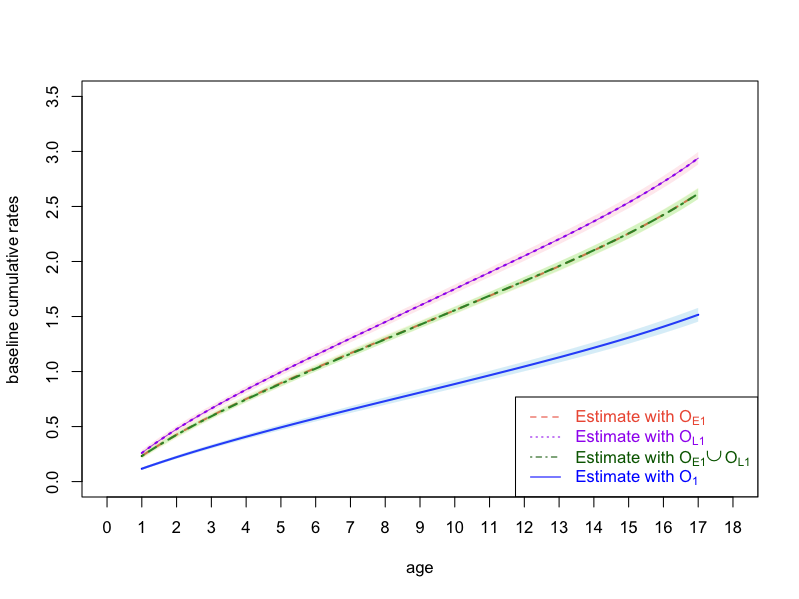}
\caption{\footnotesize Case 2}
   \label{fig:Lambda002A2}
 \end{subfigure}
       \caption{Simulation Setting 1: estimates for $\Lambda_0(\cdot)$ concerning with $\mathcal{P}_1$}
      \label{fig:Lambda002A}
\end{figure}

\begin{figure}[htbp]
 \begin{subfigure}[t]{.45\textwidth}
\includegraphics[scale=.25]
{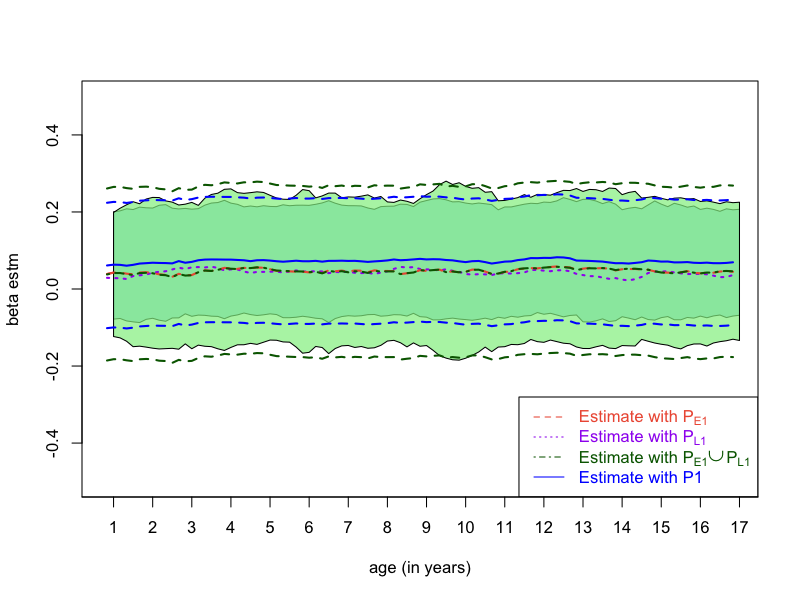}
\caption{\footnotesize concerning with $\mathcal{P}_1$}
   \label{fig:Beta002A}
 \end{subfigure}
 \hspace{.5cm}
 \begin{subfigure}[t]{.45\textwidth}
\includegraphics[scale=.25]
{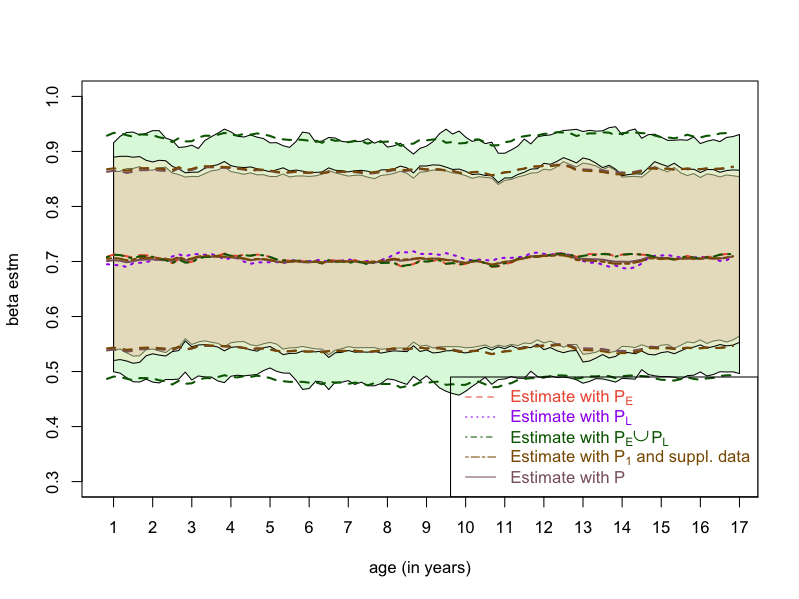}
\caption{\footnotesize concerning with $\mathcal{P}$}
   \label{fig:Beta002B}
 \end{subfigure}
       \caption{Simulation Setting 1 Case 1: estimates for $\beta(\cdot)$. The shaded areas represent the $95\%$ CI obtained using the empirical standard errors over 300 runs of simulation. The long dashed lines represent the $95\%$ CI obtained using the sample mean of $\hat{se}^{(a)}(\hat{\beta})$. }
      \label{fig:simBeta002LL}
\end{figure}

\begin{table}[htbp]
\caption{Simulation Setting 1 Case 1: $\alpha=0,\beta=.7,\gamma=0$, concerning  $\mathcal{P}$.}
\label{tab:simB11}
\centering
\begin{tabular}{lllll l llll}
  \hline
&  \multicolumn{4}{c}{$\mathcal{O}_E,\mathcal{O}_L$}&&
 \multicolumn{4}{c}{$\mathcal{O}_{E1},\mathcal{O}_{L1}$ with Suppl. Data}\\
 &  \multicolumn{4}{c}{(B.1.1)}&&
 \multicolumn{4}{c}{(B.1.2)}\\
& $\hat{\beta}_E$ & $\hat{\beta}_L$&$\hat{\lambda}_{0,E}$ & $\hat{\lambda}_{0,L}$&&
  $\tilde{\beta}_E$ & $\tilde{\beta}_L$&$\tilde{\lambda}_{0,E}$ & $\tilde{\lambda}_{0,L}$\\
 \cline{2-5}\cline{7-10}
  SMean &.7017 & .6995 &.0020&.0020&& .7021 & .7004&.0020&.0020 \\
   SSE & .0353 & .0402 &.0001&.0001&& .0359 & .0405&.0001&.0001 \\
   ESE$^{(a)}$  & .0376 & .0400 &&&& .0378 & .0403& \\
  ESE$^{(b)}$ & .0358 & .0383& &&& .0360 & .0385& \\
  \hline
 \hline
 &  \multicolumn{4}{c}{$\mathcal{O}_E\cup\mathcal{O}_L$}&
   & \multicolumn{4}{c}{$\mathcal{O}_{E1}\cup\mathcal{O}_{L1}$ with Suppl. Data}\\
  &  \multicolumn{4}{c}{(B.1.3)}&&
    \multicolumn{4}{c}{(B.1.4)}\\
   & $\hat{\alpha}$ &$\hat{\beta}$&$\hat{\gamma}$ &$\hat{\lambda}_0$&&$\tilde{\alpha}$ &$\tilde{\beta}$&$\tilde{\gamma}$&$\tilde{\lambda}_0$ \\
   \cline{2-5}\cline{7-10}
 SMean &  .0009 & .7012 & -.0018&.0020& & .0038 & .7014 & -.0016&.0020 \\
  SSE & .0460 & .0349 & .0520 &.0001&& .0464 & .0355 & .0526&.0001 \\
  ESE$^{(a)}$ & .0467 & .0373 & .0544&& & .0467 & .0373 & .0544& \\
  ESE$^{(b)}$ & .0450 & .0355 & .0519 &&& .0450 & .0355 & .0519& \\
 \hline
 \hline
 & \multicolumn{4}{c}{$\mathcal{O}$}&&\multicolumn{4}{c}{$\mathcal{O}_1$ and Suppl. data}\\
  &  \multicolumn{4}{c}{(B.1.5)}&&
    \multicolumn{4}{c}{(B.1.6)}\\
 & \multicolumn{2}{c}{$\hat{\beta}$}&\multicolumn{2}{c}{$\tilde{\lambda}_0$}&&\multicolumn{2}{c}{$\tilde{\beta}$}&\multicolumn{2}{c}{$\tilde{\lambda}_0$} \\
 \cline{2-5}\cline{7-10}
SMean &\multicolumn{2}{c}{.7005}&\multicolumn{2}{c}{.0020}&&\multicolumn{2}{c}{.7010}& \multicolumn{2}{c}{.0020}\\
SSE & \multicolumn{2}{c}{.0265 }& \multicolumn{2}{c}{.0001 }&&\multicolumn{2}{c}{.0268}& \multicolumn{2}{c}{.0001 } \\
ESE$^{(a)}$ & \multicolumn{2}{c}{.0282}&&&&\multicolumn{2}{c}{.0283} \\
ESE$^{(b)}$ & \multicolumn{2}{c}{.0262}&&&&\multicolumn{2}{c}{.0262} \\

   \hline

\end{tabular}
\end{table}

%% file: supp2023floatsE.tex
\begin{table}[htbp]
\caption{Simulation Setting 2: $\lambda_0=.002, \alpha=.6, \beta=.7,\gamma=.35$,  concerning with  $\mathcal{P}_1$.}
\label{tab:simA22}
\hspace{-1cm}
\begin{tabular}{ll@{  }l@{  }l@{  }l@{  }l l@{  }l@{  }l@{  }ll@{  } l@{  }l@{  }l@{  }l}
  \hline
&  \multicolumn{5}{c}{$\mathcal{O}_{E1},\mathcal{O}_{L1}$}& \multicolumn{5}{c}{$\mathcal{O}_{E1}\cup\mathcal{P}_{L1}$}&\multicolumn{4}{c}{$\mathcal{O}_1$}\\
&  \multicolumn{5}{c}{(A.2.1)}& \multicolumn{5}{c}{(A.2.2)}&\multicolumn{4}{c}{(A.2.3)}\\
   & $\hat{\beta}_E$ & $\hat{\beta}_L$&$\hat{\lambda}_{0,E}$ & $\hat{\lambda}_{0,L}$& & $\hat{\alpha}$ &$\hat{\beta}$&$\hat{\gamma}$ &$\hat{\lambda}_0$ && $\hat{\alpha}$ &$\hat{\beta}$&$\hat{\gamma}$&$\hat{\lambda}_0$ \\
\cline{2-5}\cline{7-10}\cline{12-15}
SMean &.0834 & .1099 &.0261 & .0290 && .1055 & .0865 & .0243 & .0261 && .0632 & .0685 & .1463 & .0166 \\
 SSE & .0114 & .0103  & .0003 & .0003&& .0130 & .0115& .0155 & .0002 & & .0189 & .0155 & .0187& .0002  \\
  ESE$^{(a)}$ &.0340 & .0321&&& & .0408 & .0340 & .0467 &&& .0451 & .0392 & .0500 \\
   ESE$^{(b)}$ & .0319 & .0297 &&&& .0387 & .0318 & .0435 &&& .0424 & .0363 & .0453 \\
   \hline

\end{tabular}
\end{table}

\begin{figure}[htbp]
 \begin{subfigure}[t]{.45\textwidth}
\includegraphics[scale=.25]
{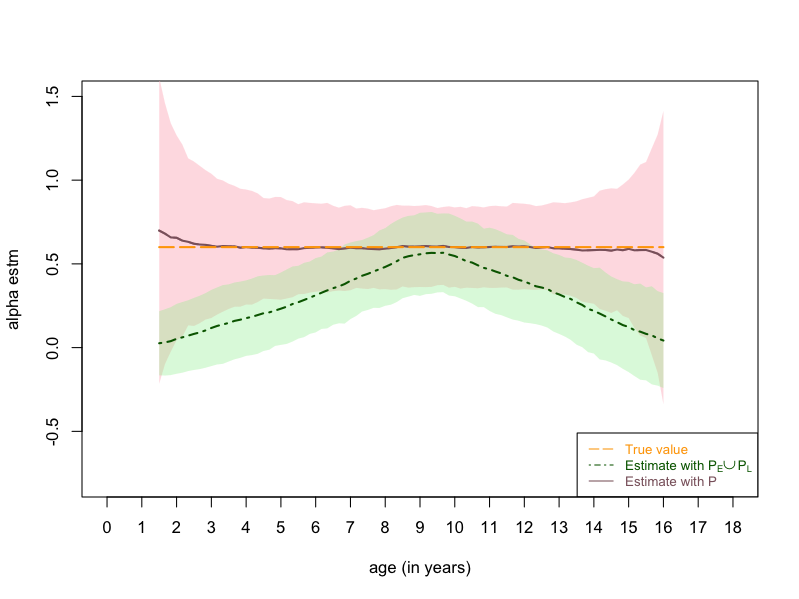}
\caption{\footnotesize $\hat{\alpha}(\cdot)$ with (B.2.3),(B.2.5)}
   \label{fig:alpha002B2All}
 \end{subfigure}
  \begin{subfigure}[t]{.45\textwidth}
\includegraphics[scale=.25]
{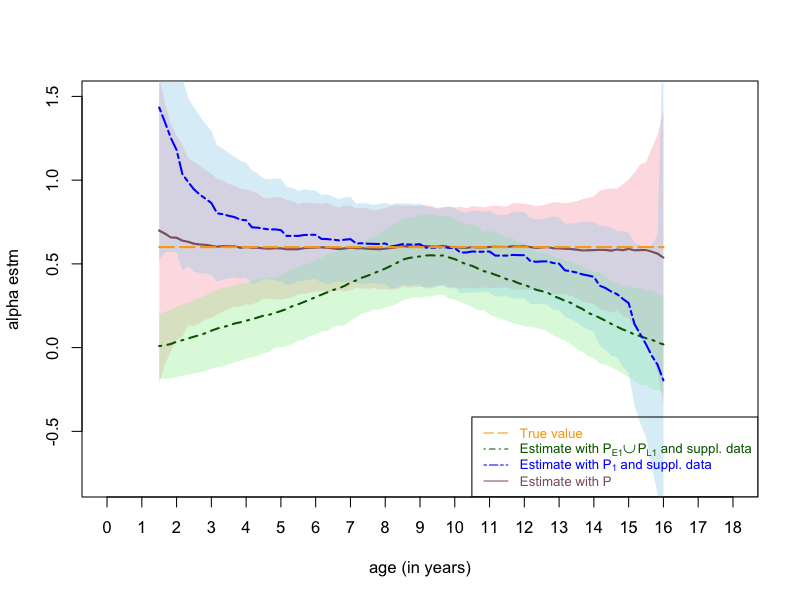}
\caption{\footnotesize $\tilde{\alpha}(\cdot)$ with (B.2.4),(B.2.6)}
   \label{fig:alpha002B2Supp}
 \end{subfigure}

 \begin{subfigure}[t]{.45\textwidth}
\includegraphics[scale=.25]
{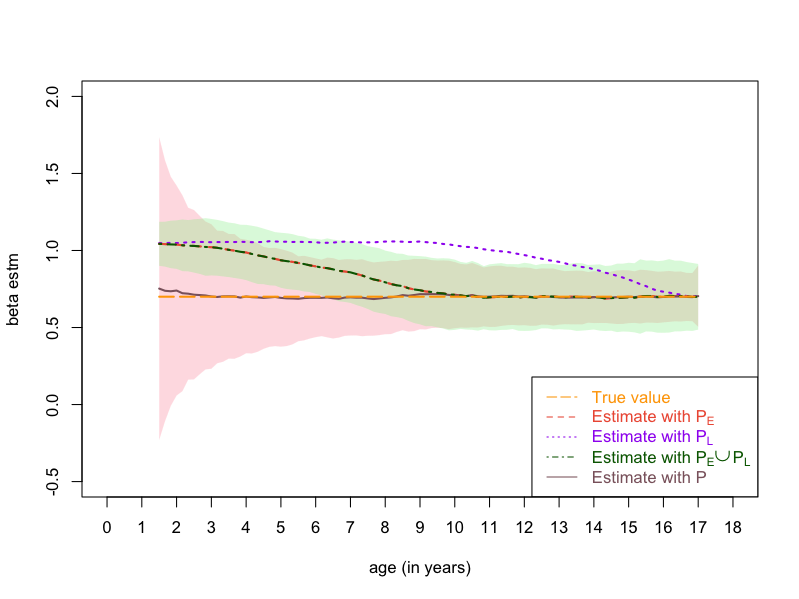}
\caption{\footnotesize $\hat{\beta}(\cdot)$ with (B.2.1),(B.2.3),(B.2.5)}
   \label{fig:Beta002B2All}
 \end{subfigure}
  \begin{subfigure}[t]{.45\textwidth}
\includegraphics[scale=.25]
{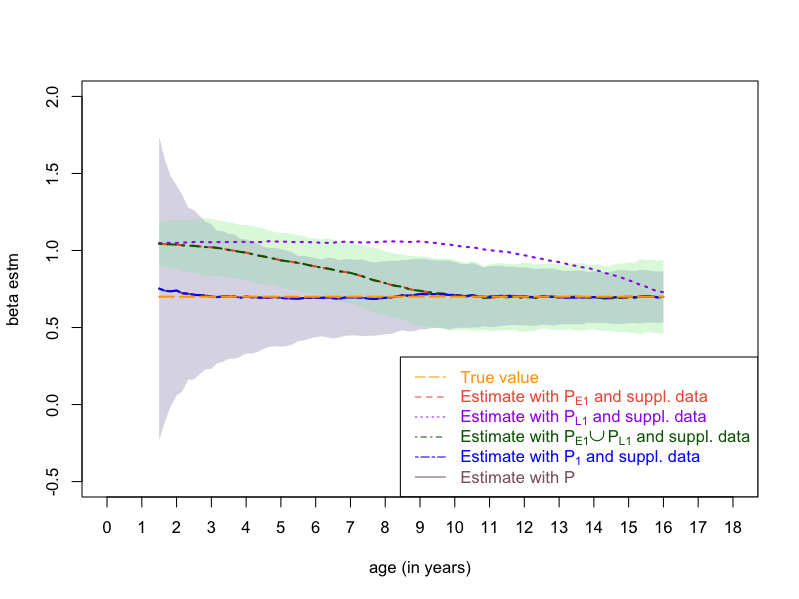}
\caption{\footnotesize $\tilde{\beta}(\cdot)$ with (B.2.2),(B.2.4),(B.2.6)}
   \label{fig:Beta002B2Supp}
 \end{subfigure}

  \begin{subfigure}[t]{.45\textwidth}
\includegraphics[scale=.25]
{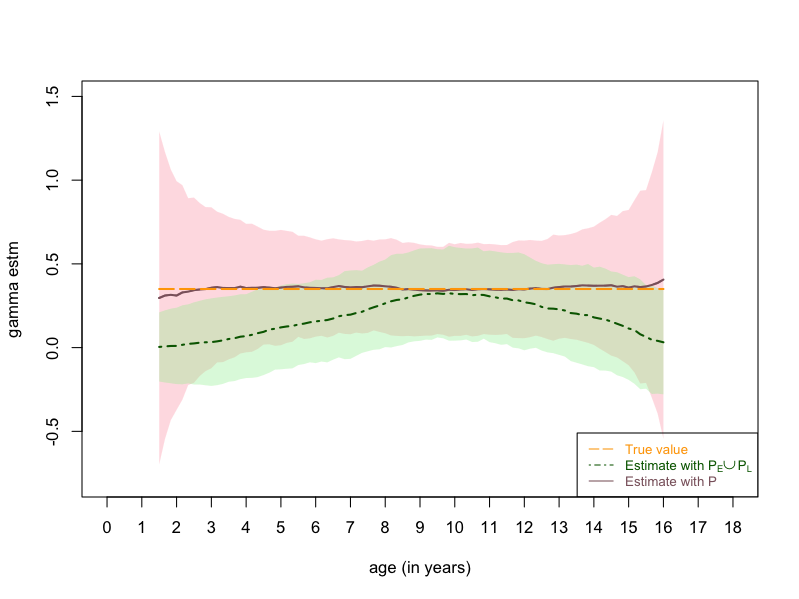}
\caption{\footnotesize $\hat{\gamma}(\cdot)$ with (B.2.3),(B.2.5)}
   \label{fig:gamma002B2All}
 \end{subfigure}
  \begin{subfigure}[t]{.45\textwidth}
\includegraphics[scale=.25]
{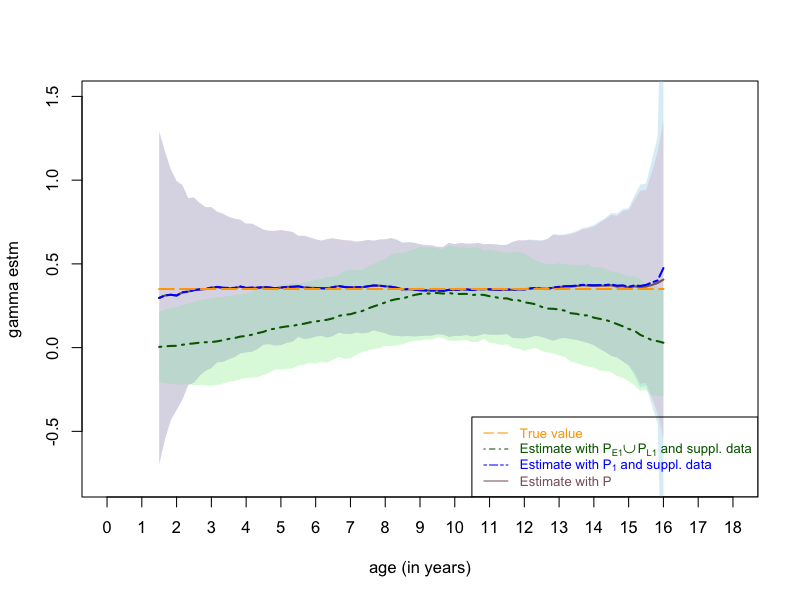}
\caption{\footnotesize $\tilde{\gamma}(\cdot)$ with (B.2.4),(B.2.6)}
   \label{fig:gamma002B2Supp}
 \end{subfigure}
       \caption{Simulation Setting 2: estimates for age-varying coefficients concerning with $\mathcal{P}$. The shaded areas represent the $95\%$ CI obtained using the empirical standard errors over 300 runs of simulation. }
      \label{fig:simLLB2}
\end{figure}

\begin{figure}[htbp]
 \begin{subfigure}[t]{.45\textwidth}
\includegraphics[scale=.25]
{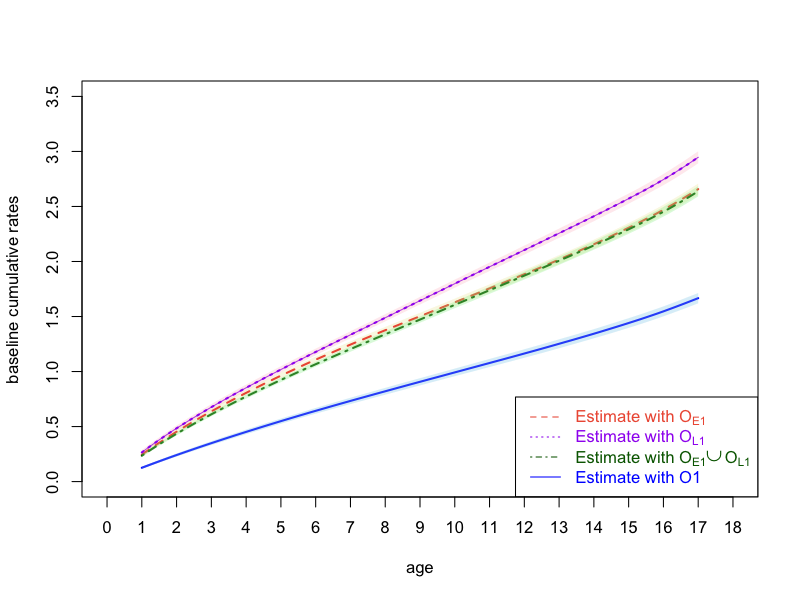}
\caption{\footnotesize concerning with $\mathcal{P}_1$}
   \label{fig:Lambda002A2}
 \end{subfigure}
 \hspace{.5cm}
 \begin{subfigure}[t]{.45\textwidth}
\includegraphics[scale=.25]
{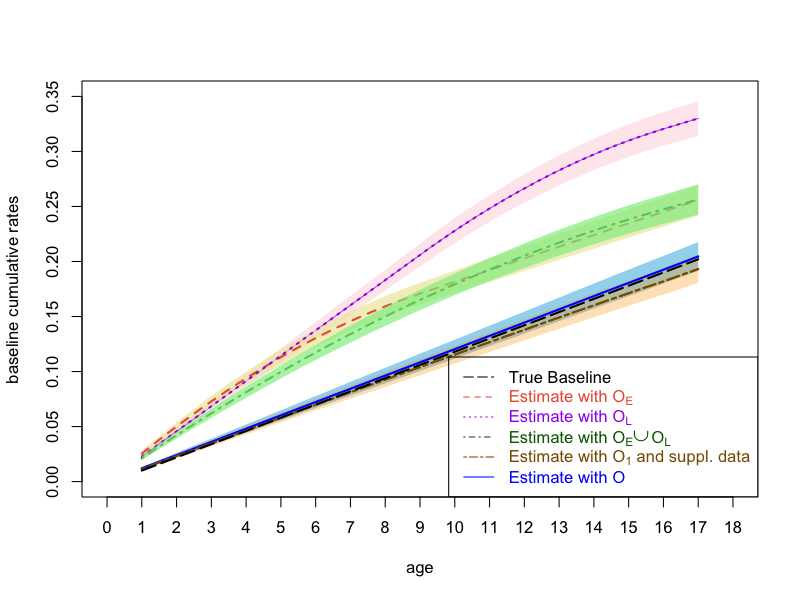}
\caption{\footnotesize concerning with $\mathcal{P}$ }
   \label{fig:Lambda002B2}
 \end{subfigure}
       \caption{Simulation Setting 2: estimates for $\Lambda_0(\cdot)$}
      \label{fig:simBaseline002AG2}
\end{figure}

%% file: archive20240712.bbl
\begin{thebibliography}{32}
\providecommand{\natexlab}[1]{#1}
\providecommand{\url}[1]{\texttt{#1}}
\expandafter\ifx\csname urlstyle\endcsname\relax
  \providecommand{\doi}[1]{doi: #1}\else
  \providecommand{\doi}{doi: \begingroup \urlstyle{rm}\Url}\fi

\bibitem[Andersen and Gill(1982)]{andersen1982cox}
P.~K. Andersen and R.~D. Gill.
\newblock Cox's regression model for counting processes: a large sample study.
\newblock \emph{The Annals of Statistics}, 10\penalty0 (4):\penalty0
  1100--1120, 1982.

\bibitem[Bitsko et~al.(2022)Bitsko, Claussen, Lichstein, Black, Jones,
  Danielson, Hoenig, Davis~Jack, Brody, Gyawali, Maenner, Warner, Holland,
  Perou, Crosby, Blumberg, Avenevoli, Kaminski, and Ghandour]{Bitsko}
R.~H. Bitsko, A.~H. Claussen, J.~Lichstein, L.~I. Black, S.~E. Jones, M.~L.
  Danielson, J.~H. Hoenig, S.~P. Davis~Jack, D.~J. Brody, S.~Gyawali, M.~J.
  Maenner, M.~Warner, K.~M. Holland, R.~Perou, A.~E. Crosby, S.~J. Blumberg,
  S.~Avenevoli, J.~W. Kaminski, and R.~M. Ghandour.
\newblock Mental health surveillance among children - united states,
  2013--2019.
\newblock \emph{Morbidity and Mortality Weekly Report}, 71\penalty0
  (Suppl2):\penalty0 1--42, 2022.

\bibitem[Breslow(1972)]{breslow1972discussion}
N.~E. Breslow.
\newblock Discussion of professor {C}ox's paper.
\newblock \emph{Journal of the Royal Statistical Society, Series B},
  34:\penalty0 216--217, 1972.

\bibitem[Burgun et~al.(2017)Burgun, Bernal-Delgado, Kuchinke, van Staa,
  Cunningham, Lettieri, Mazzali, Oksen, Estupiñan, Barone, and Chène]{Burgun}
A.~Burgun, E.~Bernal-Delgado, W.~Kuchinke, T.~van Staa, J.~Cunningham,
  E.~Lettieri, C.~Mazzali, D.~Oksen, F.~Estupiñan, A.~Barone, and G.~Chène.
\newblock Health data for public health: Towards new ways of combining data
  sources to support research efforts in {E}urope.
\newblock \emph{Yearbook of Medical Informatics}, 26\penalty0 (1):\penalty0
  235--240, 2017.
\newblock ISSN 0943-4747 (Print) 0943-4747.
\newblock \doi{10.15265/iy-2017-034}.

\bibitem[Cai and Sun(2003)]{cai2003local}
Z.~Cai and Y.~Sun.
\newblock Local linear estimation for time-dependent coefficients in {C}ox's
  regression models.
\newblock \emph{Scandinavian Journal of Statistics}, 30\penalty0 (1):\penalty0
  93--111, 2003.

\bibitem[Chevinsky et~al.(2021)Chevinsky, Tao, Lavery, Kukielka, Click, Malec,
  Kompaniyets, Bruce, Yusuf, Goodman, Dixon, Nakao, Datta, MacKenzie, Kadri,
  Saydah, Giovanni, and Gundlapalli]{Chevinsky}
J.~R. Chevinsky, G.~Tao, A.~M. Lavery, E.~A. Kukielka, E.~S. Click, D.~Malec,
  L.~Kompaniyets, B.~B. Bruce, H.~Yusuf, A.~B. Goodman, M.~G. Dixon, J.~H.
  Nakao, S.~D. Datta, W.~R. MacKenzie, S.~S. Kadri, S.~Saydah, J.~E. Giovanni,
  and A.~V. Gundlapalli.
\newblock Late conditions diagnosed 1-4 months following an initial coronavirus
  disease 2019 ({COVID-19}) encounter: A matched-cohort study using inpatient
  and outpatient administrative data-{U}nited {S}tates, 1 {M}arch-30 {J}une
  2020.
\newblock \emph{Clinical Infectious Diseases}, 73\penalty0 (Suppl 1):\penalty0
  S5--s16, 2021.
\newblock ISSN 1058-4838 (Print) 1058-4838.
\newblock \doi{10.1093/cid/ciab338}.

\bibitem[Cook and Lawless(2007)]{cook2007statistical}
R.~J. Cook and J.~F. Lawless.
\newblock \emph{The {S}tatistical {A}nalysis of {R}ecurrent {E}vents}.
\newblock New York: Springer, 2007.

\bibitem[Damon(2004)]{damon2004positive}
W.~Damon.
\newblock What is positive youth development?
\newblock \emph{The Annals of the American Academy of Political and Social
  Science}, 591\penalty0 (1):\penalty0 13--24, 2004.

\bibitem[Eddelbuettel and Fran\c{c}ois(2011)]{dirk}
D.~Eddelbuettel and R.~Fran\c{c}ois.
\newblock {Rcpp}: Seamless {R} and {C++} integration.
\newblock \emph{Journal of Statistical Software}, 40\penalty0 (8):\penalty0
  1--18, 2011.
\newblock \doi{10.18637/jss.v040.i08}.

\bibitem[Ferrante et~al.(2013)Ferrante, Lee, McCarthy, Fisher, Chen, Gonzalez,
  Love-Jackson, and Roetzheim]{Ferrante}
J.~M. Ferrante, J.~H. Lee, E.~P. McCarthy, K.~J. Fisher, R.~Chen, E.~C.
  Gonzalez, K.~Love-Jackson, and R.~G. Roetzheim.
\newblock Primary care utilization and colorectal cancer incidence and
  mortality among {M}edicare beneficiaries: a population-based, case-control
  study.
\newblock \emph{Annals of Internal Medicine}, 159\penalty0 (7):\penalty0
  437--446, 2013.
\newblock ISSN 0003-4819 (Print) 0003-4819.
\newblock \doi{10.7326/0003-4819-159-7-201310010-00003}.

\bibitem[Gao and Chan(2021)]{gao2021noniterative}
F.~Gao and K.~C.~G. Chan.
\newblock Noniterative adjustment to regression estimators with
  population-based auxiliary information for semiparametric models.
\newblock \emph{Biometrics}, 2021.
\newblock \mbox{doi}:\url{10.1111/biom.13585}.

\bibitem[Gavrielov-Yusim and Friger(2014)]{Gavrielov}
N.~Gavrielov-Yusim and M.~Friger.
\newblock Use of administrative medical databases in population-based research.
\newblock \emph{Journal of Epidemiology and Community Health}, 68\penalty0
  (3):\penalty0 283, 2014.
\newblock \doi{10.1136/jech-2013-202744}.
\newblock URL \url{http://jech.bmj.com/content/68/3/283.abstract}.

\bibitem[Hu and Lawless(1996)]{hu1996estimation}
X.~J. Hu and J.~F. Lawless.
\newblock Estimation of rate and mean functions from truncated recurrent event
  data.
\newblock \emph{Journal of the American Statistical Association}, 91\penalty0
  (433):\penalty0 300--310, 1996.

\bibitem[Hu and Rosychuk(2016)]{hu2016marginal}
X.~J. Hu and R.~J. Rosychuk.
\newblock Marginal regression analysis of recurrent events with coarsened
  censoring times.
\newblock \emph{Biometrics}, 72\penalty0 (4):\penalty0 1113--1122, 2016.

\bibitem[Janssen et~al.(1999)Janssen, Dechesne, and
  Van~Knippenberg]{janssen1999psychological}
J.~Janssen, M.~Dechesne, and A.~Van~Knippenberg.
\newblock The psychological importance of youth culture: A terror management
  approach.
\newblock \emph{Youth \& Society}, 31\penalty0 (2):\penalty0 152--167, 1999.

\bibitem[Lawless and Nadeau(1995)]{lawless1995some}
J.~F. Lawless and C.~Nadeau.
\newblock Some simple robust methods for the analysis of recurrent events.
\newblock \emph{Technometrics}, 37\penalty0 (2):\penalty0 158--168, 1995.

\bibitem[Li et~al.(2022)Li, Lu, Shu, Toh, and Wang]{li2022distributed}
D.~Li, W.~Lu, D.~Shu, S.~Toh, and R.~Wang.
\newblock Distributed {C}ox proportional hazards regression using summary-level
  information.
\newblock \emph{Biostatistics}, 2022.
\newblock \mbox{doi}:\url{10.1093/biostatistics/kxac006}.

\bibitem[Li and Liang(2008)]{li2008variable}
R.~Li and H.~Liang.
\newblock Variable selection in semiparametric regression modeling.
\newblock \emph{Annals of Statistics}, 36\penalty0 (1):\penalty0 261, 2008.

\bibitem[Lin et~al.(2000)Lin, Wei, Yang, and Ying]{lin2000semiparametric}
D.~Y. Lin, L.-J. Wei, I.~Yang, and Z.~Ying.
\newblock Semiparametric regression for the mean and rate functions of
  recurrent events.
\newblock \emph{Journal of the Royal Statistical Society: Series B (Statistical
  Methodology)}, 62\penalty0 (4):\penalty0 711--730, 2000.

\bibitem[Navaratnam et~al.(2021)Navaratnam, Gray, Day, Wendon, and
  Briggs]{Navaratnam}
A.~V. Navaratnam, W.~K. Gray, J.~Day, J.~Wendon, and T.~W.~R. Briggs.
\newblock Patient factors and temporal trends associated with {COVID-19}
  in-hospital mortality in {E}ngland: an observational study using
  administrative data.
\newblock \emph{Lancet Respiratory Medicine}, 9\penalty0 (4):\penalty0
  397--406, 2021.
\newblock ISSN 2213-2600 (Print) 2213-2600.
\newblock \doi{10.1016/s2213-2600(20)30579-8}.

\bibitem[Penfold et~al.(2018)Penfold, Burgess, Lee, Li, Miller, Nealon~Seibert,
  Semla, Mohr, Kazis, and Bauer]{Penfold}
R.~B. Penfold, J.~Burgess, J.~F., A.~F. Lee, M.~Li, C.~J. Miller,
  M.~Nealon~Seibert, T.~P. Semla, D.~C. Mohr, L.~E. Kazis, and M.~S. Bauer.
\newblock Space-time cluster analysis to detect innovative clinical practices:
  A case study of aripiprazole in the {D}epartment of {V}eterans {A}ffairs.
\newblock \emph{Health Services Research}, 53\penalty0 (1):\penalty0 214--235,
  2018.
\newblock ISSN 0017-9124 (Print) 0017-9124.
\newblock \doi{10.1111/1475-6773.12639}.

\bibitem[Pepe and Cai(1993)]{pepe1993some}
M.~S. Pepe and J.~Cai.
\newblock Some graphical displays and marginal regression analyses for
  recurrent failure times and time dependent covariates.
\newblock \emph{Journal of the American Statistical Association}, 88\penalty0
  (423):\penalty0 811--820, 1993.

\bibitem[Pietrosanu et~al.(2021)Pietrosanu, Rosychuk, and
  Hu]{pietrosanu2021handling}
M.~Pietrosanu, R.~J. Rosychuk, and X.~J. Hu.
\newblock Handling missing birthdates in marginal regression analysis with
  recurrent events.
\newblock \emph{Communications in Statistics-Simulation and Computation},
  50\penalty0 (1):\penalty0 142--152, 2021.

\bibitem[Qin et~al.(2011)Qin, Ning, Liu, and Shen]{qin2011maximum}
J.~Qin, J.~Ning, H.~Liu, and Y.~Shen.
\newblock Maximum likelihood estimations and em algorithms with length-biased
  data.
\newblock \emph{Journal of the American Statistical Association}, 106\penalty0
  (496):\penalty0 1434--1449, 2011.

\bibitem[Rosychuk et~al.(2020)Rosychuk, Bachman, Chen, and
  Hu]{rosychuk2020handling}
R.~J. Rosychuk, J.~W.~N. Bachman, A.~Chen, and X.~J. Hu.
\newblock Handling coarsened age information in the analysis of emergency
  department presentations.
\newblock \emph{BMC Medical Research Methodology}, 20\penalty0 (1):\penalty0
  1--11, 2020.

\bibitem[Shariff et~al.(2022)Shariff, Richard, Hwang, Kwong, Forchuk, Dosani,
  and Booth]{Shariff}
S.~Z. Shariff, L.~Richard, S.~W. Hwang, J.~C. Kwong, C.~Forchuk, N.~Dosani, and
  R.~Booth.
\newblock {COVID-19} vaccine coverage and factors associated with vaccine
  uptake among 23247 adults with a recent history of homelessness in {O}ntario,
  {C}anada: a population-based cohort study.
\newblock \emph{Lancet Public Health}, 7\penalty0 (4):\penalty0 e366--e377,
  2022.
\newblock \doi{10.1016/s2468-2667(22)00037-8}.

\bibitem[Shen et~al.(2009)Shen, Ning, and Qin]{shen2009analyzing}
Y.~Shen, J.~Ning, and J.~Qin.
\newblock Analyzing length-biased data with semiparametric transformation and
  accelerated failure time models.
\newblock \emph{Journal of the American Statistical Association}, 104\penalty0
  (487):\penalty0 1192--1202, 2009.

\bibitem[Tamariz et~al.(2012)Tamariz, Harkins, and Nair]{Tamariz}
L.~Tamariz, T.~Harkins, and V.~Nair.
\newblock A systematic review of validated methods for identifying ventricular
  arrhythmias using administrative and claims data.
\newblock \emph{Pharmacoepidemiology and Drug Safety}, 21\penalty0 (Suppl
  1):\penalty0 148--53, 2012.
\newblock ISSN 1053-8569.
\newblock \doi{10.1002/pds.2340}.

\bibitem[Tian et~al.(2005)Tian, Zucker, and Wei]{tian2005cox}
L.~Tian, D.~Zucker, and L.~Wei.
\newblock On the {C}ox model with time-varying regression coefficients.
\newblock \emph{Journal of the American Statistical Association}, 100\penalty0
  (469):\penalty0 172--183, 2005.

\bibitem[Wang(1996)]{wang1996hazards}
M.-C. Wang.
\newblock Hazards regression analysis for length-biased data.
\newblock \emph{Biometrika}, 83\penalty0 (2):\penalty0 343--354, 1996.

\bibitem[Zhang and Li(1996)]{zhang1996linear}
C.-H. Zhang and X.~Li.
\newblock Linear regression with doubly censored data.
\newblock \emph{The Annals of Statistics}, 24\penalty0 (6):\penalty0
  2720--2743, 1996.

\bibitem[Zheng et~al.(2021)Zheng, Zheng, and Hsu]{zheng2021risk}
J.~Zheng, Y.~Zheng, and L.~Hsu.
\newblock Risk projection for time-to-event outcome leveraging summary
  statistics with source individual-level data.
\newblock \emph{Journal of the American Statistical Association}, 2021.
\newblock \mbox{doi}:\url{10.1080/01621459.2021.1895810}.

\end{thebibliography}
